\documentclass[12pt,psfig]{article}

\usepackage{fancyheadings}
\usepackage{amsmath}
\usepackage{rotate}
\usepackage{rotating}
\usepackage{epsfig}
\usepackage{multirow}
\usepackage{graphicx}
\usepackage{color}
\usepackage{tabularx}

\def \simless {\mathbin{\lower 3pt\hbox{$\rlap{\raise 5pt
              \hbox{$\char'074$}}\mathchar"7218$}}}
\def \simgreat {\mathbin{\lower 3pt\hbox{$\rlap{\raise 5pt
              \hbox{$\char'076$}}\mathchar"7218$}}}

\newdimen\rotdimen
\def\vspec#1{\special{ps:#1}}
\def\rotstart#1{\vspec{gsave currentpoint currentpoint translate
   #1 neg exch neg exch translate}}
\def\rotfinish{\vspec{currentpoint grestore moveto}}
\def\rotr#1{\rotdimen=\ht#1\advance\rotdimen by\dp#1%
   \hbox to\rotdimen{\hskip\ht#1\vbox to\wd#1{\rotstart{90 rotate}%
   \box#1\vss}\hss}\rotfinish}
\def\rotl#1{\rotdimen=\ht#1\advance\rotdimen by\dp#1%
   \hbox to\rotdimen{\vbox to\wd#1{\vskip\wd#1\rotstart{270 rotate}%
   \box#1\vss}\hss}\rotfinish}%
\def\rotu#1{\rotdimen=\ht#1\advance\rotdimen by\dp#1%
   \hbox to\wd#1{\hskip\wd#1\vbox to\rotdimen{\vskip\rotdimen
   \rotstart{-1 dup scale}\box#1\vss}\hss}\rotfinish}%
\def\rotf#1{\hbox to\wd#1{\hskip\wd#1\rotstart{-1 1 scale}%
   \box#1\hss}\rotfinish}%

\begin{document}

\newpage
\setlength{\topmargin}{-30mm} \setlength{\footskip}{-0mm}
\thispagestyle{empty} \vspace*{-0.0in} \centerline{\normalsize
April 15,
2003~~~~~~~~~~~~~~~~~~~~~~~~~~~~~~~~~~~~~~~~~~~~~~~~~~~~~~~~~~~~~~~~~~~BNL-71228-2003-IR}
\vspace{0.25in}
 \Huge \centerline{The AGS-Based Super Neutrino Beam Facility}
 \centerline{The BNL Neutrino
Working Group Report-II} \vspace*{0.1in} \large
\centerline{Coordinators: M. Diwan, W. Marciano, W. Weng }
\centerline{Editor: D. Raparia} \vspace{0.1in}
\input epsf
\begin{figure}[htbp]
\begin{center}
\centerline{\includegraphics[width=4.92in,height=2.385in]{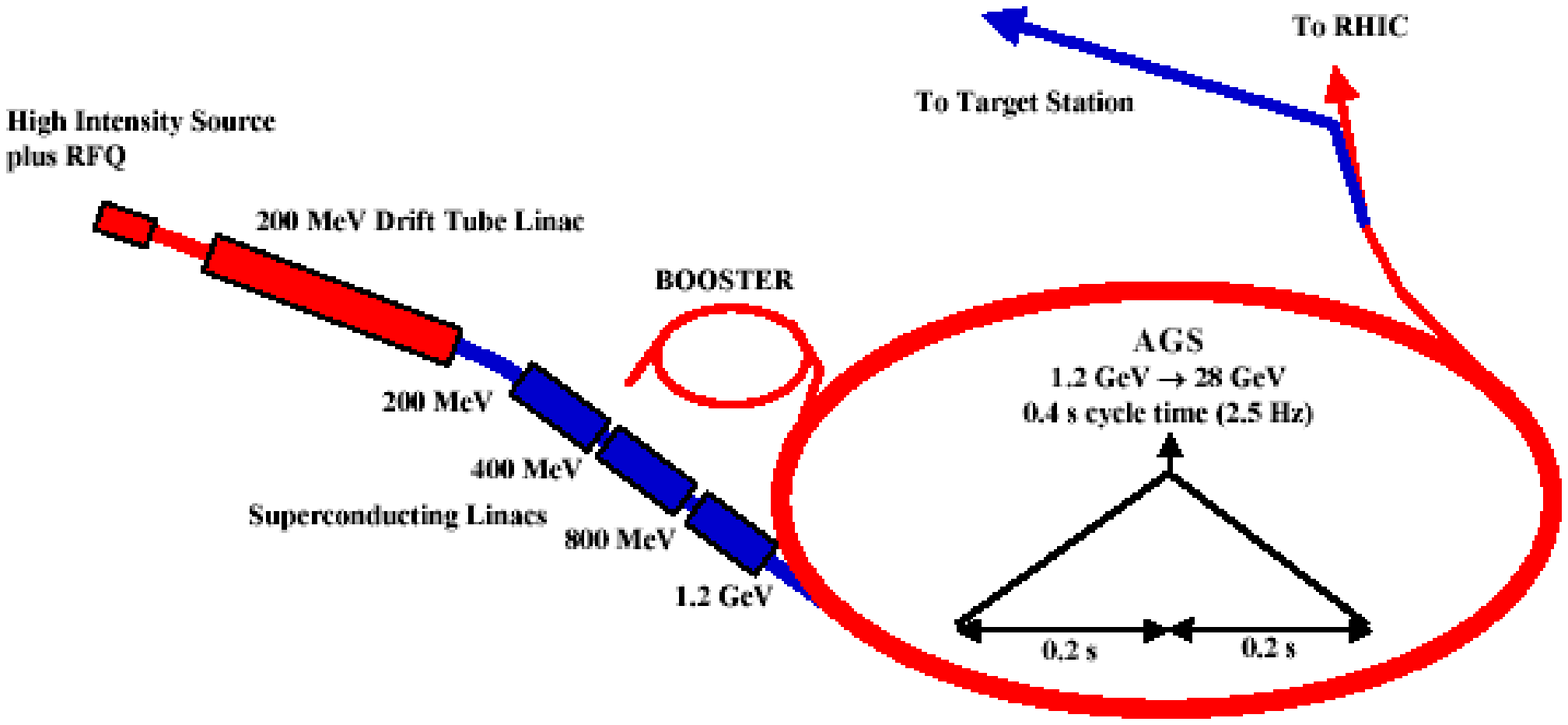}}
\end{center}
\end{figure}

\begin{figure}[htbp]
\centerline{\includegraphics[width=3.9in,height=2.7in]{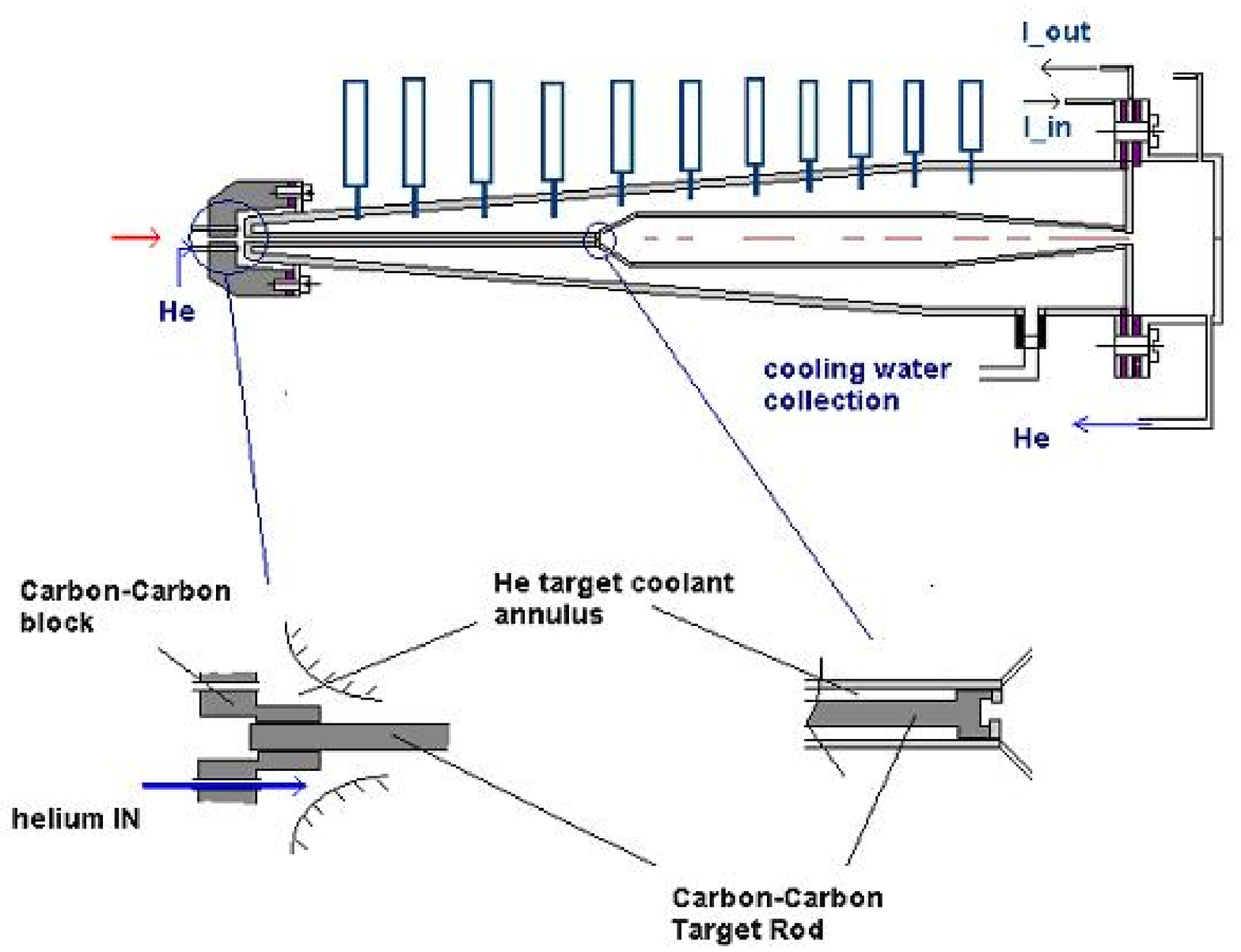}}
\end{figure}
\vspace{0.0in}
 \large \centerline{Brookhaven National Laboratory}
\normalsize \centerline{April 15, 2003}

\newpage
\thispagestyle{empty}\setlength{\topmargin}{-15mm}\Large \sf
\vspace*{-0.5in} \centerline{\normalsize April 15,
2003~~~~~~~~~~~~~~~~~~~~~~~~~~~~~~~~~~~~~~~~~~~~~~~~~~~~~~~~~~~~~~~~~~~BNL-71228-2003-IR}
\vspace{0.25in}
 \centerline{\bf AGS Super Neutrino Beam Facility}
 \centerline{\bf Accelerator and Target
System Design}
 \vspace{0.1in}
 \centerline{\bf  (Neutrino Working Group Report-II)}
  \vspace{0.1in} \normalsize \centerline{Coordinators: M. Diwan, W. Marciano, W. T. Weng }
  \centerline{Editor: D. Raparia}
   \vspace{0.1in} \centerline{Contributors and Participants}
\vspace{0.1in}
 \centerline{J. Alessi,  D. Barton, D. Beavis, S. Bellavia, J.
Brennan, B. Bromley, M. C. Chen, M. Diwan,  }\centerline{R.
Fernow, J. Gallardo, R. Hahn, S.Kahn, H. Kirk, Y. Y. Lee, D.
Lowenstein, H. Ludewig,  }\centerline{ W. Marciano, I. Marneris,
R. Palmer, Z. Parsa, A. Pendzick, C. Pearson, D. Raparia,
}\centerline{T. Roser, A. Ruggiero, J. Sandberg, N. P. Samios, C.
Scarlet, N. Simos, N. Tsoupas, }\centerline{ J. Tuozzolo, J.
Beebe-Wang, W. T. Weng, B. Viren, P. Yamin, M. Yeh, Wu Zhang}
\vspace{0.1in}\centerline{\it Brookhaven National Laboratory, P.
O. Box 5000, Upton, NY 11973-5000} \vspace{0.2in}\centerline{ W.
Frati, J. R.Klein, K. Lande, A. K. Mann, R. Van Berg, and P.
Wildenhain}\vspace{0.1in} \centerline{ \it University of
Pennsylvania Philadelphia, PA 19104-6396} \vspace{0.2in}
\centerline{R. Corey}\vspace{0.1in}\centerline{ \it South Dakota
School of Mines and Technology, Rapid City, S. D.
57701}\vspace{0.2in}\centerline{D. B. Cline, K. Lee, B. Lisowski,
P. F. Smith}\vspace{0.1in}\centerline{\it Department of Physics
and Astronomy, University of California, Los Angeles, CA
90095}\vspace{0.2in}\centerline{I. Mocioiu, R.
Shrock}\vspace{0.1in}\centerline{ \it C. N. Yang Institute for
Theoretical Physics State University of New York,}\centerline{
Stony Brook, NY 11974} \vspace{0.2in}\centerline{C. Lu and K. T.
McDonald}\vspace{0.1in} \centerline{\it Joseph Henrry
Laboratories, Princeton University, Princeton, NJ 08544,
USA}\vspace{0.2in}\centerline{ R.
Potenza}\vspace{0.1in}\centerline{ \it Instituto Nazionale di
Fisica Nucleare,Dipartimento de Fisica e
Astronomia,}\centerline{\it  Universita di Catania, 64, Via s.
Sofia, I-95123 Catania, Italy}\vspace{0.2in}\centerline{ G.
Evangelakis}\vspace{0.1in}\centerline{\it Physics Department,
University of Ioannina,  Greece}

\newpage
\thispagestyle{empty} \setlength{\topmargin}{-0mm} \normalsize \sf

\noindent This document contains figures in color. The figures
should be viewed in color.

\noindent This work was performed under the auspices of the U. S.
Department of Energy, Contract No. DE-AC02-98CH10886.

\setlength{\topmargin}{-5mm}

\rm \normalsize \pagestyle{fancy}
\newpage
\lhead{AGS Super Neutrino Beam Facility} \rhead{Preface}
\rfoot{April 15, 2003}

{\noindent\Large\bf Preface}
\vspace*{1cm}

This document describes the design of the accelerator  and target
systems  for the AGS  Super Neutrino Beam Facility. Under the
direction of the Associate Laboratory Director Tom Kirk, BNL has
established a Neutrino Working Group to explore the scientific
case and facility requirements for a very long baseline neutrino
experiment. Results of a study of the physics merit and detector
performance was published in BNL-69395 in October
2002\cite{ref:prefone}\cite{ref:preftwo}, where it was shown that
a wide-band neutrino beam generated by a 1 MW proton beam from the
AGS, coupled with a half megaton water Cerenkov detector located
deep underground in the former Homestake mine in South Dakota
would be able to measure the complete set of neutrino oscillation
parameters:
\begin{itemize}
\item precise determination of the oscillation
parameters $\Delta $m$_{32}^{2}$ and sin$^{2}$2$\theta_{32}$

\item detection of the oscillation of $\nu_{\mu}
 \to   \nu_{e}$ and measurement of sin$^{2}$ 2$\theta_{13}$

\item measurement of $\Delta $m$_{21}^{2}$
sin2$\theta_{12}$ in a $\nu_{\mu } \to  \nu _{e}$ appearance mode,
independent of the value of $\theta_{13}$

\item verification of matter enhancement and the
sign of $\Delta $m$_{32}^{2}$

\item determination of the CP-violation parameter
$\delta _{CP}$ in the neutrino sector
\end{itemize}

    This report details the performance requirements and
 conceptual design of the accelerator and the target systems for
the production of a neutrino beam by a 1.0 MW proton beam from the
AGS. The major components of this facility include a new 1.2 GeV
superconducting linac,  ramping the AGS at 2.5 Hz, and the new
target station for 1.0 MW beam. It also calls for moderate
increase, about 30\%, of the AGS intensity per pulse. Special care
is taken to account for all sources of proton beam loss plus
shielding and collimation of  stray beam halo particles to ensure
equipment reliability and personal safety. A preliminary cost
estimate and schedule for the accelerator upgrade and target
system are also included.

\newpage
\lhead{AGS Super Neutrino Beam Facility} \rhead{Contents}
\rfoot{April 15, 2003}

\tableofcontents

\clearpage
\newpage
\lhead{AGS Super Neutrino Beam Facility} \rhead{Introduction}
\rfoot{April 15, 2003}
\renewcommand{\theequation}{\thesection.\arabic{equation}}
\renewcommand{\thefigure}{\thesection.\arabic{figure}}
\renewcommand{\thetable}{\thesection.\arabic{table}}

\section{Introduction and Accelerator Performance}

\label{sec:int}

After more than 40~years of operation, the AGS is still at the
heart of the Brookhaven hadron accelerator complex. This system of
accelerators presently comprises a 200~MeV linac for the
pre-acceleration of high intensity and polarized protons, two
Tandem Van der Graaffs for the pre-acceleration of heavy ion
beams, a versatile Booster that allows for efficient injection of
all three types of beams into the AGS and, most recently, the two
RHIC collider rings that produce high luminosity heavy ion and
polarized proton collisions. For several years now, the AGS has
held  the world intensity record with more than $7\times 10^{13}$
protons accelerated in a single pulse \cite{ref:intone}.

\begin{figure}[h]
\begin{center}
\epsfxsize=5.0in \epsfysize=3.0in
\centerline{\includegraphics[width=4.84in,height=2.7in]{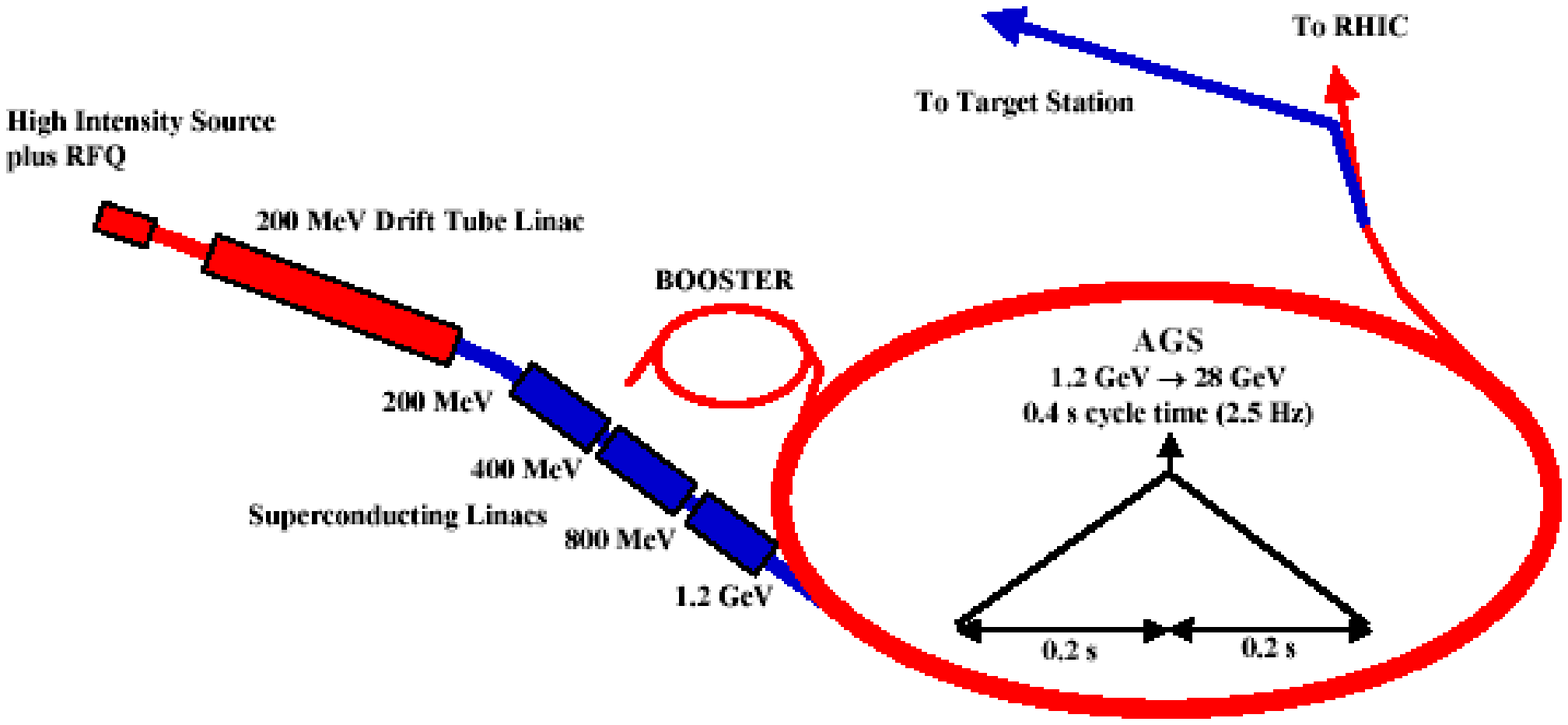}}
\end{center}
\caption{\label{fig:agspdlayout}AGS proton driver layout.}
\end{figure}

The requirements for the proton beam for the super neutrino beam
 are summarized in
Table ~\ref{tab:intone} and a layout of the upgraded AGS is shown
in Figure ~\ref{fig:agspdlayout}. Since the present number of
protons per fill is already close to the required number, the
upgrade focuses on increasing the repetition rate and reducing
beam losses (to avoid excessive shielding requirements and to
maintain activation of the machine components at workable level).
It is also important to preserve all the present capabilities of
the AGS, in particular its role as injector to RHIC.

\begin{table}[h]
\begin{center}
\caption{\label{tab:intone}Performance of the present and upgrade
AGS.}
\begin{tabular}{|@{}l|l|l|}
\hline & Present & Upgrade \\
 \hline Average Beam Power & 0.14 MW&1.0 MW \\
 \hline Beam Energy & 24 GeV & 28 GeV \\
 \hline Average Beam Current & 6 $\mu$A & 36 $\mu$A \\
 \hline Cycle Time & 2 sec&400 ms \\
\hline Number of Protons per Fill &$7.0\times 10^{13}$& $8.9\times
10^{13}$ \\
 \hline Number of Bunches per Fill & 12 & 23 \\
 \hline Protons per Bunch & $5.8\times 10^{12}$& $3.72\times 10^{12}$ \\
 \hline Number of Injected Turns & 190& 240 \\
 \hline Repetition Rate &0.5 Hz & 2.5 Hz \\
\hline Pulse Length & 0.35 ms& 0.72 ms \\
 \hline Chopping Rate &0.75 & 0.75 \\
\hline Linac Average/Peak Current &26/35 mA & 21/28 mA \\
 \hline
\end{tabular}
\end{center}
\end{table}

The AGS Booster was built not only to allow the injection of any
species of heavy ion into the AGS but to allow a fourfold increase
of the AGS intensity. It is one-quarter the circumference  of the
AGS with the same aperture. However, the accumulation of four
Booster loads in the AGS takes about 0.6 sec, and is therefore not
well suited for high average beam power operation. To minimize the
injection time to about 1 msec, a 1.2 GeV linac will be used
instead. This linac is consists the existing warm linac of 200 MeV
and a new superconducting linac of 1.0 GeV. The multi-turn H$^{-}$
injection from a source of 30 mA and 720 $\mu$ sec pulse width is
sufficient to accumulate 9 x 10$^{13}$ particle per pulse in the
AGS\cite{ref:inttwo}.

\input epsf

 The minimum ramp
time of the AGS to full energy is presently 0.5 s; this must be
upgraded to 0.2 s to reach the required repetition rate of 2.5~Hz.
The required upgrade of the AGS power supply, The RF system, and
other rate dependent accelerator issues will be discussed in
Chapter ~\ref{sec:agsup}.

\begin{figure}[htbp]
\centerline{\includegraphics[width=3.9in,height=2.7in]{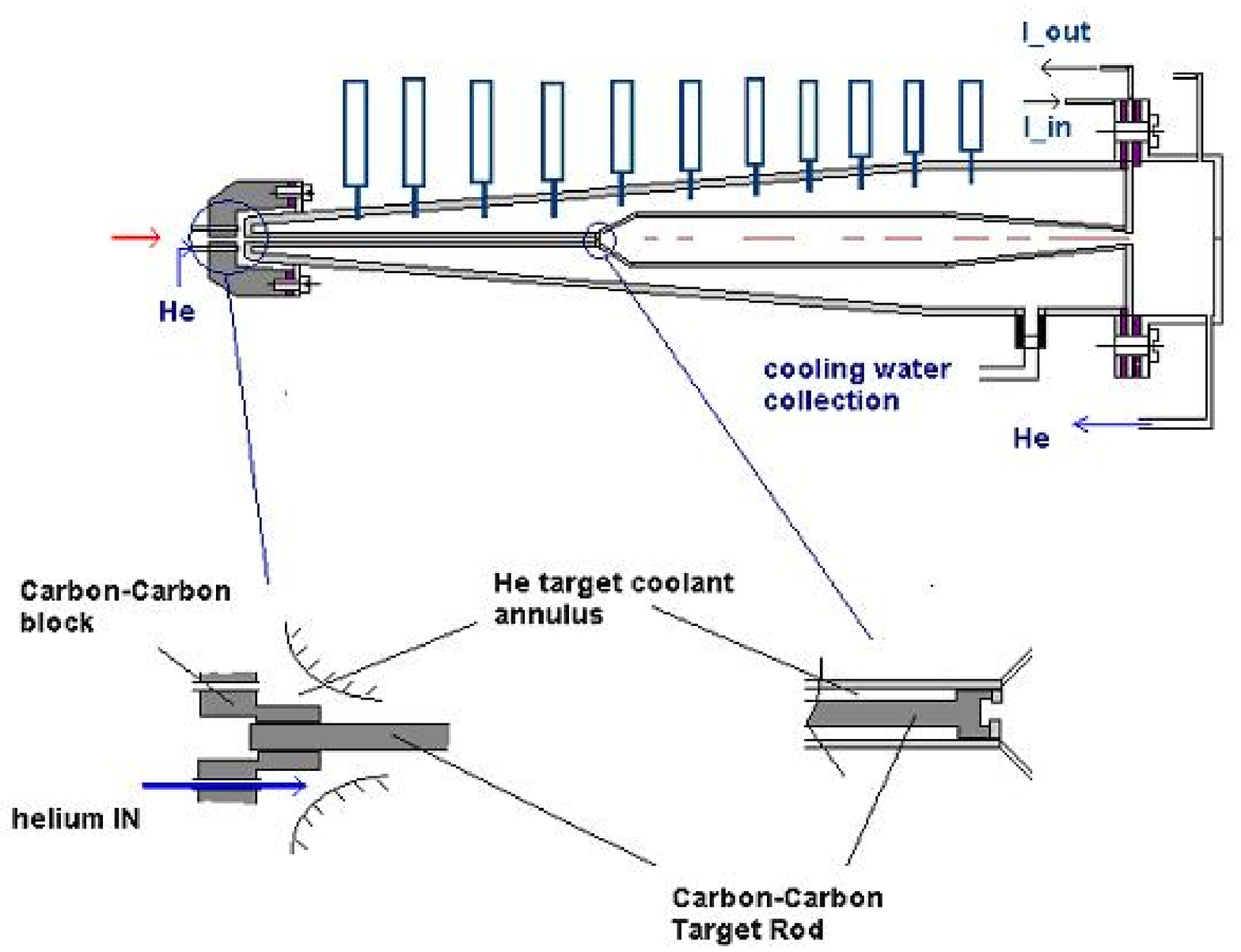}}
\caption{ Graphite target and horn configuration.}
\label{fig:intfour}
\end{figure}

The design of the target/horn configuration is shown in Figure
~\ref{fig:intfour} The material selected for the proton target is
a Carbon-Carbon composite. It is a 3-dimensional woven material
that exhibits extremely low thermal expansion for temperatures up
to 1000$^{0}$C; for higher temperatures it responds like graphite.
This property is important for greatly reducing the thermo-elastic
stresses induced by the beam, thereby extending the life of the
target.

The target consists of an 80 cm long cylindrical rod of 12 mm
diameter. The target intercepts a 2 mm rms proton beam of
10$^{14}$ protons/pulse. The total energy deposited as heat in the
target is 7.3 kJ with peak temperature rise of about 280$^{0}$C.
Heat will be removed from the target through forced convection of
helium gas across its outside surface.

\begin{figure}[htbp]
\centerline{\includegraphics[width=4.92in,height=2.9in]{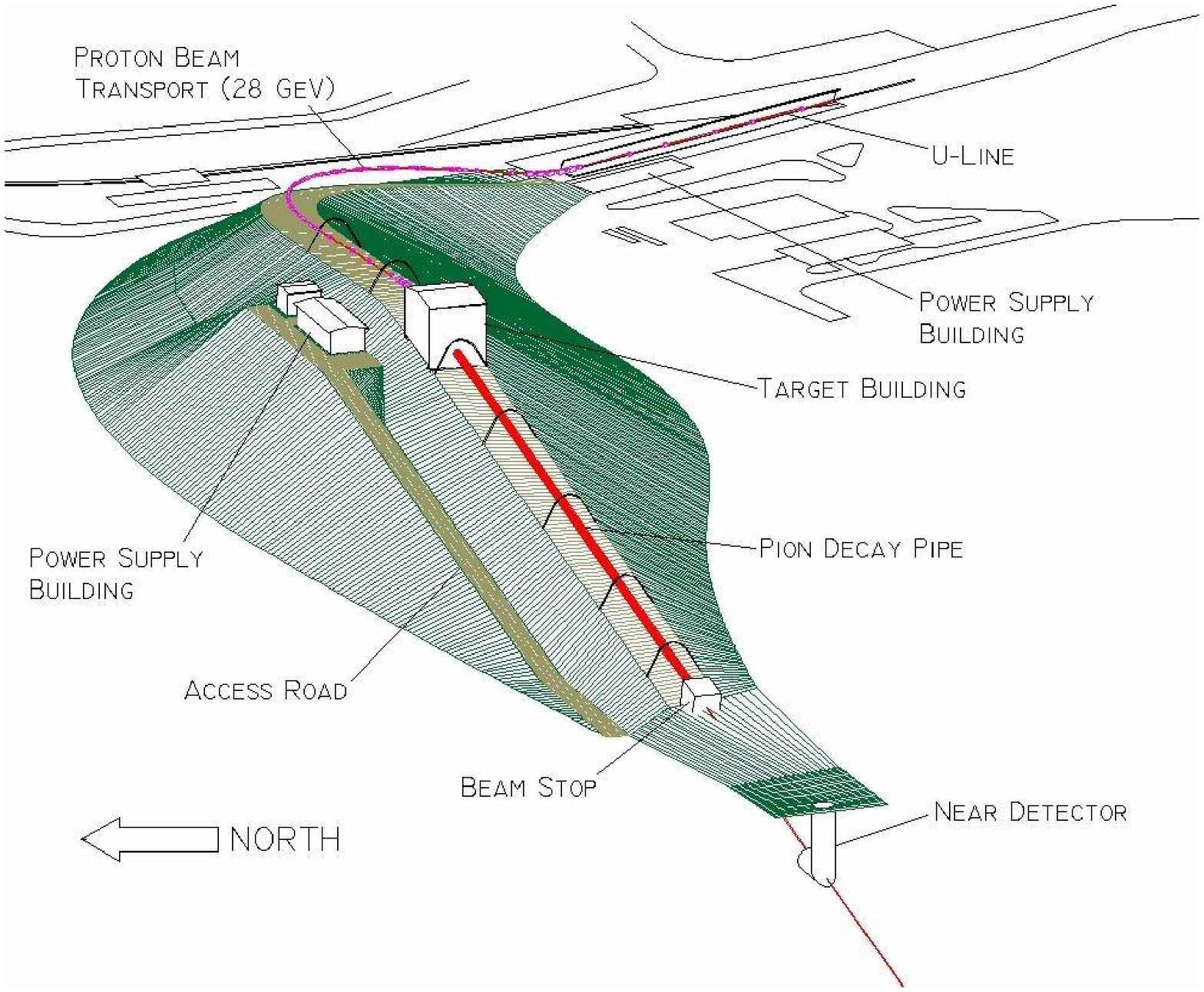}}
\caption{3-dimensional view of the neutrino beamline. The beamline
is shown without shielding on top of the beam-line magnets and the
decay tunnel.} \label{fig:intsix}
\end{figure}

The extracted proton beam uses an existing beamline at the AGS,
but is then directed to a target station atop a constructed
earthen hill. The target is followed by a downward sloping pion
decay channel. This vertical arrangement keeps the target and
decay pipe well above the water table in this area. The 11 degrees
slope aims the neutrino beam at a water Cerenkov  neutrino
detector to be located in the Homestake mine at Lead, South
Dakota. A plan view of the AGS facility is shown in Figure
~\ref{fig:ultbt}. A 3-dimensional view of the neutrino beam is
provided in Figure ~\ref{fig:intsix}.

To assist the reading of this report and enforce consistency
across chapters, the design parameters of each subsystem are
provided in  Appendix A.

As explained in Chapter ~\ref{sec:costsched}, the resultant total
direct cost of the 1 MW AGS Super Neutrino Beam Facility, not
including both near and main detectors, is  \$218.5 M. The
preliminary total estimated cost (TEC) is {\$}369 M in FY03
dollars, including EDIA {\@} 15{\%}; contingency {\@} 30{\%}; BNL
project overhead {\@} 13{\%}. Escalation cannot be estimated
without a project start year.

 It is estimated that
three years of R \& D are needed to build prototypes and complete
a detailed  engineering design that will reduce cost and improve
operational reliability. This will be followed by 4.5 years of
construction and 0.5 year of commissioning to prepare the facility
 for physics research operations.

\clearpage
\newpage
\lhead{AGS Super Neutrino Beam Facility} \rhead{Injector Linac }
\rfoot{April 15, 2003}

\section{Injector Linac}
\label{sec:linacup}
\setcounter{equation}{0}
 \setcounter{table}{0}
 \setcounter{figure}{0}

To provide 1 MW proton beam power, the AGS has to operate at 2.5
Hz with little time allowed for injection. The present injector
consists of the 200 MeV room temperature linac and 1.5 GeV
Booster. It takes about 0.6 seconds to inject four Booster pulses
to fill the AGS, which is not suitable for the upgrade operation.
If a 1.2 GeV linac is used instead, the injection time can be
reduced to less than 1 ms, allowing the AGS to cycle at the
desired rate of 2.5 Hz. A 1.2 GeV linac injection can
simultaneously fulfill the requirements of keeping the space
charge tune shift in the AGS to be less than 0.25 and the
injection losses down for reliable operation.

The distance between the exit of the 200 MeV linac and the AGS
injection point is about 120 m. Only a superconducting linac (SCL)
with sufficiently high gradient can meet the requirement of
acceleration to 1.2 GeV within that distance. The superconducting
linac technology has been used in many electron accelerators, such
as LEP, CEBAF, and Tesla. The SNS project has successfully
designed a 1.0 GeV proton SCL system with an accelerating gradient
of about 18 MeV/m. The design of the new AGS injector linac
follows closely that developed at SNS. This will substantially
reduce the design cost and increase confidence in the design.

 The project described corresponds to an average SCL
beam current of 37.6 $\mu $A, that yields the required average
beam power of 1 MW at the top energy of 28 GeV, including also a
controlled beam loss of about 5{\%} during multi-turn injection
into the AGS. The average beam power in exit is 45 kW,
considerably less than the 1-MW level of the equivalent 1.0-GeV
SCL for the Spallation Neutron Source (SNS) \cite{snsdr}. Thus the
concern about component activation by the induced radiation from
uncontrolled beam losses is greatly reduced. The repetition rate
of 2.5  beam pulses per second gives a beam intensity of 0.89 x
10$^{14}$ protons accelerated per AGS cycle; this is about 30{\%}
higher than the intensity routinely obtained with the present
injector. At the end of an injection phase that takes about 240
turns, the space-charge tune depression is $\Delta \nu $ = 0.2,
assuming a bunching factor (the ratio of beam peak current to
average current), of 3. Also, with the normalized beam emittance
of 100 $\pi $ mm-mrad, the actual beam emittance at 1.2 GeV is
$\varepsilon $ = 50 $\pi $ mm-mrad. Obviously, the effective
vertical acceptance of the AGS at injection is to match the final
beam emittance value. The SCL beam pulse length is 0.72 ms, and
the beam duty cycle 0.18{\%}.
\input epsf
\begin{figure}[h]
\begin{center}
\resizebox{6.0in}{3.0in}{\includegraphics*[0.0in,0.00in][8.0in,3.0in]{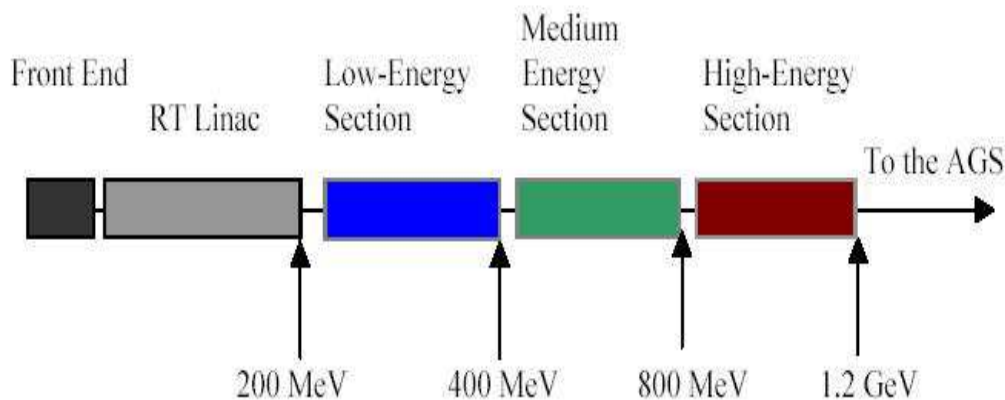}}
\end{center}
\caption{\label{fig:scllayout}Layout of the 1.2-GeV
Superconducting linac injector for the AGS.}
\end{figure}

\begin{sidewaysfigure}[htbp]
\centerline{\includegraphics[width=7.84in,height=3.96in]{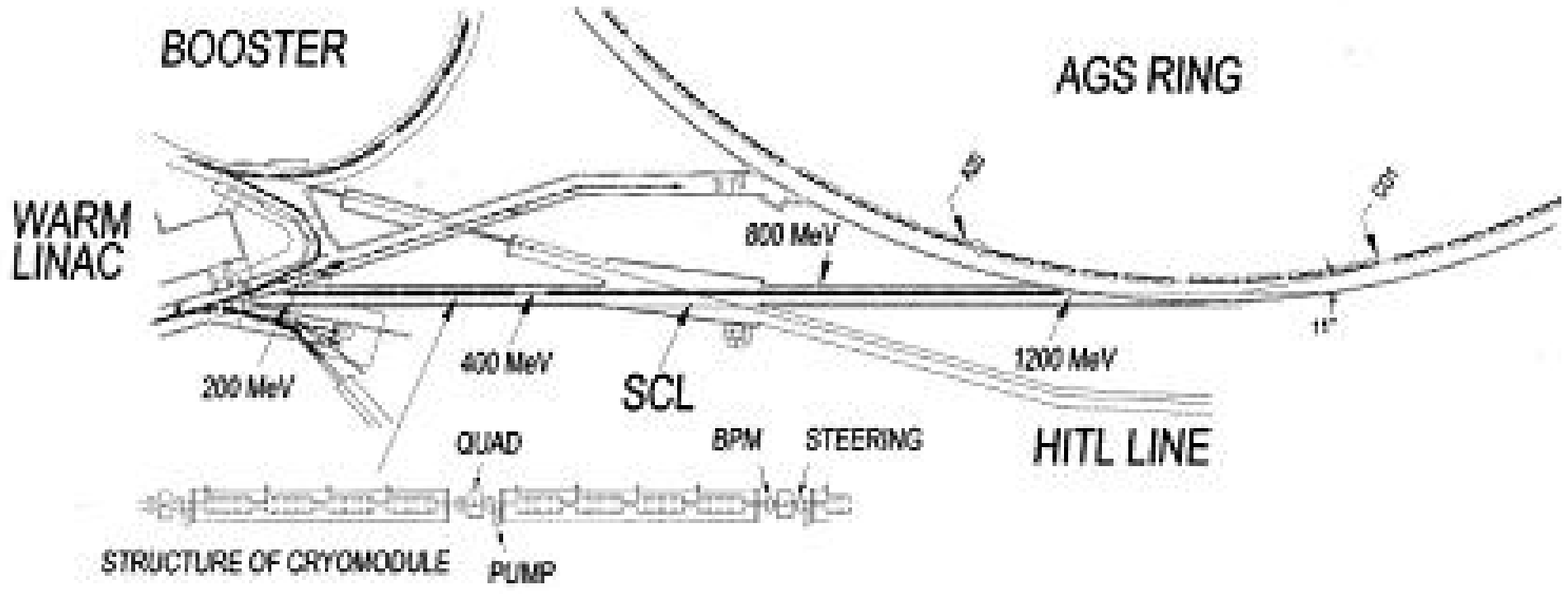}}
\caption{ Layout of the superconducting linac. }
\label{fig:sec3_one}
\end{sidewaysfigure}

The preliminary design of the SCL consists of three parts: (i)
Low-Energy (LE), (ii) Medium Energy (ME), and (iii) High Energy
(HE). A schematic view of the new injector is given in Figure
~\ref{fig:scllayout}. The actual location of the SCL on the BNL
site is shown in Figure ~\ref{fig:sec3_one}. The beam leaves the
present room temperature linac at the energy of 200 MeV and, after
a bend of 17.5 degrees, enters a new 120 m long tunnel, where the
SCL  is located, and joins the AGS beamline at the location of
magnet C01. The design parameters of the SCL and the AGS are given
in Table ~\ref{tab:iagsp}.

\begin{table}[htbp]
\begin{center}
\caption{\label{tab:iagsp}Injector and AGS parameters for 1-MW
upgrade (28 GeV).}
\begin{tabular}
{|p{162pt}|p{81pt}|} \hline Increm. Linac Ave. Power, kW& 37.5 \\
\hline \hline Kinetic Energy, GeV& 1.2 \\ \hline $\beta $& 0.8986
\\ \hline Momentum, GeV/c& 1.92 \\ \hline Magnetic Rigidity, T-m&
6.41 \\ \hline Repetition Rate, Hz& 2.5 \\  \hline Linac No. of
Protons / pulse& 9.38 x 10$^{13}$ \\\hline Linac Duty Cycle, {\%}&
0.179 \\ \hline AGS Circumference, m& 807.076 \\ \hline Revol.
Frequency, MHz& 0.3338 \\ \hline Revolution Period, $\mu $s& 2.996
\\ \hline Bending Radius, m& 85.378 \\ \hline Injection Field, kG&
0.7507 \\  \hline Injection Loss, {\%}& 5.0
\\ \hline Injected Protons per Turn& 3.74 x 10$^{11}$ \\ \hline
  Norm. Emitt., $\pi $
mm-mrad& 100 \\ \hline Emittance, $\pi $ mm-mrad& 48.8 \\ \hline
Space-Charge $\Delta \nu $& 0.187 \\ \hline
\end{tabular}
\end{center}
\end{table}

\lhead{AGS Super Neutrino Beam Facility} \rhead{Room Temperature
Linac Upgrade} \rfoot{April 15, 2003}
\subsection{Room Temperature Linac Upgrade}
\label{sec:rtlu}

The Brookhaven 200 MeV H- linac typically operates $\sim $ 5000
hours/year in support of high intensity proton operation of the
AGS, polarized protons for RHIC, and medical isotope production.
Some nominal operating parameters of the linac are given in Table
~\ref{tab:nopbnllinac}. A detailed description of the linac can be
found in references \cite{bnllinac} and \cite{bnllinacupgrade}.
One can see from this table that the present linac can meet the
requirements for beam current and repetition rate, and the
required duty factor is less than the typical operating value. As
will be discussed below, upgraded power supplies will be required
for several systems to achieve the desired beam pulse width.
However, since this upgrade is straightforward, the linac
operation remains reliable, and there is room following this linac
for the addition of a SCL; it was most cost effective to continue
to use the full 200 MeV warm linac.

\begin{table}[htbp]
\begin{center}
\caption{ Nominal operating parameters for the BNL linac.}
\begin{tabular}
{|p{194pt}|p{135pt}|} \hline Output Energy& 200 MeV \\ \hline
Frequency& 201.25 MHz \\ \hline Repetition Rate& $ \le \quad 10$Hz
\\ \hline Beam Pulse Width& $ \le \quad 500\mu $s \\ \hline
Nominal Duty Factor& $\sim \quad 0.3\% $ \\ \hline Output Beam
Current& $ \le \quad 35$ mA \\ \hline Output Emittance (rms,
Normalized)& 2 $\pi $ mm mrad \\ \hline Output Energy Spread
(rms)& 0.2\%
 \\
\hline
\end{tabular}
\label{tab:nopbnllinac}
\end{center}
\end{table}

\begin{figure}[h]
\begin{center}
\resizebox{6.0in}{3.0in}{\includegraphics*[0.0in,0.00in][8.0in,3.0in]{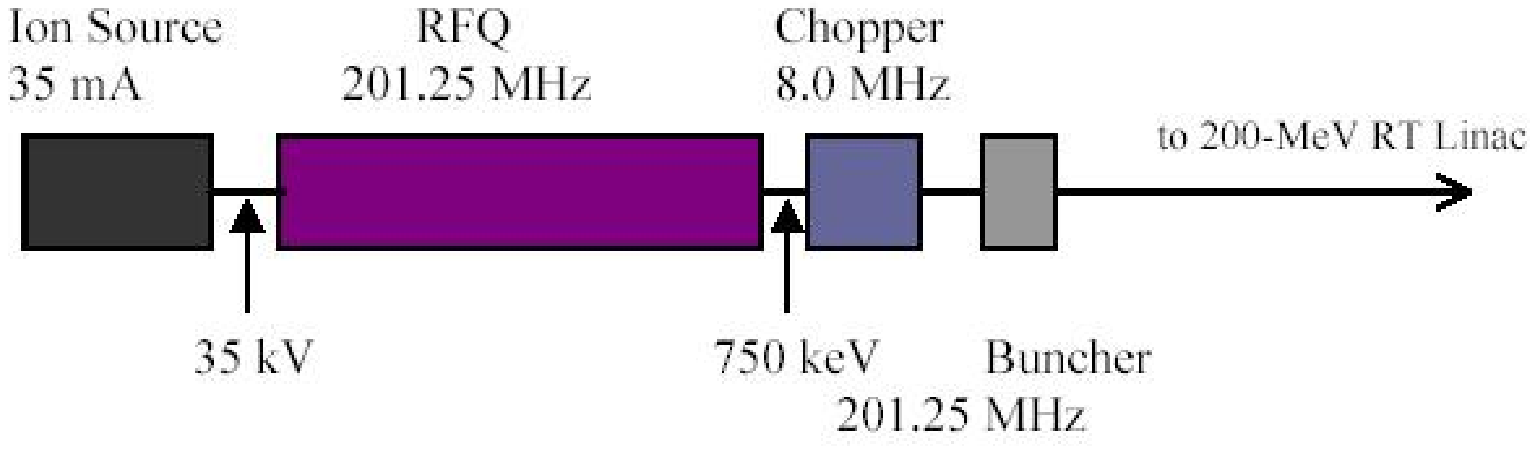}}
\end{center}
\caption{\label{fig:felayout}Layout of the Front-End.}
\end{figure}

The front end of the linac is shown schematically in Figure
~\ref{fig:felayout}. It starts with a magnetron H- ion source,
which produces in excess of 80 mA H$^{-}$ beam at 35 keV.
Following this, two magnetic solenoids are used to transport the
H$^{-}$ beam $\sim $ 1 m and match it into an RFQ. The RFQ,
operating at the linac frequency of 201.25 MHz, accelerates the
beam from 35 keV to 750 keV. After the RFQ, the beam is
transported $\sim $ 6 m to the linac. This 750 keV transport
includes 10 quadrupoles and 3 bunchers for beam matching into the
linac, and a fast beam chopper which allows beam chopping with
$\sim $ 10 ns rise and fall times. This chopper is a travelling
wave structure, and the beam is chopped at a frequency to match
into the Booster RF accelerating bucket at injection energy. A
201.25 MHz drift tube linac (DTL) accelerates the beam from 750
keV to 200 MeV. This linac has 9 cavities, each powered by a 5 MW
peak power RF system. There are a total of 286 drift tubes, with a
focusing electromagnetic quadrupole in each drift tube.

\input epsf

As mentioned above, except for the beam pulse width, the present
operation fulfills all requirements. In the future, we may
consider the advantages of eliminating the long transport line
from the RFQ to linac, in order to reduce the linac output
emittance, but the present design assumes that this line is
unchanged. Limitations in beam pulse width exist not due to
mechanical limits on the ion source or linac, but rather as a
result of limits in ion source power supplies, RF system power
supplies, and pulsed transport line and tank quadrupole power
supplies. Therefore, in order to serve as the injector to the SCL,
the following improvements will be required:

  \begin{enumerate}

 \item The magnetron ion source discharge, extractor, and pulsed gas power supplies
must be replaced by components with wider pulse capability.

\item On the 400 kW driver stage amplifier RF systems, one must increase the 4616
plate capacitor bank. This entails the purchase and installation
of new 35 kV capacitors in the existing frame. The the crowbar
ignitrons and sockets must be replaced. Finally, we need to
increase the size of the capacitors on grid power supplies.

\item In the RF modulator system, the capacitor bank has to be increased for the
4cw25000 anode supply, and the 8618 grid and cathode deck power
supplies will need to be reworked.

\item On the 5 MW RF system, we need to replace the 60 kV capacitor banks with
banks having more capacity;  again these can be installed within
existing frames. The crowbar ignitrons and sockets have to be
replaced for 100 kA tubes.

\item On the low level RF systems, all 400 watt solid state amplifiers will need
replacement.

\item All pulsed transport line and tank quadrupoles will have to be replaced with
solid state units having wider pulse capabilities.

  \end{enumerate}

The above upgrades are all straightforward, and we are confident
that the required performance can be achieved.

\lhead{AGS Super Neutrino Beam Facility} \rhead{The
Superconducting Linac(SCL)} \rfoot{April 15, 2003}

\subsection{The Superconducting Linac(SCL)}
\label{sec:tscl}
\begin{figure}[h]
\begin{center}
\resizebox{6.0in}{3.5in}{\includegraphics*[0.5in,0.0in][8.0in,5.5in]{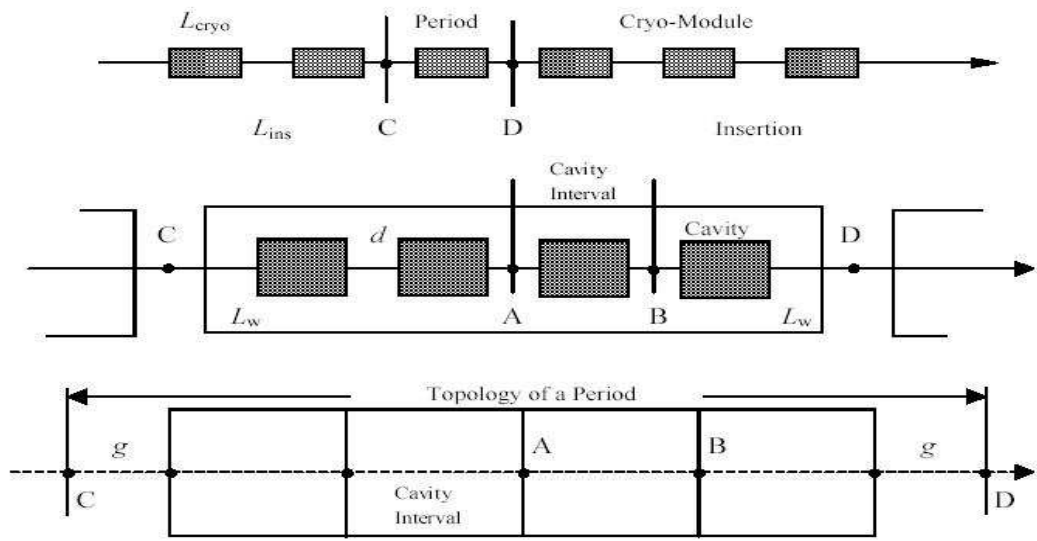}}
\end{center}
\caption{\label{fig:confscl}Configuration of a proton
superconducting linear accelerator.}
\end{figure}

 The SCL accelerates the proton beam from 200 MeV to
1.2 GeV. The configuration and the design procedure of the SCL are
described in detail in reference  \cite{ir62312}. A typical
sequence of identical periods is shown in Figure
~\ref{fig:confscl}. Each period consists of a cryo-module of
length $L_{cryo}$ with an insertion of length $L_{ins}$. The
insertion is needed for the placement of focusing quadrupoles,
vacuum pumps and valves, steering magnets, beam diagnostic
devices, bellows, and flanges. It can be either at room
temperature or in a cryostat. Here we assume that the insertions
are at room temperature. The cryo-module includes $M$ identical
cavities, each of $N$ identical cells, and each having a length
{\it NL}$_{cell}$, where $L_{cell}$ is the length of a cell. To
avoid coupling by the leakage of the field, cavities are separated
from each other by a sufficiently long drift space, $d$. An extra
drift of length $L_{w}$ may be added internally on both sides of
the cryo-module to provide a transition between cold and warm
regions. Thus, the length of a cryo-module is

\begin{equation}
L_{cryo} = {\it MN L}_{cell} + (M - 1) d + 2 L_{w}.
\end{equation}

There are two symmetric intervals: a minor one, between the two
middle points A and B, as shown in Figure ~\ref{fig:confscl}, that
is the interval of a cavity of length $N L_{cell}+d$; and a major
one, between the two middle points C and D, that defines the range
of a period of total length $L_{cryo}+L_{ins}$. Thus, the topology
of a period can be represented as a drift of length $g$, followed
by $M$ cavity intervals, and a final drift of length $g$, where

\begin{equation}
g=L_{w} + (L_{ins} - d ) / 2.
\end{equation}

The choice of cryo-modules with identical geometry and with the
same cavity/cell configuration is economical and convenient for
construction. There is, nonetheless, a penalty due to the reduced
transit-time-factors when a particle crosses cavity cells with
length adjusted to a common central value $\beta _{0}$ that does
not correspond to the particle's instantaneous velocity. To
minimize this affect the SCL is divided into three sections, each
designed around a different central value $\beta _{0}$, and with a
different cavity/cell configuration. The cell length in a section
is fixed to be

\begin{equation}
L_{cell}=\lambda \beta _{0}/ 2,
\end{equation}

\noindent where $\lambda $ is the RF wavelength. We adopted an
operating frequency of 805 MHz for the LE-section of the SCL, and
1,610 MHz for the subsequent two sections, ME and HE. The choice
of the large RF frequency in the last two sections has been
dictated by the need to achieve as a large accelerating gradient
as possible so the SCL would fit entirely within the available
space. The major parameters of the three sections of the SCL are
given in Tables ~\ref{tab:gpscl} and ~\ref{tab:sscl}.

\input epsf

The length of the SCL depends on the average accelerating
gradient. The local gradient has a maximum value that is limited
by three causes: (1) The surface field limit, in the frequency
range of interest between 805 and 1,610MHz, is around 40 MV/m. For
a realistic cell shape, we set a limit on the axial electric field
to 15 MV/m at 805 MHz, and 30 MV/m at 1,610 MHz. (2) There is a
limit on the peak power provided by RF couplers that we take here
not to exceed 400 kW, including a contingency of 50{\%} to avoid
saturation effects. (3) To make the longitudinal motion stable, we
can only apply an energy gain per cryo-module that is a relatively
small fraction of the beam energy at the exit of the cryo-module.
The conditions for stability of motion have been derived in
reference \cite{ir62312}.

The proposed mode of operation is to operate each section of the
SCL with the same energy increment. This requires the same axial
field from one cryo-module to the next. To achieve this, and to
compensate for the transit time variation from one cryo-module to
next, it may be necessary to adjust locally the RF phase, taken
here to be 30$^{\circ}$. Also the coupling power may have to be
adjusted according to the local transit time factor.
 The number of cells and cavities may vary
from section to section, but we have found convenient here to
adopt the same distribution in all sections. There is one klystron
feeding a single coupler to a single cavity. The total length of
the SCL injector proper from end to end is about 130 meters,
including a 4.5-m long matching section between LE and ME
sections. When averaged over the real estate, the actual
acceleration rate is about 5 MeV/m in the LE section and 10 MeV/m
in the ME and HE sections. Efficiencies, defined as the ratio of
beam power to required total AC power, is relatively high for a
pulsed linac, ranging between 9 and 15{\%}.

\begin{table}[htbp]
\begin{center}
\caption{\label{tab:gpscl}General Parameters of the SCL.}
\begin{tabular}
{|p{199pt}|p{50pt}|p{56pt}|p{56pt}|} \hline Linac Section& LE& ME&
HE \\ \hline \hline Ave. {\it Incremental} Beam Power, kW& 7.52&
15.0& 15.0 \\ \hline Average Beam Current, $\mu $A& 37.6& 37.6&
37.6 \\ \hline Initial Kinetic Energy, MeV& 200& 400& 800 \\
\hline Final Kinetic Energy, MeV& 400& 800& 1200 \\ \hline
Frequency, MHz& 805 & 1610& 1610 \\ \hline No. of Protons / Bunch
x 10$^{8}$& 8.70 & 8.70& 8.70 \\ \hline Temperature, $^{o}$K& 2.1&
2.1& 2.1 \\\hline \hline Cells / Cavity& 8& 8& 8 \\ \hline
Cavities / Cryo-Module& 4& 4& 4
\\ \hline Cavity Separation, cm& 32.0 & 16.0& 16.0 \\ \hline
Cold-Warm Transition, cm& 30 & 30& 30 \\ \hline Cavity Internal
Diameter, cm& 10& 5& 5 \\ \hline Length of Warm Insertion, m&
1.079& 1.379& 1.379 \\ \hline \hline Accelerating Gradient, MeV/m&
10.5 & 22.9& 22.8 \\ \hline {\it Ave. (real-estate) Gradient,
MeV/m}& {\bf 5.29}& {\bf 9.44}& {\bf 10.01} \\ \hline Cavities /
Klystron& 1& 1& 1 \\ \hline No. of RF Couplers / Cavity& 1& 1& 1
\\ \hline RF Phase Angle& 30$^{o}$& 30$^{o}$& 30$^{o}$ \\ \hline
Method for Transverse Focusing& FODO& Doublets& Doublets \\ \hline
Betatron Phase Advance / FODO cell& 90$^{o}$& 90$^{o}$& 90$^{o}$
\\ \hline \hline Norm. rms Emittance, $\pi $ mm-mrad& 2.0 & 2.0 &
2.0  \\ \hline Rms Bunch Area, $\pi ~^{o}$MeV (805 MHz)& 0.5 & 0.5
& 0.5
\\ \hline
\end{tabular}
\end{center}
\end{table}

Negative ion stripping during transport along the SCL has been
found to be very negligible, never exceeding a rate of 2x10$^{ -
8}$ per ion. But the final 11$^{o}$ bend, before injection into
the AGS, could be  a concern \cite{ir62310}. To control the rate
of beam loss by stripping to a 10$^{ - 5}$ level or less, the
bending field should not exceed 1.25 kGauss over a total
integrated bending length of 12 m.

A superconducting linac is most advantageous for a continuous mode
of operation (CW). There are two problems in the case of the
pulsed-mode of operation. First, the pulsed thermal cycle
introduces Lorentz forces that deform the cavity cells out of
resonance. This can be controlled with a thick cavity wall
strengthened on the outside by mechanical supports. The actual
design of a cavity cell is described in detail in reference
\cite{cavitycell}. Second, there is an appreciable period of time
to fill the cavities with RF power before the maximum gradient is
reached \cite{ir62312}. During the filling time, extra power is
dissipated  before the beam is injected into the linac. The extra
amount of power required is the ratio of the filling time to the
beam pulse length. The filling times are also shown in Table
~\ref{tab:sscl}.

\begin{table}[htbp]
\begin{center}
\caption{\label{tab:sscl}Summary of the SCL Design.}
\begin{tabular}
{|p{209pt}|p{50pt}|p{50pt}|p{50pt}|} \hline Linac Section& LE& ME&
HE
\\ \hline \hline  Velocity, $\beta $: In \par~~~~~~~~~~~~~~~ Out& 0.5662 0.7131& 0.7131
\par 0.8418& 0.8418 \par 0.8986 \\ \hline {\bf Cell Reference
$\beta $}$_{0}$ & {\bf 0.615}& {\bf 0.755}& {\bf 0.851}
\\ \hline Cell Length, cm& 11.45& 7.03& 7.92 \\
\hline Total No. of Periods& 6& 9& 8 \\ \hline Length of a Period,
m& 6.304& 4.708& 4.994 \\ \hline FODO-Cell Ampl. Func., $\beta
_{Q}$, m& 21.52& 8.855& 8.518 \\ \hline Total Length, m& $37.82$&
$42.38$& $39.96$ \\ \hline Coupler RF Power, kW (*)& 263& 351& 395
\\ \hline Energy Gain/Period, MeV& 33.33& 44.57& 50.10 \\ \hline
Total No. of Klystrons& 24& 36& 32 \\ \hline Klystron Power, kW
(*)& 263& 351& 395 \\ \hline Z$_{0}$T$_{0}^{2}$, $^{ }$ohm/m&
378.2& 570.0& 724.2 \\ \hline Q$_{0 }$ x 10$^{10}$& 0.97 & 0.57&
0.64
\\
\hline Transit Time Factor, T$_0$ & 0.785 & 0.785 & 0.785\\
 \hline Ave. Axial Field, E$_a$, MV/m & 13.4 & 29.1 & 29.0
\\ \hline Filling Time, ms& 0.337& 0.273& 0.239 \\ \hline Ave.
Dissipated Power, W& 2& 11& 8 \\ \hline Ave. HOM-Power, W& 0.2&
0.5& 0.4 \\ \hline Ave. Cryogenic Power, W& 65& 42& 38 \\ \hline
Ave. Beam Power, kW& 7.52& 15& 15 \\ \hline Total Ave. RF Power,
kW (*)& 17& 31& 30 \\ \hline Ave. AC Power for rf, kW (*)& 37& 69&
67
\\ \hline Ave. AC Power for Cryo., kW& 46& 30& 27 \\ \hline {\bf
Total Ave. AC Power, kW (*)}& 83& 99& 94 \\ \hline {\bf
Efficiency, {\%} (*)}& 9.05& 15.21& 16.08 \\ \hline
\multicolumn{4}{l}{\footnotesize (*) Including 50{\%} RF power
contingency.}
\\
\end{tabular}
\end{center}
\end{table}

\begin{figure}[htbp]
\begin{center}
\centerline{\includegraphics[width=5.03in,height=8.01in]{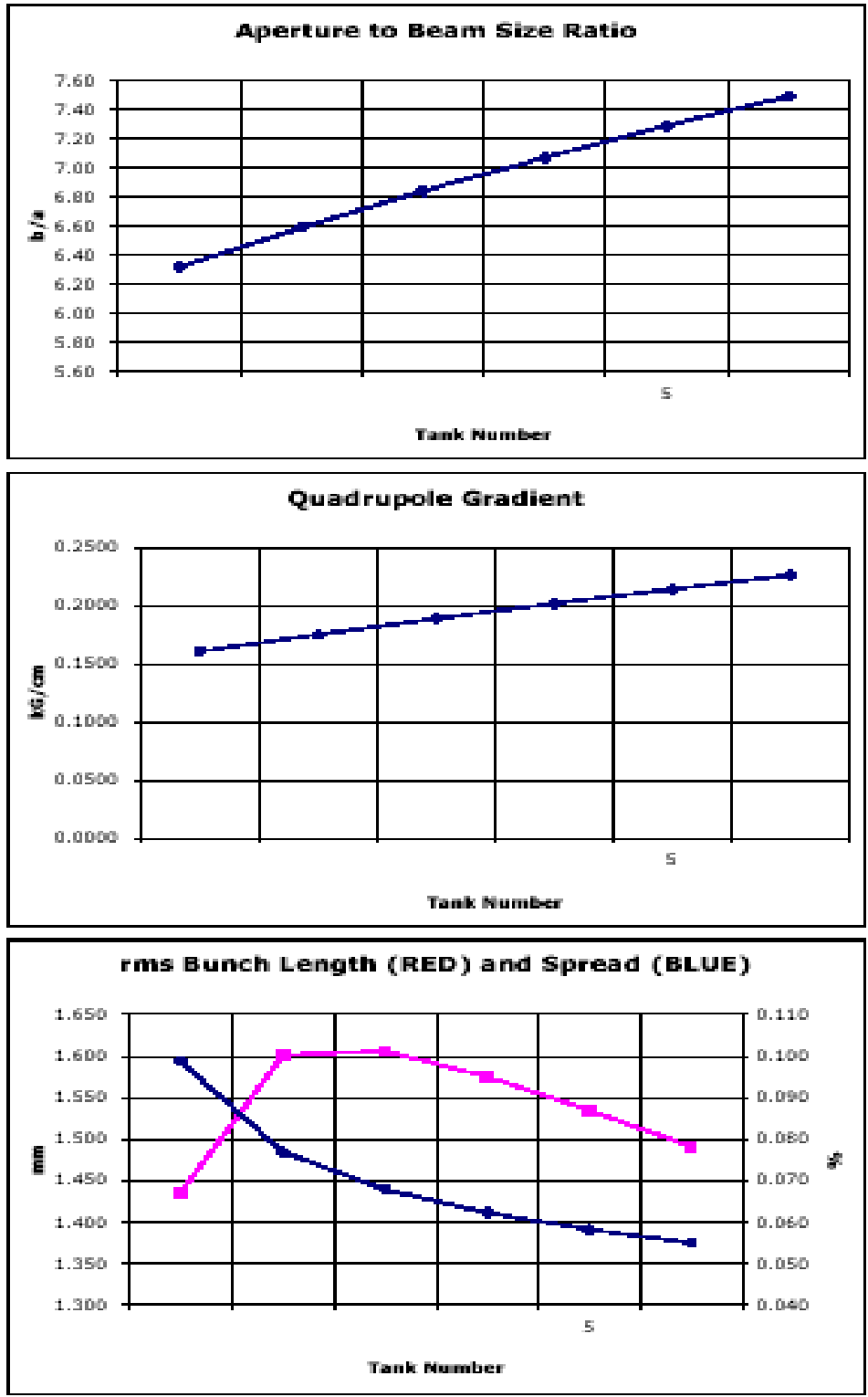}}
\caption{Plots (1) of behavior vs. \textit{period} (tank) number
of LE Section.} \label{fig:les1}
\end{center}
\end{figure}

\begin{figure}[htbp]
\begin{center}
\centerline{\includegraphics[width=5.04in,height=8.04in]{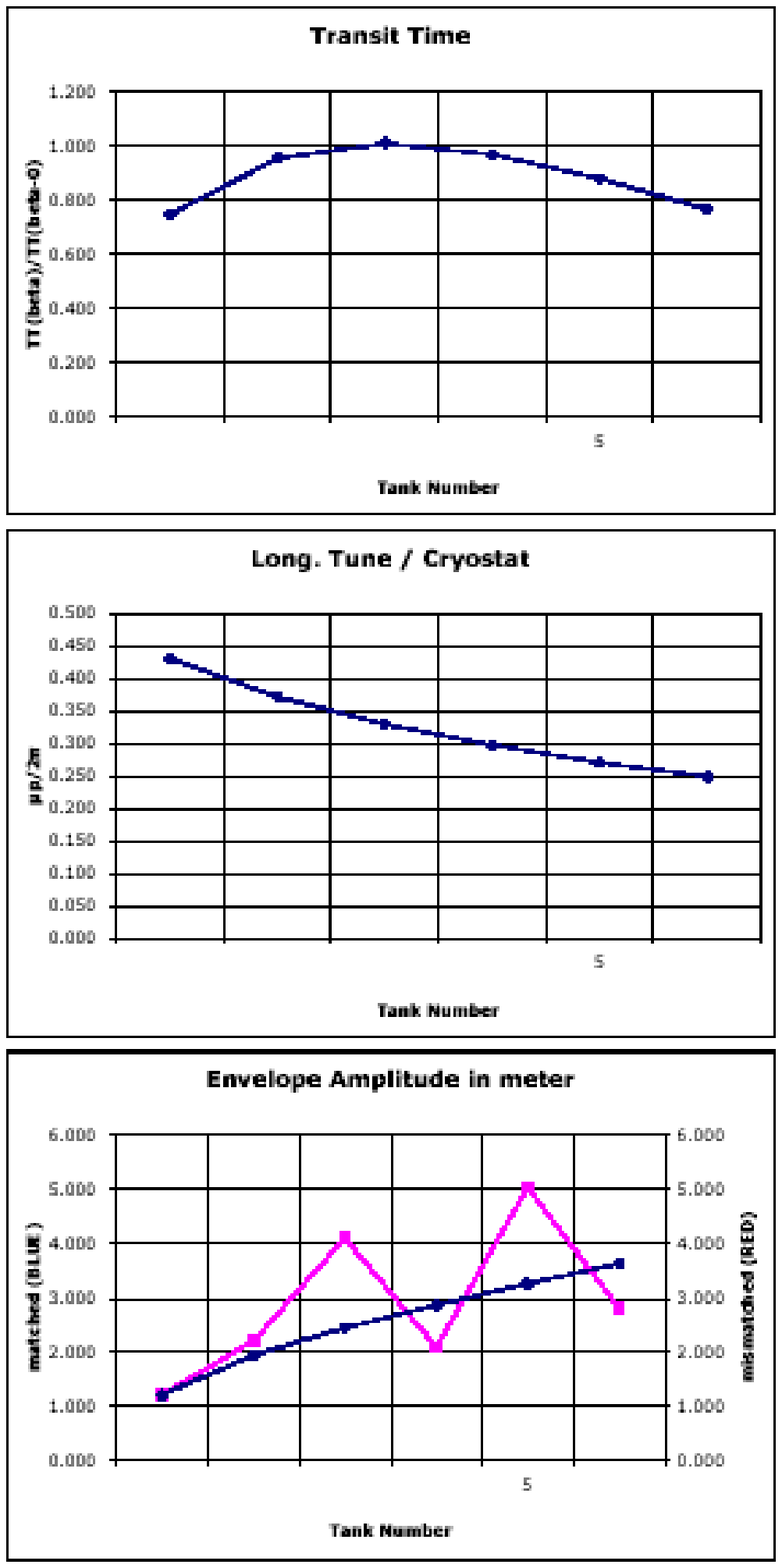}}
\caption{Plots (2) of behavior vs. \textit{period} (tank) number
of LE Section.} \label{fig:les2}
\end{center}
\end{figure}

\begin{figure}[htbp]
\begin{center}
\centerline{\includegraphics[width=5.06in,height=8.30in]{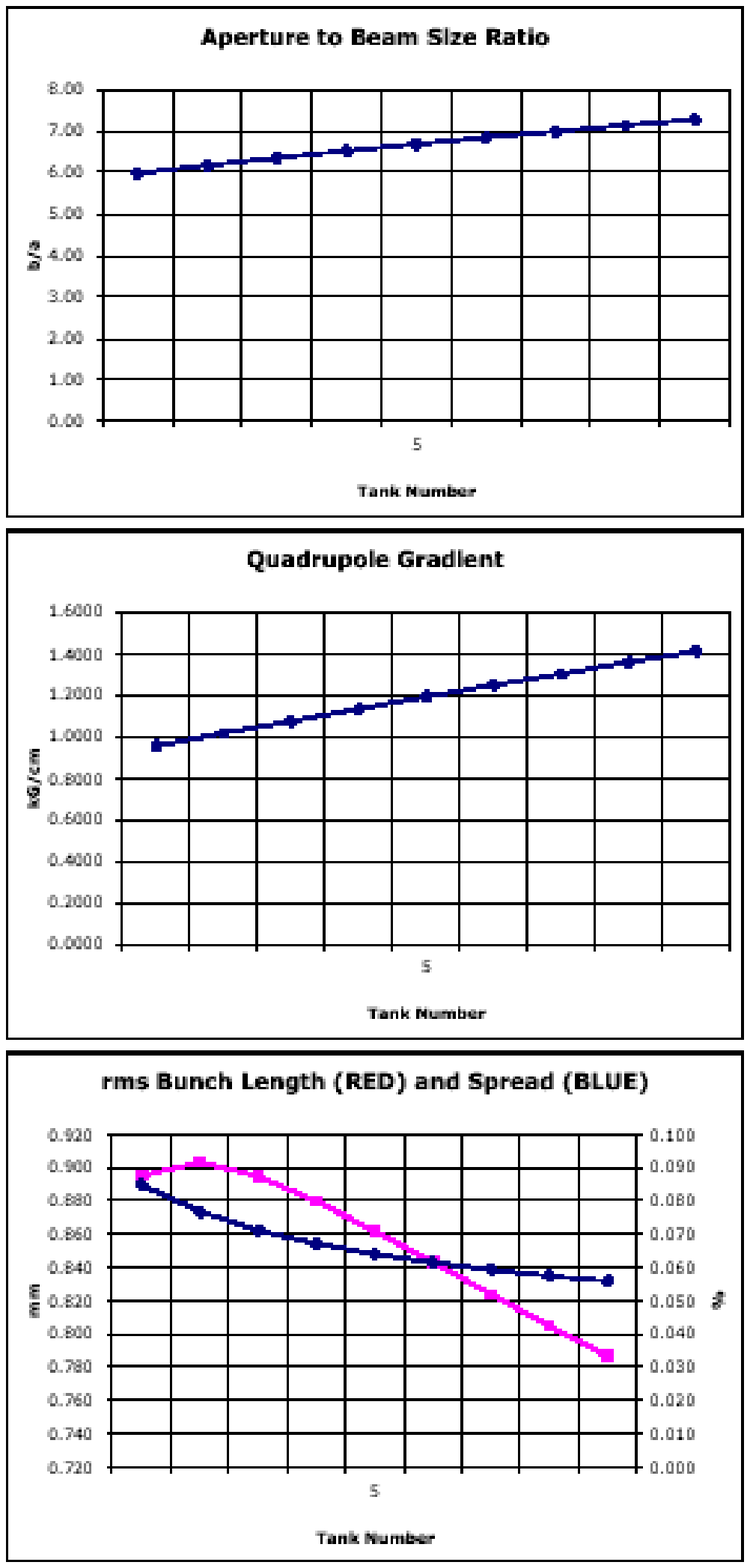}}
\caption{Plots (1) of behavior vs. \textit{period} (tank) number
of ME Section.} \label{fig:mes1}
\end{center}
\end{figure}

\begin{figure}[htbp]
\begin{center}
\centerline{\includegraphics[width=5.06in,height=8.34in]{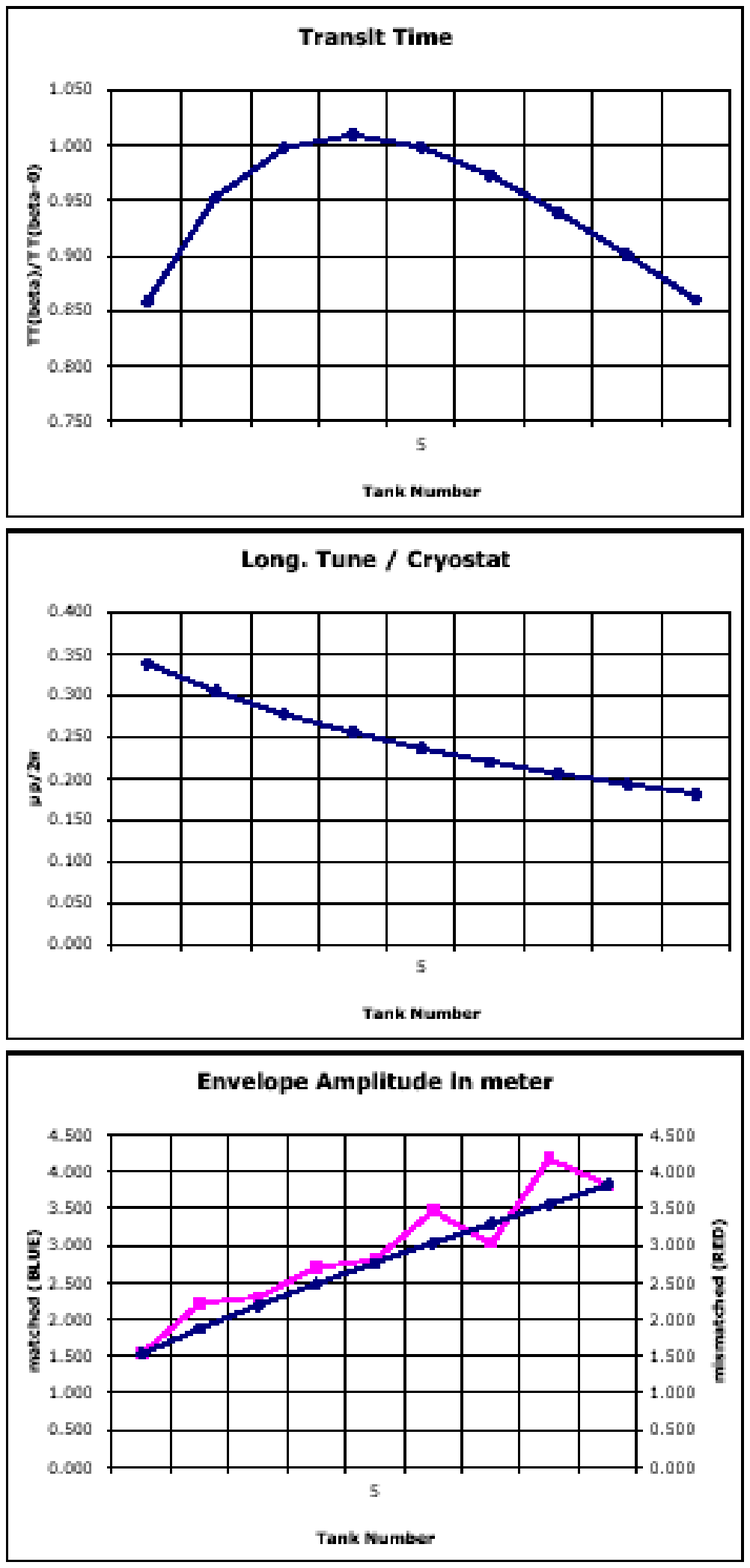}}
\caption{Plots (2) of behavior vs. \textit{period} (tank) number
of ME Section.} \label{fig:mes2}
\end{center}
\end{figure}

\begin{figure}[htbp]
\begin{center}
\centerline{\includegraphics[width=5.06in,height=8.34in]{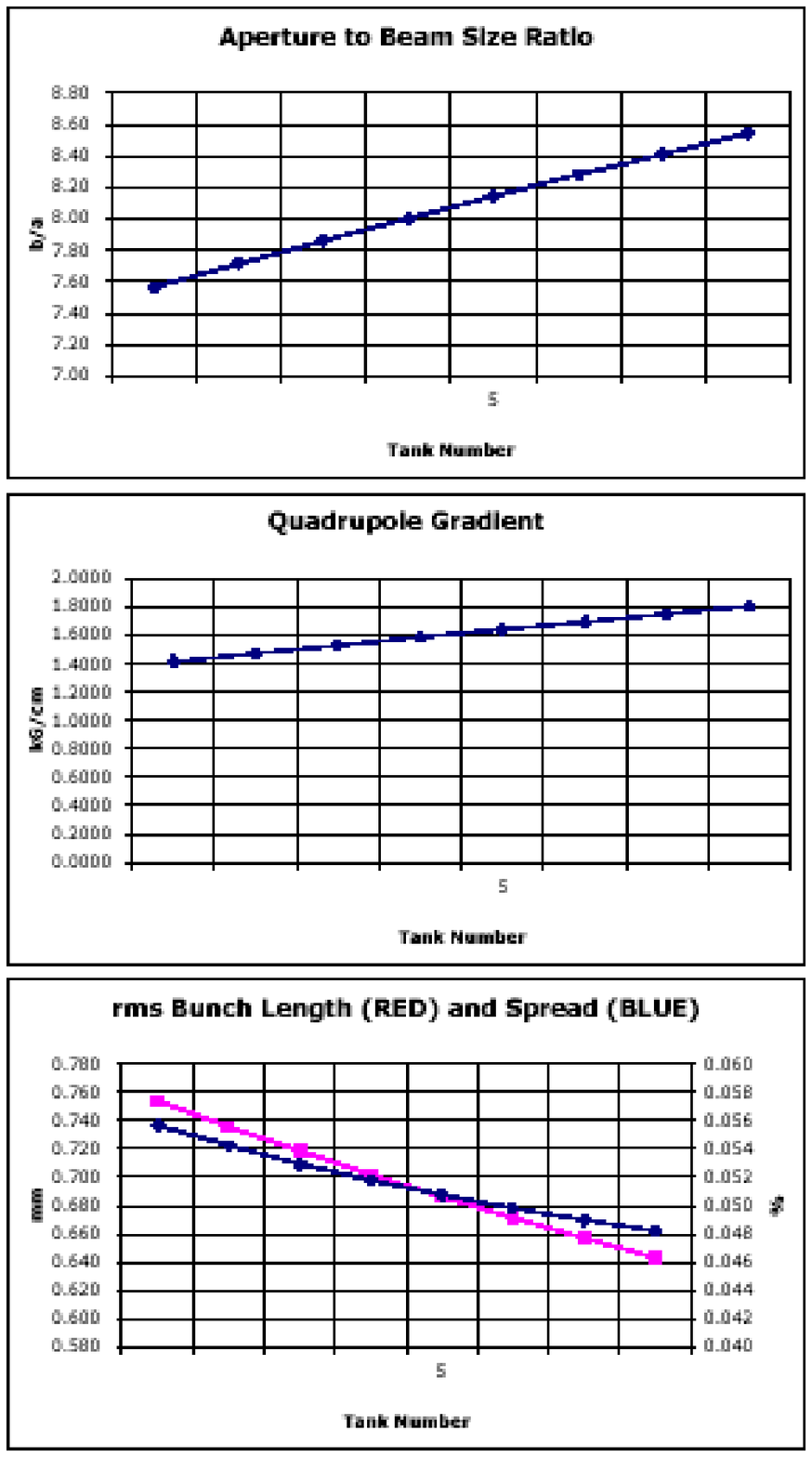}}
\label{fig:hes1} \caption{Plots (1) of behavior vs.
\textit{period} (tank) number of HE Section.}
\end{center}
\end{figure}

\begin{figure}[htbp]
\begin{center}
\centerline{\includegraphics[width=5.06in,height=8.34in]{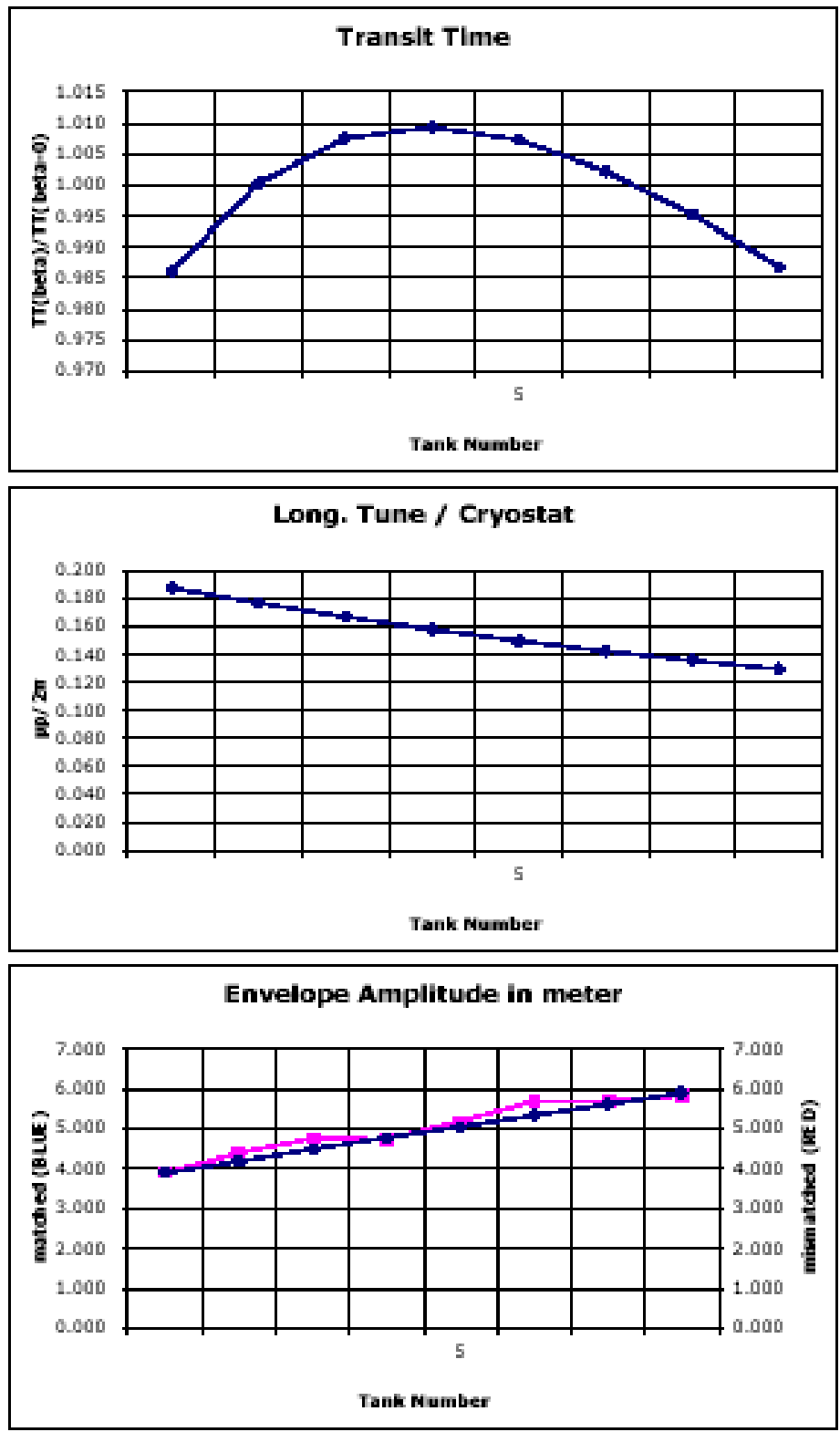}}
\caption{Plots (2) of behavior vs. \textit{period} (tank) number
of HE Section.} \label{fig:hes2}
\end{center}
\end{figure}

A program \cite{sandrop} was written in Visual Basic ( included
with the MS Excel application), to calculate the beam and RF
dynamics during acceleration in each of the three sections of the
SCL. The results are displayed in Figures ~\ref{fig:les1} to
~\ref{fig:hes2}.

\lhead{AGS Super Neutrino Beam Facility} \rhead{Transverse
Focusing} \rfoot{April 15, 2003}

\subsubsection{Transverse Focusing}
\label{sec:scltf}

The upgrade makes use of the present 200 MeV room temperature
linac, with proper modifications as described. This linac provides
a negative ion beam with an emittance of 2.0 $\pi $ mm-mrad. To
avoid uncontrolled beam losses that may cause radiation
activation, we require that the ratio of inner cavity radius to
rms beam size is at least a factor of 6  over the length of the
SCL. This is difficult to achieve in the ME and HE section where
the inner aperture is of only 5 cm because of the larger RF
frequency. We have thus adopted in these two sections transverse
focusing with doublets of quadrupoles, whereas a FODO singlet
sequence was  found to be adequate in the LE section.
\clearpage
\newpage
\lhead{AGS Super Neutrino Beam Facility} \rhead{AGS Upgrade}
\rfoot{April 15, 2003}

\section{AGS Upgrade}
\label{sec:agsup}
 \setcounter{table}{0}
 \setcounter{figure}{0}
 \setcounter{equation}{0}

As explained in the introduction, a 1.2 GeV linac will be used for
the injection into the AGS directly to reduce the injection time
to about 1.0 msec. The results of the study of the multi-turn
direct injection are presented in Section ~\ref{sec:multiturninj}.
The approache to increase the AGS repetition rate from current 0.5
Hz to 2.5 Hz is discussed in Section ~\ref{sec:agspsupg}. In
parallel with improvement to the power supply system is the
upgrade of the RF system to raise the accelerating voltage from
400 kV to 1.0 MV. Finally, the eddy current effects due to
increased repetition rate are covered in section
~\ref{sec:eddycurrenteffects}.

All the above improvements employ proven technology and a
successful  implementation  of this design can be assured.

\lhead{AGS Super Neutrino Beam Facility} \rhead{Multi-Turn
Injection into the AGS } \rfoot{April 15, 2003}

\subsection{Multi-Turn Injection into the AGS}
\label{sec:multiturninj}

The front end ion source has to operate  with a 1\% duty cycle at
the repetition rate of 2.5~Hz as explained in Chapter
~\ref{sec:linacup}. The combination of the chopper and of the RFQ
pre­bunches the beam with a sufficiently small bunch length so
that each beam bunch fits in the accelerating RF buckets of the
downstream DTL, which operates at 201.25~MHz. The DTL is a
room-temperature conventional linac that accelerates to 200~MeV.
The proposed new injector for the AGS adds a 1.2~GeV SCL with an
average output beam power of about 45~kW. The injection energy is
still low enough to control beam losses due to stripping of the
negative ions that are used for multi­turn injection into the AGS.
The duty cycle is about 0.5\%. Injection into the AGS is modeled
after the SNS scheme~\cite{snsdr}. However, the repetition rate,
and consequently the average beam power, is much lower here. The
larger circumference of the AGS also reduces the number of foil
traversals. Beam losses at injection into AGS are estimated to be
about 3\% controlled losses and 0.3\% uncontrolled losses. This is
based on a comparison with the actual experience in the AGS
Booster and the LANL PSR and the predicted losses at the SNS using
the quantity (N$_{\rm{}P}$ /$\beta^2\gamma^3$ A), which is
proportional to the Laslett tune shift, as a scaling factor. This
is summarized in Table ~\ref{tab:ch-ip}. As can be seen, the
predicted 3\% beam loss is consistent with both the AGS Booster
and the PSR experience and the SNS prediction.
\begin{table}[!htb]
\begin{center}
\caption[~Comparison of H minus injection parameters] {Comparison
of $\textrm{H}^-$ injection parameters.} \label{tab:ch-ip}
\begin{tabular}{|l|c|c|c|c|}
\hline $\textrm{H}^-$ Injection Parameters &AGS Booster & SNS &
PSR &1 MW AGS\\
 \hline
 Beam Power,Linac Exit (kW) & 3 & 1000 & 80 &45\\
\hline Kinetic Energy (MeV) & 200 & 1000 & 800 & 1200\\
 \hline
 No.of Protons N$_{\rm{}P}$ $(10^{12})$&15&100&31&100\\
\hline
 Vertical Acceptance, A  $(\pi \textrm{mm mrad})$ &89 &480 &
140 &55\\
 \hline
  $\beta^{2} \gamma^{3}$ &0.57 &6.75  &4.50 &
9.56\\ \hline $N_{P}/(\beta^{2}\gamma^{3} A)$ $(10^{12}/\pi
\textrm{mm mrad})$&0.296&0.031&0.049&0.190\\
 \hline
 Total Beam Losses \%  & 5 &0.1 &0.3& 3\\
\hline
 Total Lost Beam Power W  & 150 &1000 & 240 & 1440\\
\hline

Circumference m  & 202 &248 & 90& 807\\
 \hline

Lost Beam Power per Meter W/m  & 0.8 &4.0 & 2.7 & 1.8\\ \hline
\end{tabular}
\end{center}
\end{table}

The AGS injection parameters are summarized in Table
~\ref{tab:agsip}. A relatively low RF voltage of 450~kV at the
injection is necessary to limit the beam momentum spread at the
multi-turn injection to be about 0.48~\%, and the longitudinal
emittance to be about 1.2~eVs per bunch. Such a small emittance is
important to limit beam losses during transition crossing in the
AGS.

\begin{figure}
\begin{center}
\includegraphics[width=5.5in]{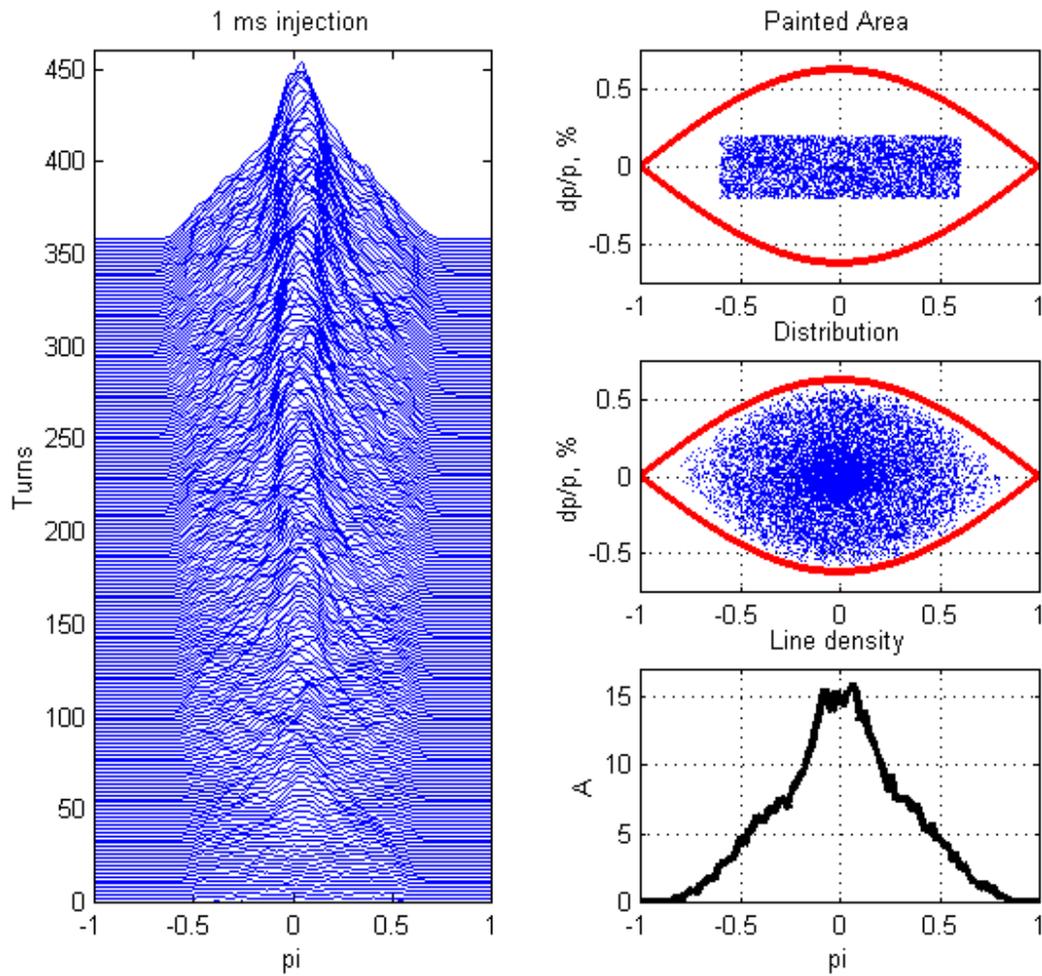}
\caption{~AGS injection simulation. The abscissa is phase.}
\label{fig:agsinjsim}
\end{center}
\end{figure}

A preliminary simulation of the 360 turns injection process is
shown in Figure ~\ref{fig:agsinjsim}. Without the second harmonic
RF, some dilution of the injected particles in the phase space is
inevitable. The bunch shape is similar to the one at the PSR of
Los Alamos, with a noticeable sharp peak, however, a possible
linac beam momentum ramping could improve this.
\begin{table}[!bth]
\begin{center}
\caption{~AGS injection parameter.} \label{tab:agsip}
\begin{tabular}{|l|c|}
\hline Injected Turns &360\\ \hline

Repetition Rate (Hz) &2.5 \\ \hline

Pulse Length (ms)&1.08 \\ \hline

Chopping Rate \%&0.65\\ \hline

Linac Average/Peak Current (mA)& 20/30 \\ \hline

Momentum Spread & $\pm 0.0015$ \\ \hline

Norm. 95\% Emittance $(\pi \mu m)$& 12\\ \hline

RF Voltage (kV)& 450 \\ \hline

Bunch Length (ns) &85\\ \hline \hline

 Longitudinal Emittance (eVs)&1.2 \\
\hline

Momentum Spread &$\pm 0.0048$\\ \hline

Norm. 95\% Emittance $(\pi \mu m)$ & 100\\ \hline
\end{tabular}
\end{center}
\end{table}

The beam instability consideration is focused on two aspects.
These are, as usual for the AGS, at high energy, the longitudinal instability
around  transition, and the transverse instability  above
transition.

The fractional beam momentum spread at transition has to be
less than 0.0075 because of the limited momentum aperture during
the transition energy jump. With the transition jump, the slippage
factor can be controlled to be greater than 0.002. With a bunch
rms length of 4.25~ns and the peak current of 85~A at transition,
the longitudinal impedance needs to be less than $11~\Omega$ to
avoid longitudinal microwave instabilities.

The measured AGS broadband impedance is about $30~\Omega$. The
broadband impedance mainly comes from the unshielded bellows, the
vacuum chamber connections and steps and cavities, and also with
possible contribution from BPMs and ferrite kickers. With a
modest effort, this impedance can be reduced to be less than
$10~\Omega$, which is consistent with newly designed proton
machines.

In fact, if only the longitudinal microwave instability were of
concern, a larger broadband impedance could be tolerated since
the longitudinal space charge impedance of about $10~\Omega$ at
transition, which is capacitive, has the effect of cancelling the
inductive broadband impedance. However, the transverse instability
at the high energy is more serious, even with a broadband
impedance of $10~\Omega$.
In summary, since the intensity of $1\times10^{14}$ is only
marginally higher than the present intensity of $7\times10^{13}$,
the beam instability during acceleration and transition crossing
can be avoided.
\lhead{AGS Super Neutrino Beam Facility} \rhead{AGS Main Power
Supply Upgrade } \rfoot{April 15, 2003}

\subsection{AGS Main Power Supply Upgrade}
\label{sec:agspsupg}
\subsubsection{Present Mode of Operation}
\label{sec:pmofo}
The present AGS Main Magnet Power Supply (MMPS) is a fully
programmable 6000~A, $\pm$9000 V SCR power supply. A 9~MW Motor
Generator (MG), made by Siemens, is a part of the main magnet power supply
of the accelerator. The MG permits pulsing the main magnets up to 50~MW
peak power, while the input power of the MG itself
 remains constant. The highest power into the MG ever utilized is 7~MW,
that is, the maximum average power dissipated in the AGS magnets
has never exceeded 5~MW.

The AGS ring comprises 240~magnets connected in series. The total
resistance, R, is $0.27~\Omega$ and the total inductance, L, is
0.75~H. There are 12 superperiods, designed \textit{A} through
\textit{L}, of 20 magnets each, divided in two identical sets of
10~magnets per superperiod.

Two stations of power supplies are each capable of delivering up
to 4500~V and 6000~A. Every station consists of two power supplies
connected in parallel. One power supply is a 12 pulse SCR  rated
at +/-5000~V, 6000~A unit (P type) that is typically used for fast
ramping during acceleration and energy recovery. The other is a
lower voltage 24 pulse unit (F type), rated at +/-1000~V, 6000~A,
that is used for flattop or slow ramping operation.  The two
stations are connected in series, with the magnet coils arranged
to have a total resistance R/2 and a total inductance of L/2. The
grounding of the power supply is done only in one place, in the
middle of station 1 or 2, through a resistive network. With this
grounding configuration, the maximum voltage to ground in the
magnets does not exceed 2500~V. The magnets are tested at 3~kV to
ground prior to each startup of the AGS MMPS after long
maintenance periods.

\subsubsection{Super Neutrino Beam Mode of Operation}
\label{sec:nsbmofo}
\begin{figure}[!hbt]
\begin{center}
\includegraphics[width=6in]{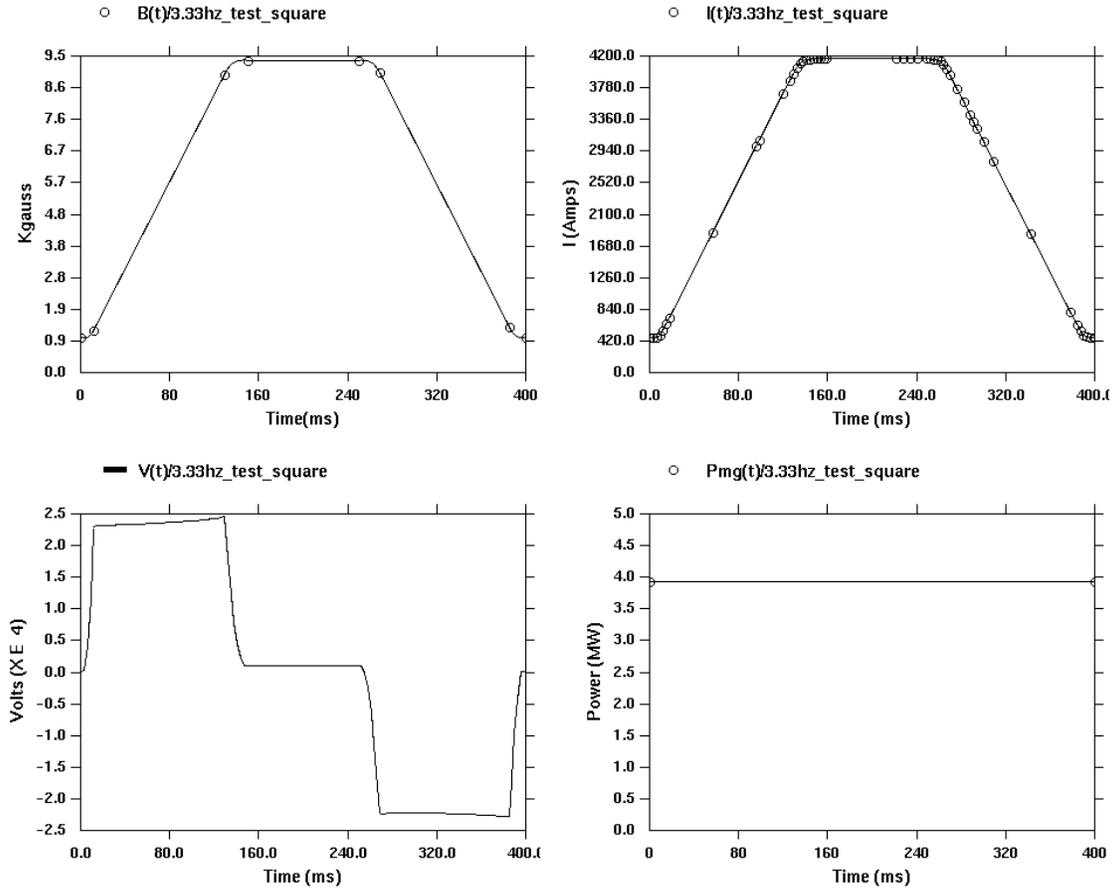}
\caption[~Current and voltage cycle for 2.5 Hz operation ]{Current
and voltage cycle for 2.5~Hz operation. Also shown are the AGS
dipole field and average power.} \label{fig:cvcfor25hz}
\end{center}
\end{figure}

\begin{figure}[!htb]
\begin{center}
\includegraphics[width=4in]{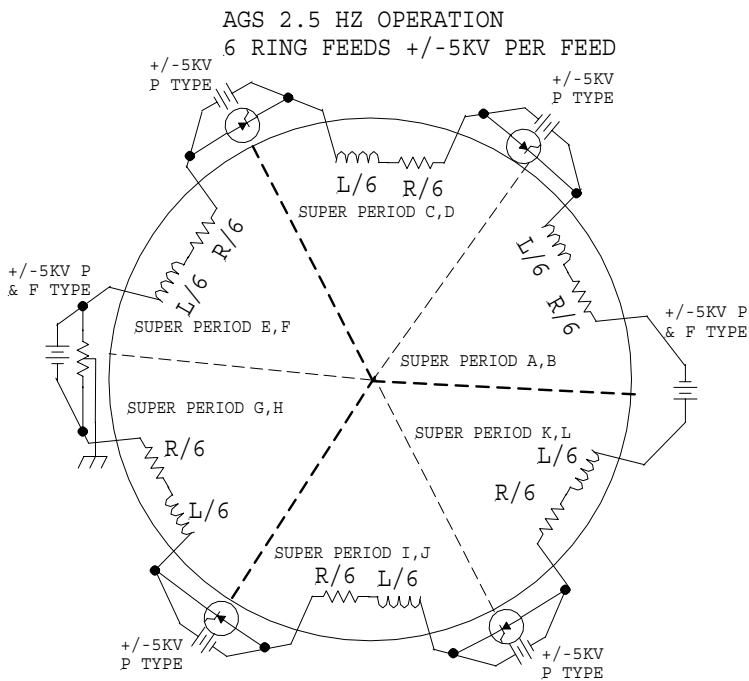}
\caption[~Schematic of power supply connections to the AGS
magnets]{Schematic of power supply connections to the AGS magnets
for 2.5-Hz operation.} \label{fig:spsctoagsm}
\end{center}
\end{figure}

 To cycle the AGS ring to 28~GeV at 2.5~Hz and with
a ramp time of 200~ms, the magnet peak current is 4300~A and the
peak voltage is 25~kV.  Figure ~\ref{fig:cvcfor25hz} displays the
magnet current and voltage of a 2.5~Hz cycle. The total average
power dissipated in the AGS magnets is estimated to be 3.7~MW. To
limit the AGS coil voltage to ground to 2.5~kV, the AGS magnets
must be divided into three identical sections, each powered
similarly to the present AGS except that now the magnet loads
represent only
 1/6 of the total resistance and inductance. Every section will
 be powered separately with its own feed to the ring magnets and an
identical system of power supplies, as shown in Figure
~\ref{fig:spsctoagsm}. Bypass SCR's will be used across the four
new P type stations, to bypass these units during the flattop, and
ensure minimum ripple. Note that only station 1 will be grounded
as it is done presently.

Although the average power will not be higher than now, the peak
power required is approximately 110~MW, exceeding the 50~MW rating
of the existing MG. A new MG capable of providing 100~MW, may
operate with 12~phases to limit, or even eliminate, the need for
phase shifting transformers so that every power supply system
would generate 24~pulses. The generator voltage will be about
15~kV line-to-line, to limit the generator current to less than
6000~A during pulsing. The generator will be rated at a slip
frequency of 2.5~Hz.

Running the AGS at 2.5~Hz requires that the acceleration ramp
period decreases from 0.5~s down to 0.20~s. That is, the magnet
current variation \textsl{dI/dt} is about 3 times larger than
at present. Eddy current losses in the vacuum chamber are
proportional to the square of \textsl{(dI/dt)}, that is they are
10~times larger. However, this is still significantly below the
present ramp rate of the AGS Booster which does not require
active cooling. The increased eddy currents give rise to increased
sextupole fields during the ramp, and will add about 20 units of
chromaticity. The present chromaticity sextupoles will be
upgraded to correct this.
\lhead{AGS Super Neutrino Beam Facility} \rhead{AGS RF System
Upgrade} \rfoot{April 15, 2003}

\subsection{AGS RF System Upgrade}
\label{sec:agsrfupg}

At 2.5 Hz the peak acceleration rate is three times the present
value for the AGS. With 10 accelerating stations each station will
need to supply 270~kW peak power to the beam. The present power
amplifier design, employing a 300~kW power tetrode will be
suitable to drive the cavities and supply power to the beam. The
number of power amplifiers will be doubled so that each station
will be driven by two amplifiers of the present design. This
follows not so much from the power consideration but from the
necessity to supply 2.5 times the RF voltage.

An AGS RF station comprises four acceleration gaps surrounded by
0.35~m of ferrite stacks. The maximum voltage capability of a gap
is not limited by the sparking threshold of the gap but by the
ability of the ferrite to supply the magnetic induction. When the
AGS operates with 0.5~Hz ramp the gap voltage is 10~kV. At 2.5~Hz
we will need up to 25~kV per gap (roughly equal to the voltage
from the same gap design used at the Booster, 22.5~kV) and this
taxes the properties of the ferrite. Above a certain threshold
value of $B_{rf}$ (20 mT for AGS ferrite 4L2) a ferrite becomes
unstable and excessively lossy. The gap voltage at this
$B_{rf,max}$ is simply given by

\begin{equation}
  V=-\frac{d}{dt}\int \omega B_{rf} dA = \omega a l B_{rf,max}
  \ln\frac{b}{a},
\end{equation}

\noindent where $\omega$ is the RF radian frequency and the
variables $a$, $b$, and $l$ are the inner and outer radius and
length of the ferrite stack, respectively.

The only free variable is $\omega$. If we operate the RF system at
the 24th harmonic of the revolution frequency (9~MHz) then the
required voltage of 25~kV can be achieved with a safe value for
$B_{rf,max}$ of 18~mT.

The next concern is the power dissipation in the ferrite and the
thermal stress that is created by differential heating due to rf
losses in the bulk of the material. We know from experience that
below 300~mW/cm$^3$ the ferrites can be adequately cooled. The
power density is also proportional to $B^2_{rf}$ and is given by

\begin{equation}
  \frac{P}{V}=\frac{\omega B^2_{rf}}{2 \mu_0(\mu Q)},
\end{equation}

\noindent where $\mu Q$ is the quality factor of the ferrite.

The $\mu Q$ product is a characteristic of the ferrite material
and depends on frequency and $B_{rf}$. We have data on ferrite 4M2
(used in the Booster and SNS) at 9~MHz and 20~mT where the power
dissipation is 900~mW/cm$^3$. The details of the acceleration
cycle determine the RF voltage program that is needed. For the
cycle shown in Figure ~\ref{fig:cvcfor25hz} a peak voltage of 1~MV
(40 gaps each with 25~kV) is needed but for only 20~ms during
acceleration. This is a duty factor of less than 0.05
 giving an average power dissipation much less than the
limit. We do not yet have data on the present AGS ferrite, 4L2 at
9~MHz. Characterizing 4L2 in this parameter regime is identified
as an R\&D issue but at least we know that retrofitting the AGS
cavities with 4M2 is a fallback option.

\begin{figure}[!hbt]
\begin{center}
\includegraphics[width=4in]{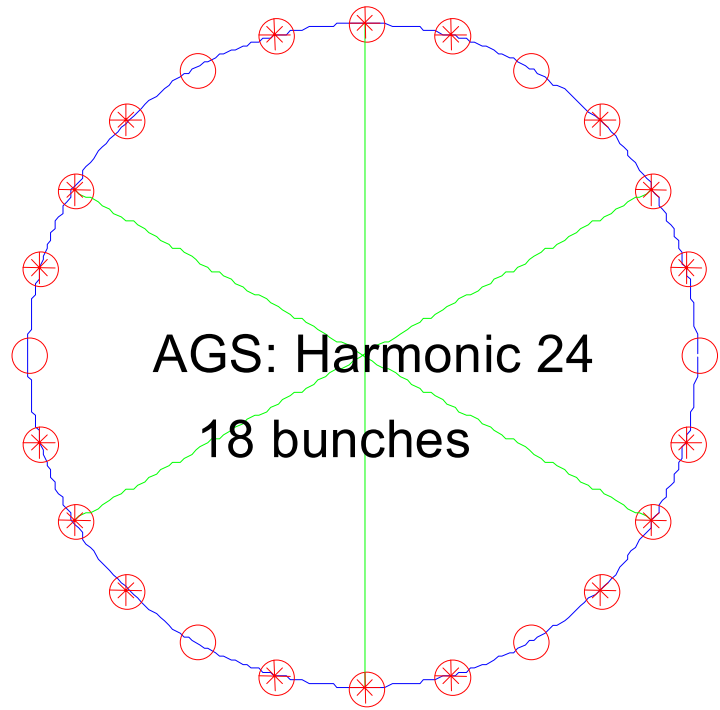}
\caption[~Bunch pattern for using harmonic 24 to create 6 bunches
]{Bunch pattern for using harmonic 24 to create 6 bunches.}
\label{fig:bunchpatt}
\end{center}
\end{figure}

\clearpage
\newpage
\lhead{AGS Super Neutrino Beam Facility} \rhead{Eddy Current
Effects} \rfoot{April 15, 2003}
\subsection{Eddy Current Effects }
 \label{sec:eddycurrenteffects}
 Presently the magnet cycle of the AGS
accelerator has a period of $\sim $3.5 sec with rise time of $\sim
$200 msec between the injection energy and top energy.The proposed
magnet cycle of 2.5 Hz for the neutrino production operation will
reduce the time between the injection and extraction to $\sim $90
msec. The 75 mils thick vacuum beam pipe of the circulating beam
of AGS (see Figure ~\ref{fig:b20cs}) is made of inconel conducting
material. The increased rate of change of the magnetic flux
($\Delta \Phi $/dt) during the 2.5 Hz operation will generate eddy
currents in both, the current caring coils and the vacuum chamber,
that may introduce the following adverse effects:
\begin{description}
\item[a.]  Ohmic heating of the coils. These Ohmic losses are added to the Ohmic losses
due to the current that is used to excite the magnet and are due
to the eddy currents generated in the conductors because of the
rate of change of the magnetic flux ($\Delta \Phi $/dt).
\item[b.] Ohmic heating of the chamber wall.
\item[c.] Magnetic multipoles at the region of the circulating beam. Such magnetic
multipoles that will vary during the acceleration of the beam may
bring the circulating beam in a strong resonance that will results
in large beam losses.
\end{description}
The Ohmic losses may require cooling of the vacuum chamber wall
and additional cooling of the magnet coil and may increase
dramatically the requirements of the power supply of the main
magnet.

The above effects a) b) and c) have been studied by modeling the
AGS magnet and the vacuum chamber in a 2-dimentional model, using
the computer code OPERA\cite{vecfields} for electromagnetic fields
calculations, and the results are reported in the following
sections.

 The iron of the AGS main magnet is of silicon-steel
magnetic material comprised of 1/8'' thick laminations, which are
electrically insulated and held together (sandwiched) by two 1''
thick plates of the same magnetic material, one at each end of the
magnet. The effects of the eddy currents generated in the iron of
the magnet can also be studied by 3-dimentional modeling of the
magnet and performing 3-dimentional electromagnetic calculations.
\lhead{AGS Super Neutrino Beam Facility} \rhead{Eddy Current
Effects} \rfoot{April 15, 2003}
\subsubsection{Effect of the Eddy Currents in the Current Carrying Conductors of the
Magnet.}
 \label{sec:eddycurr}

\begin{figure}[htbp]
\begin{center}
\centerline{\includegraphics[width=3.22in,height=2.63in]{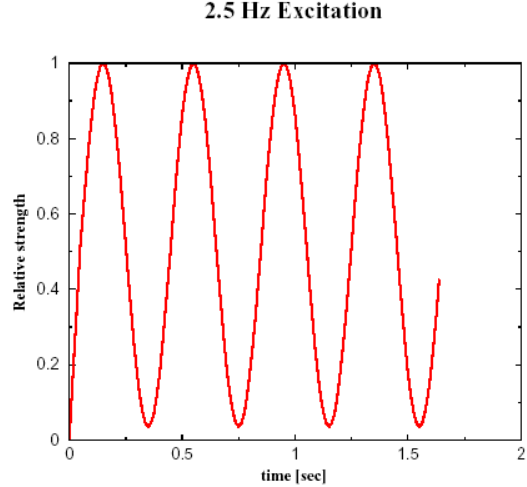}}
\caption{The relative excitation strength of the main AGS magnet
(I vs time), used in the transient electromagnetic field
calculations.} \label{fig:agsmes}
\end{center}
\end{figure}

An approximate function of the excitation of the magnet (I vs
time) that has been used in the electromagnetic calculations is
shown in Figure ~\ref{fig:agsmes}. The minima of the function
I(t)=I$_{dc}$+I$_{0}$sin($\omega $t) correspond to the injection
of the H$^{ - }$ beam and the maxima to the extraction of the 28
GeV proton beam. The time varying magnetic flux generated by the
excitation of the main magnet generates an EMF and therefore eddy
currents on the magnets coils. The cross section of a coil in a
main magnet is shown in Figure ~\ref{fig:b20cs} and consists of a
water cooled conductor with 32 turns (16 top and 16 bottom).

The effect of the EMF has been calculated\cite{imarneris} and can
be counter balanced by increasing the voltage of the main magnet
power supply.

\begin{figure}[htbp]
\begin{center}
\centerline{\includegraphics[width=4.112in,height=3.352in]{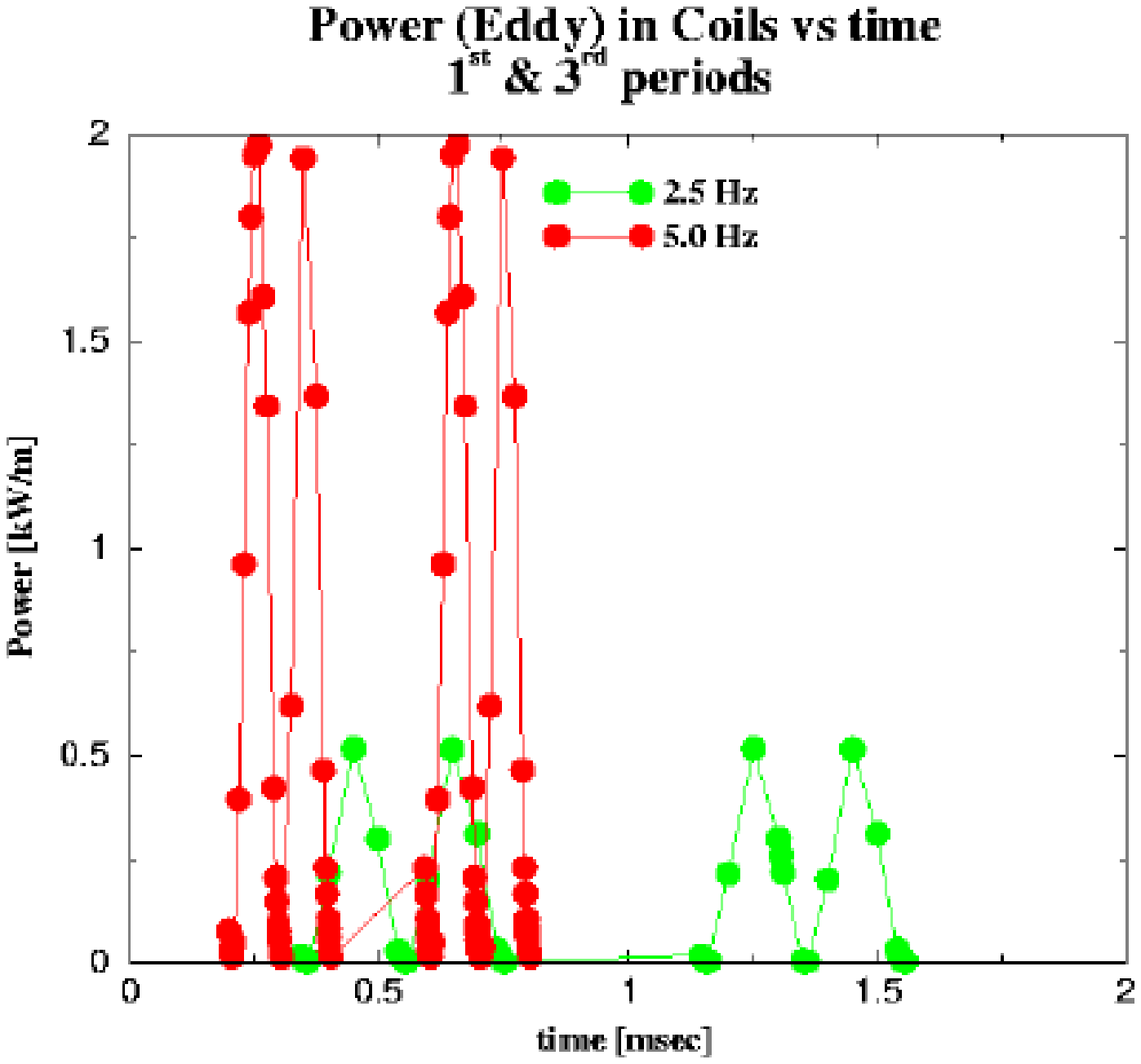}}
\caption{Ohmic losses per unit length due to eddy currents in the
coil of the main magnet for 2.5 Hz operation (green curve) and 5.0
Hz operation (red curve). Only the Ohmic losses of the 1$^{st}$
and the 3$^{rd}$ periods are plotted in order to ascetaine that
the eddy currents acquired a steady-state.} \label{fig:eddypower}
\end{center}
\end{figure}

The effect of the eddy currents generated on the coil have been
calculated and the results are shown in Figure
~\ref{fig:eddypower}. The Ohmic losses have been calculated for
two excitations of the main magnet 2.5 Hz (green curve) and 5.0 Hz
(red curve). The fact that the Ohmic losses during the 1$^{st}$
period are comparable to those of the 3$^{rd}$ period signifies
that eddy currents acquire a steady-state very early in the magnet
cycle.

The consecutive minima of the Ohmic losses curves shown in Figure
~\ref{fig:eddypower}, correspond to the consecutive minima and
maxima of the excitation curve shown in Figure ~\ref{fig:agsmes}.

The maxima of the Ohmic losses curves shown in Figure
~\ref{fig:eddypower}, correspond to the points of the excitation
curve of Figure ~\ref{fig:agsmes} where the rate of change
(derivatives) is a maximum.

As expected, the calculated integrated Ohmic losses for the 5.0 Hz
operation are $\sim $4 times larger than those for 2.5 Hz
operation.

\begin{figure}[htbp]
\begin{center}
\centerline{\includegraphics[width=2.924in,height=2.4256in]{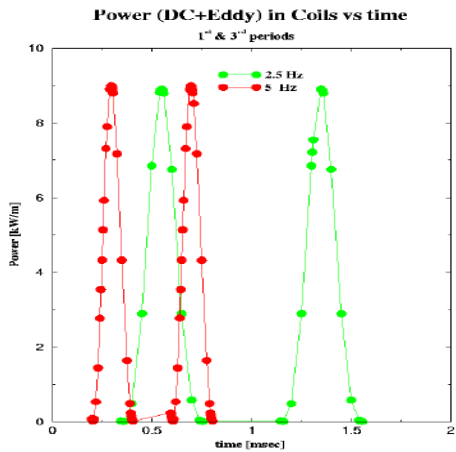}}
\caption{Total Ohmic losses per unit length due to eddy currents
in the coil of the main magnet for 2.5 Hz operation (green curve)
and 5.0 Hz operation (red curve). Only the Ohmic losses of the
1$^{st}$ and the 3$^{rd}$ periods are plotted.}
\label{fig:totalpower}
\end{center}
\end{figure}

In Figure ~\ref{fig:totalpower} plotted are the total Ohmic losses
in the coil due to both the excitation current and the eddy
currents. Like in Figure ~\ref{fig:eddypower} only the 1$^{st}$
and the 3$^{rd}$ periods are plotted.

The Ohmic losses due to eddy currents are $\sim $3{\%} and $\sim
$12{\%} of the Ohmic losses due to the excitation current, for 2.5
Hz and 5.0 Hz operation respectively.

\lhead{AGS Super Neutrino Beam Facility} \rhead{Eddy Current
Effects} \rfoot{April 15, 2003}
\subsubsection{Eddy Currents in the Wall of the Vacuum Chamber}
\label{sec:eddycurinwall} The time varying magnetic flux generated
by the excitation of the main magnet generates eddy currents in
the wall of the vacuum chamber of the circulating beam.

\begin{figure}[htbp]
\begin{center}
\centerline{\includegraphics[width=4.68in,height=3.33in]{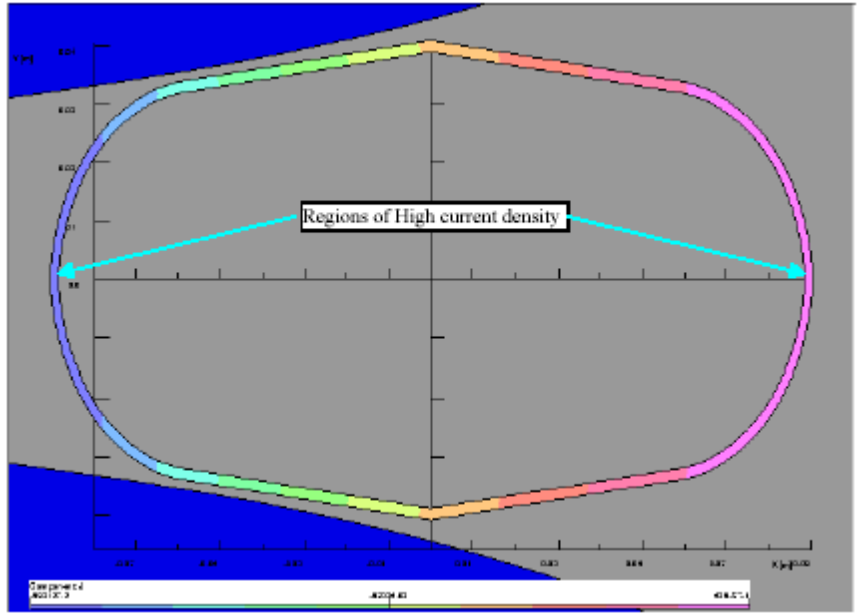}}
\caption{The cross section of the vacuum chamber with the
magnitude of the eddy current density shown on the walls of the
vacuum chamber with different colors. The regions of high current
density are at left and right edges of the vacuum chamber. The
regions of low density are at the top and bottom of the vacuum
chamber.} \label{fig:vccs}
\end{center}
\end{figure}

The eddy currents generated on the wall of the vacuum chamber have
the following adverse effect:
\begin{description}
\item[a.]Ohmic heating on the wall of the vacuum chamber.
\item[b.] Introduce magnetic multipoles including dipole field.
\end{description}
Figure ~\ref{fig:vccs} shows the cross section of the vacuum
chamber with the magnitude of the eddy current density shown on
the walls of the vacuum chamber with different colors. The regions
of high current density are on the left and right edges of the
vacuum chamber. The regions of low density are at the top and
bottom of the vacuum chamber.

\begin{figure}[htbp]
\begin{center}
\centerline{\includegraphics[width=3.2832in,height=2.9792in]{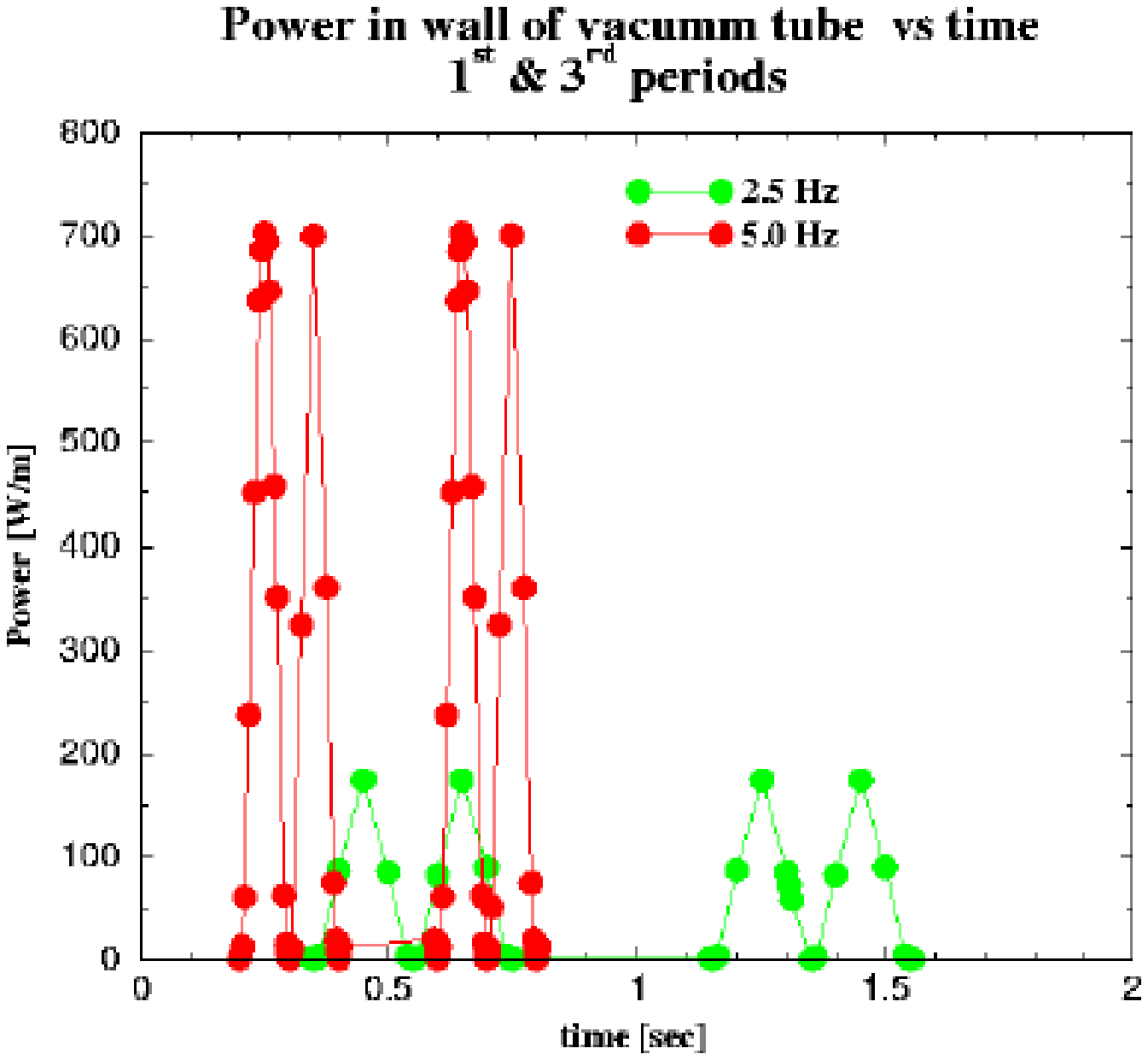}}
\caption{Ohmic losses per unit length due to eddy currents in the
walls of the vacuum chamber for 2.5 Hz operation (green curve) and
5.0 Hz operation (red curve). Only the Ohmic losses of the
1$^{st}$ and the 3$^{rd}$ periods are plotted in order to
ascertaine that the eddy currents acquired a steady-state.}
\label{fig:powerinwall}
\end{center}
\end{figure}

The calculated Ohmic losses per unit length due to the eddy
currents generated on the walls of the vacuum chamber are plotted
in Figure ~\ref{fig:powerinwall} as a function of time. The Ohmic
losses have been calculated for two excitations of the main
magnet; 2.5 Hz (green curve) and 5.0 Hz (red curve). The eddy
current Ohmic losses as a function of time follow the same pattern
as those in the coil of the main magnet (see previous section).

\begin{figure}[htbp]
\begin{center}
\centerline{\includegraphics[width=6.175in,height=4.0in]{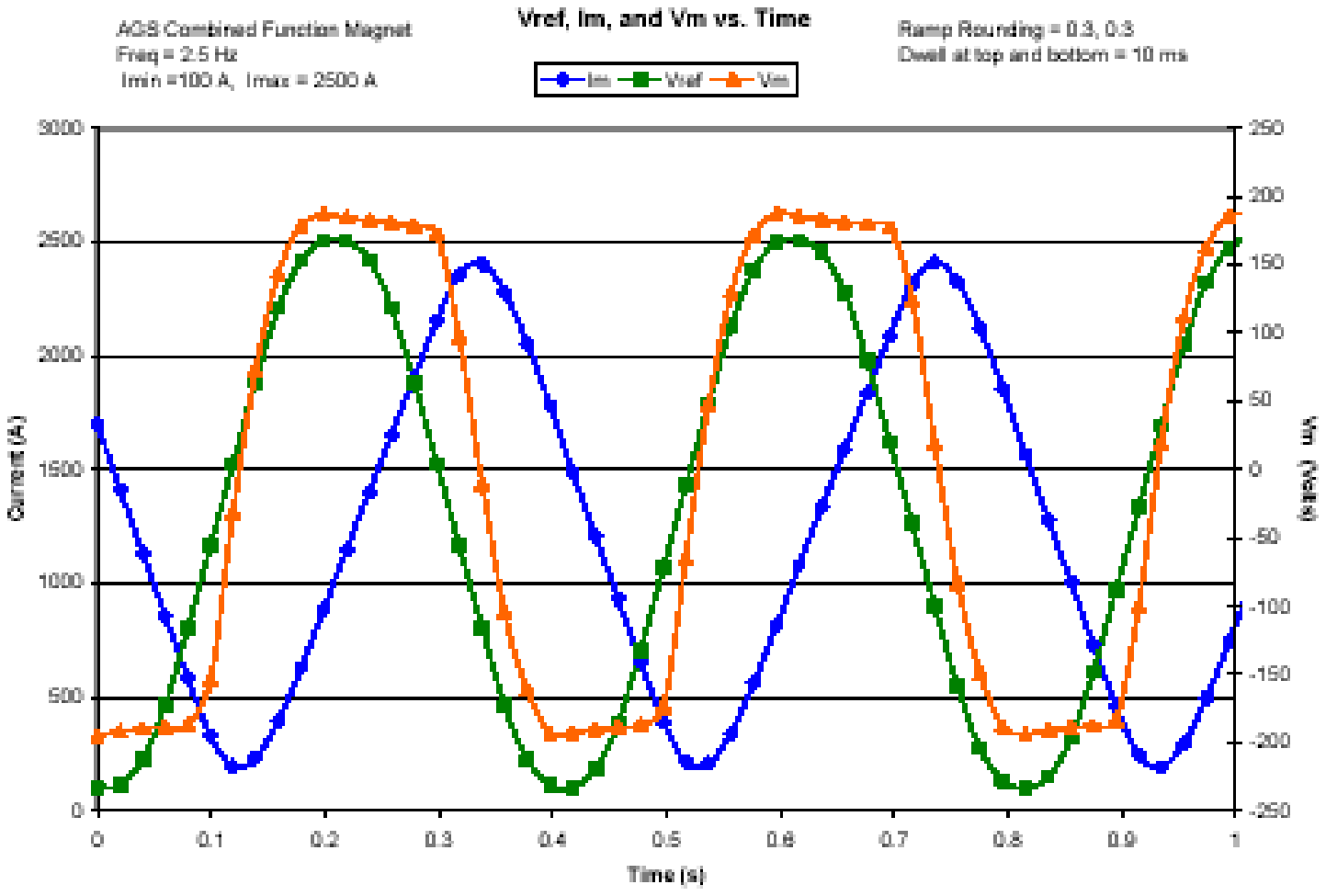}}
\caption{The current I in the coil of the AGS as a function of
time.} \label{fig:ctforags}
 \end{center}
\end{figure}

\begin{figure}[htbp]
\begin{center}
\centerline{\includegraphics[width=5.32in,height=2.92in]{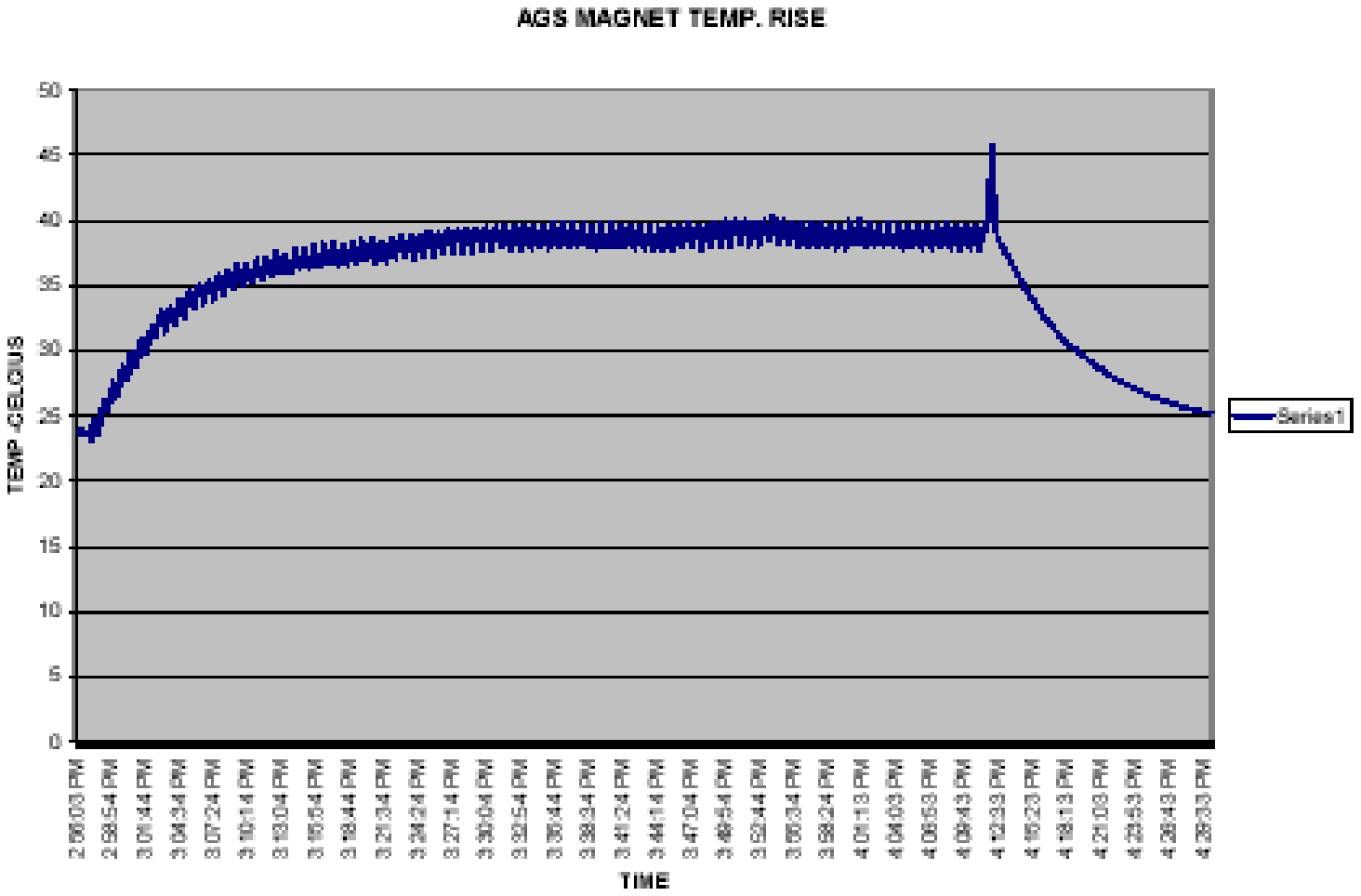}}
\caption{The temperature of the vacuum chamber as a function of
time while the current in the coil of the magnet has the
functional form shown in Figure ~\ref{fig:ctforags}.}
\label{fig:vctemp}
\end{center}
\end{figure}

Experimental measurements of the temperature rise of the vacuum
chamber of the AGS have been performed for a single AGS c-type
magnet when the coil of the magnet is subject to time varying
sinusoidal current as shown in Figures ~\ref{fig:ctforags}, and
~\ref{fig:vctemp}

\lhead{AGS Super Neutrino Beam Facility} \rhead{Eddy Current
Effects} \rfoot{April 15, 2003}
\subsubsection{Magnetic Multipoles Generated by the Eddy
Currents}

\label{sec:mpolebyeddy} The eddy currents generated on the wall of
the vacuum chamber will affect the magnetic field at the
circulating beam region. Due to the median plane symmetry of the
magnet we only consider the normal multipoles which are define in
this report with the relation:
 \vskip 0.5in

 B$_{r}$=
b$_{dip}$sin($\theta )$+b$_{quad}$r$^{1}$cos(2$\theta
)$+b$_{sex}$r$^{2}$cos(3$\theta ){\rm g}\theta
)$+b$_{oct}$r$^{3}$cos(4$\theta )$+b$_{dec}$r$^{4}$cos(5$\theta
)$+\ldots \ldots .

\vskip 0.5in

\begin{figure}[htbp]
\begin{center}
\centerline{\includegraphics[width=4.256in,height=3.168in]{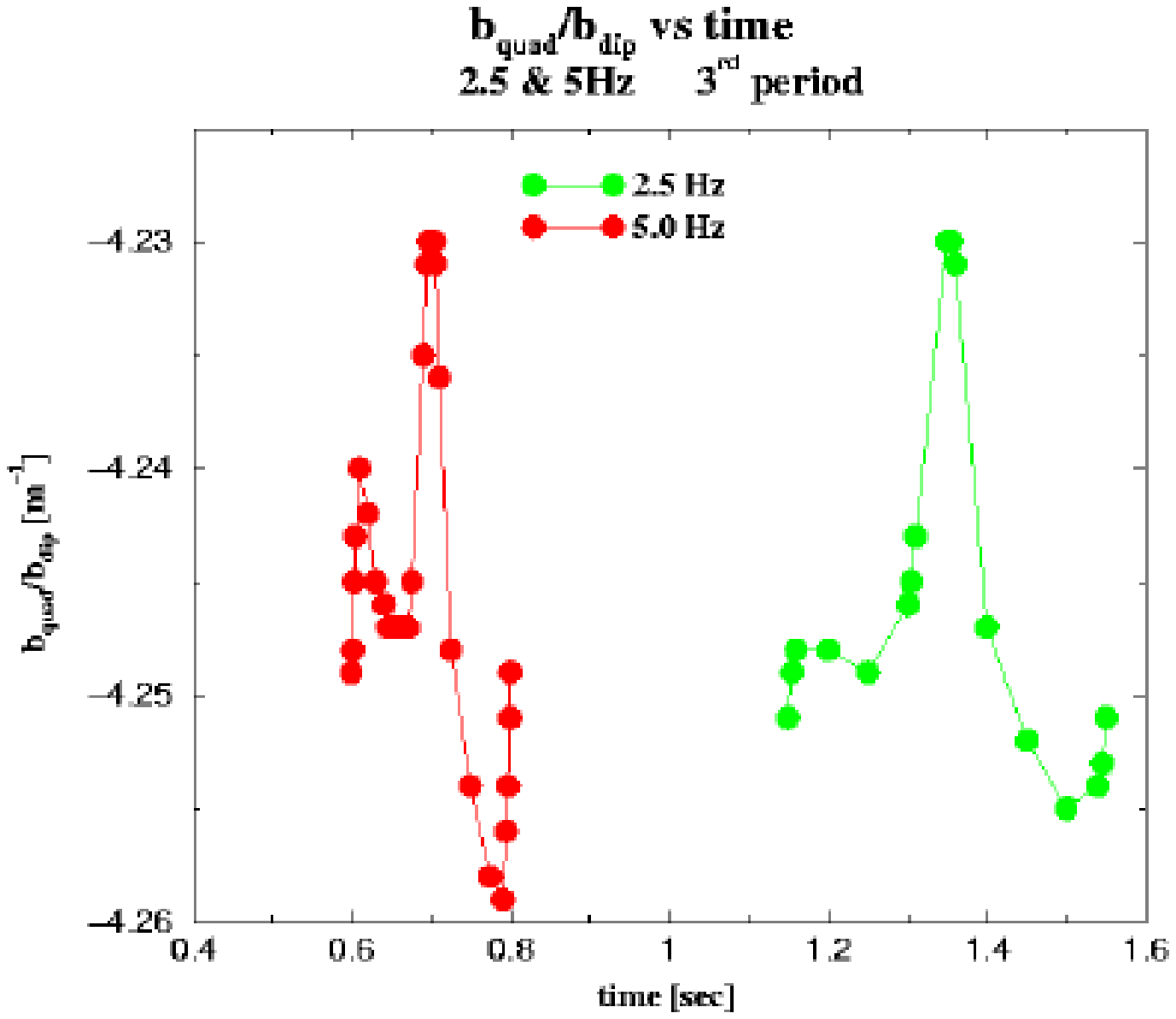}}
\caption{The ratio b$_{quad}$/b$_{dip}$ as a function of time
during the 3$^{rd}$ period. The multipoles for any subsequent
period is identical to these shown in the figure. }
\label{fig:qdratio}
\end{center}
\end{figure}

\begin{figure}[htbp]
\begin{center}
\centerline{\includegraphics[width=4.256in,height=3.088in]{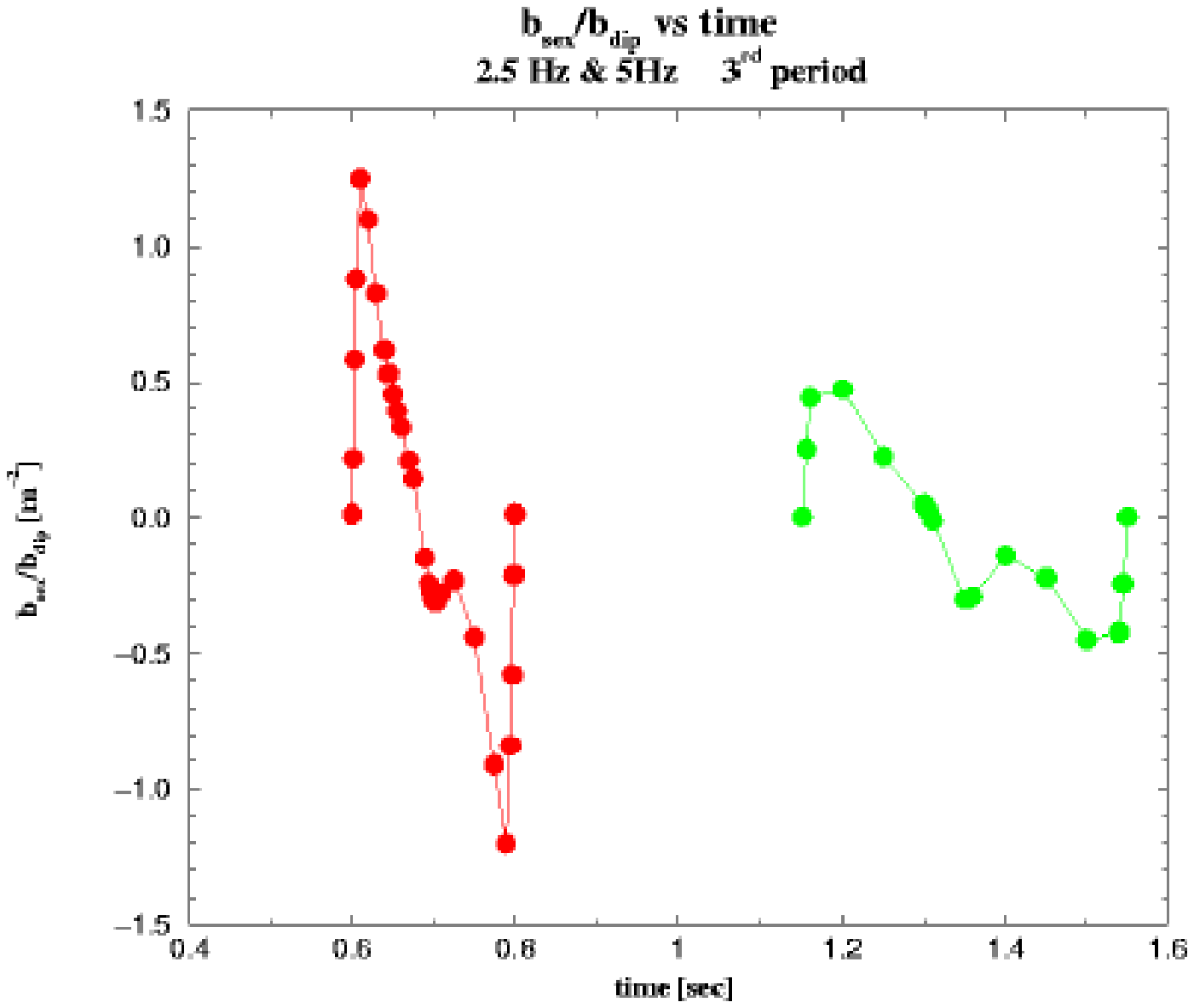}}
\caption{ The ratio b$_{sex}$/b$_{dip}$ as a function of time
during the 3$^{rd}$ period. The multipoles for any subsequent
period is identical to these shown in the figure.}
\label{fig:sdratio}
\end{center}
\end{figure}

\begin{figure}[htbp]
\begin{center}
\centerline{\includegraphics[width=4.0in,height=4.0in]{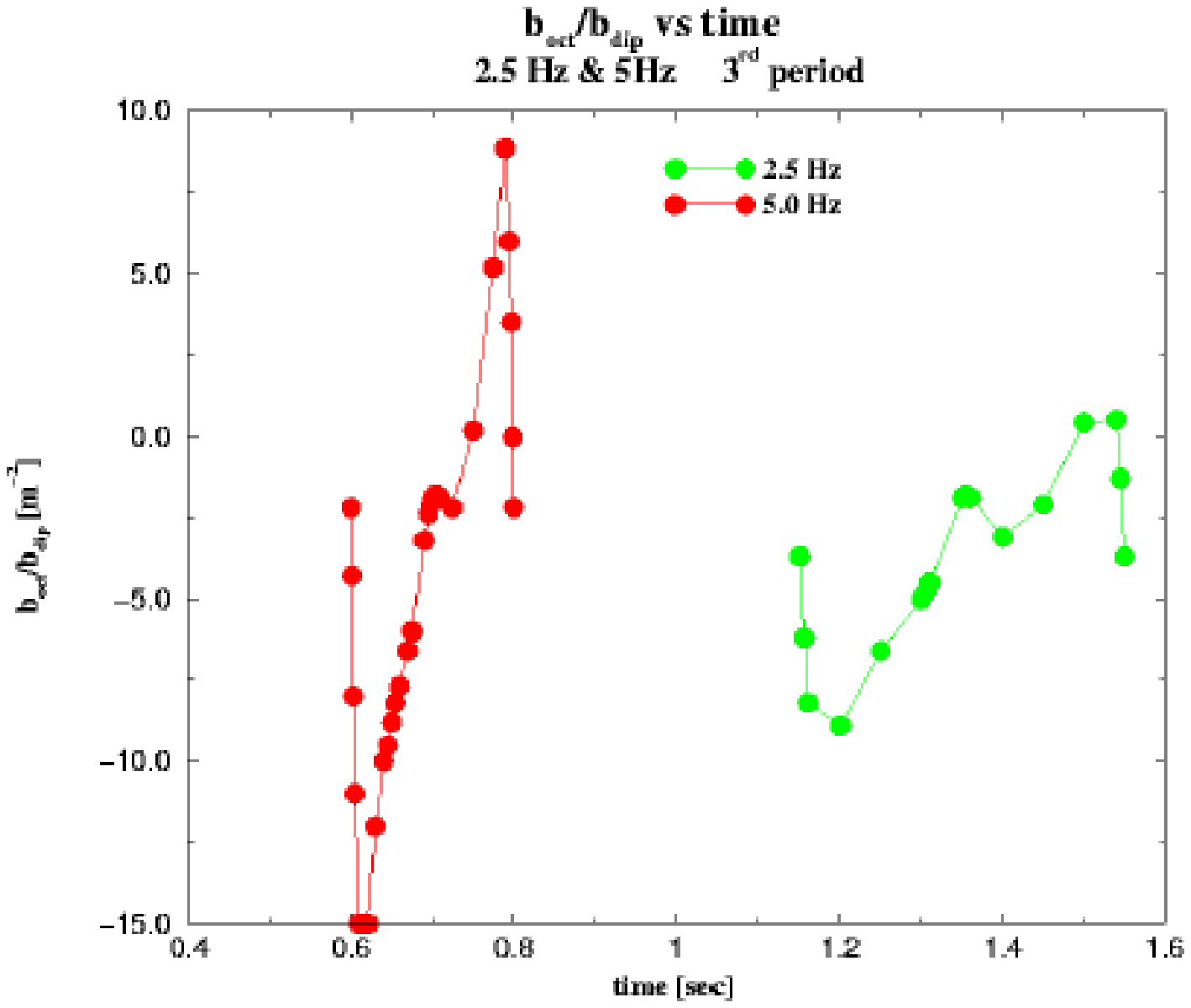}}
\caption{ The ratio b$_{oct}$/b$_{dip}$ as a function of time
during the 3$^{rd}$ period. The multipoles for any subsequent
period is identical to these shown in the figure. }
\label{fig:odratio}
\end{center}
\end{figure}

 Figures ~\ref{fig:qdratio},
~\ref{fig:sdratio} and ~\ref{fig:odratio} plot ratios
b$_{quad}$/b$_{dip}$, b$_{sex}$/b$_{dip}$, and
b$_{oct}$/b$_{dip}$, of the magnetic multipoles as a function of
time during the 3$^{rd}$ period from the start of the excitation
of the main magnet. The calculation have shown that a steady has
been achieved, therefore the functional form of the multipoles for
any subsequent period is identical to the one shown in these
figures.

The 5.0 Hz excitation of the main magnet generates larger
variation of the magnetic multipoles than the 2.5 Hz excitation.
The effect of the calculated multipoles on the circulating beam
will be studied in a separate report.

In conclusion,
the scheme for a 1~MW proton driver based on the AGS with upgraded
injection is feasible. Indeed, the AGS beam intensity is only
modestly higher than during the present high-intensity proton
operation and, therefore, beam instability is not expected to be a
problem during acceleration.


\clearpage
\newpage
\lhead{AGS Super Neutrino Beam Facility} \rhead{AGS Injection and
Extraction} \rfoot{April 15, 2003}
\section{AGS Injection and Extraction}
\label{sec:agsinjext}
 \setcounter{table}{0}
 \setcounter{figure}{0}
 \setcounter{equation}{0}
 A high intensity proton beam of $\sim $8.9x10$^{13}$ protons at an
energy of 28 GeV is required to irradiate  the production target
at a repetition rate of 2.5 Hz. This high intensity proton beam
can be produced  by injecting  1.2 GeV H$^{ - }$ ion from   a
superconducting linac (see Section ~\ref{sec:tscl}).  This section
provides  a feasibility study of injection, acceleration and
extraction of such a high intensity proton beam at a frequency of
2.5 HZ using the AGS machine. The following items have been
studied here:

 \begin{enumerate}
\item Injection of the H$^{-}$ Beam:
\begin{description}

\item[a.] The geometry of the H$^{ - }$ injection: Location of the
injection region in AGS ring and required devices in the injection
region.

\item[b.] The calculation of the  beam parameters of the circulating beam in AGS at the
H$^{-}$ injection point. These parameters are required for the
matching of the beam parameters of the H$^{-}$ injected beam to
those of the circulating beam.
\end{description}
\item "Stripping Foil" issues generated by the partially stripped
H$^{-}$ injected beam:
\item Acceleration:
\begin{description}
\item[a.] The effect that the AGS-main-magnet cycle generates on the vacuum
chamber of the AGS ring during the 2.5 Hz ramping of the main
magnet.
\item[b.] Calculations of the magnetic multipoles introduced, in the main magnetic
field of the AGS synchrotron, by the eddy currents of the vacuum
chamber.
\end{description}

\item Extraction:
\begin{description}
\item[a.] The geometry of the extraction region. Location of the extraction region in
the AGS ring, and required devices for beam extraction in the
extraction region.
 \item[b.]The calculation of the beam parameters of the circulating beam in the AGS at
the extraction point. These beam parameters are required for the
matching of the beam parameters of the extracted beam, to those of
the transport beam line to the ``neutrino production'' target
(RTBT).
\end{description}
\end{enumerate}

\lhead{AGS Super Neutrino Beam Facility} \rhead{ H$^{ - }$
Injection into AGS} \rfoot{April 15, 2003}
\subsection{H$^{ - }$   Injection into AGS}
\label{sec:h-injinags}
 Figure ~\ref{fig:sec3_one} is a drawing showing a
sections of the AGS with the main magnets and the layout of the
proposed superconducting linac that will accelerate the H$^{-}$
ions to an energy of 1.2 GeV  for injection into the AGS. A
section of the AGS-Booster ring appears in the upper left corner
of the drawing. The H$^{ - }$ Injection region has been chosen to
be   the B20 straight section of the AGS.

\lhead{AGS Super Neutrino Beam Facility} \rhead{H$^{ - }$
Injection into AGS} \rfoot{April 15, 2003}
\subsubsection{Considerations for the 1.2 GeV H$^{ - }$ Injection into the AGS}
\label{sec:consinjintoags}

\begin{description}

\item[a.] The H$^{ - }$ ions are to be injected from the
SC{\_}linac into the AGS using the ``electron stripping'' method
\cite{ir32784}. The electrons of the injected H$^{ - }$ ions will
be stripped by a foil located inside the pipe of the circulating
beam of AGS. \textit{The issue is to determine the optimum
location of the stripping foil}.

\item[b.] The integrated magnetic field {\{}$\smallint $Bdl{\}} of the
magnets comprising the High Energy Beam Transport (HEBT) line that
transports the H$^{ - }$ ions from the exit of the SCL to the
injection points (stripping-foil location) should be minimized in
order to keep the ionization probability of the H$6{-}$ ions
\cite{ir62312} below  1x10$^{ - 6}$. For 8.9x10$^{13}$ H$^{ - }$
ions injected into AGS at 2.5 Hz,  an ionization probability of
1x10$^{ - 6}$ is equivalent to 0.05 W of energy deposited into the
HEBT line. This constraint on the allowed upper limit ($<$ 1x10$^{
- 6 })$ of the ionization probability of the H$^{ - }$ ions, sets
an upper limit on the maximum value of the magnetic fields of the
magnets needed to bend and focus the H$^{ - }$ injected beam.

\item[c.] The HEBT line must provide space for Dipoles and Quads to be
used to match the beam parameters of the injected H$^{ - }$ beam
to those of the circulating proton beam of the AGS at the
injection point. (The injection point is defined as the point
where the central trajectory of the injected H$^{ - }$ beam
intersects the stripping foil).

\item[d.] In order to avoid the coherent betatron-oscillations that
protons of the injected beam could perform while circulating in
the AGS, the injected H$^{ - }$ beam and the central orbit of the
circulating beam must be collinear at the injection point. This
can be accomplished  either by, locally displacing the circulating
beam ``locally bumping the beam'' at the location of the injection
point; and/or by adjusting the injection angle of the H$^{ - }$
injected beam. Both methods will be discussed.

\item[e.] In order fill the phase space of the circulating beam with
injected beam, the central orbit of the circulating beam is
displaced ``bumped locally'' in the horizontal and vertical
direction during the H$^{ - }$ beam injection. To implement the
displacement of the central orbit of the circulating beam we use
the same or similar hardware that is used to implementing the
``local bump'' in (e). The method to ``locally bump'' the beam
will be discussed in a technical note. At time varying bump will
also help reduce the number of ``foil crossing'' of the
circulating beam during the time the AGS is being injected with
H$^{ - }$ beam.

\item[f.] The electrons generated by stripping of the H$^{ - }$ beam at
the stripping foil, as well as the H$^{0}$ and H$^{ - }$ particles
that may result from incomplete stripping, should be removed from
the injected beam with a controlled method. A proposed method of
``dumping properly'' all particles emerging from the stripping
foil( except the protons) will be discussed.
\end{description}

\lhead{AGS Super Neutrino Beam Facility} \rhead{H$^{ - }$
Injection into AGS} \rfoot{April 15, 2003}
\subsubsection{Layout of the H$^{ - }$ Injection Region}
\label{sec:layoutinjreg}
\begin{figure}[h]
\begin{center}
\resizebox{6.0in}{2.5in}{\includegraphics*[0.5in,0.0in][8.0in,4.00in]{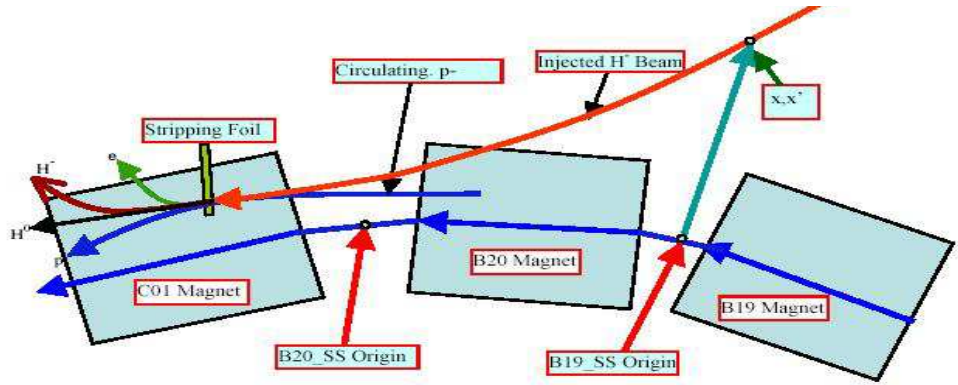}}
\end{center}
\caption{\label{fig:injreg}Schematic diagram of the H- Injection
Region.}
\end{figure}

 A schematic diagram of the Injection region
is shown in Figure ~\ref{fig:injreg}. In this diagram the
following items are shown:

\input epsf

\begin{description}

\item[a.]  Three of the main AGS combined-function main magnets
(B19,B20, and C01).
\item[b.] The stripping foil which is shown
inside the magnet C01.

\item[c.] Part of the closed orbit of the circulating proton-beam. The
closed orbit is locally displaced ``bumped'' to the outside, and
at the location of the stripping foil. The reason for ``locally
bumping'' the closed orbit is partly described in Section
~\ref{sec:consinjintoags}
 d), e) above. Section
~\ref{sec:lbbir} below describes the method to generate the
locally bumped orbit. At the end of the H$^{ - }$ injection time
interval, the ``locally-bumped'' orbit of the circulating beam
will return to a new orbit which is not bumped (labeled as
``Circulating p-beam'' in Figure ~\ref{fig:injreg} )

\item[d.] The trajectory of the injected H$^{ - }$ beam. The direction
of the injected H$^{ - }$ is defined by the coordinate point
(x,x') shown in the diagram. The coordinate x is the distance from
the center point of the straight section SS{\_}B19 and x' is the
angle between the direction of the H$^{ - }$ and the SS{\_}B19. At
this distance x, the influence of the fringe field of the AGS main
magnets on the H$^{ - }$is negligible. As the injected beam
approaches the injection point (stripping foil shown in Figure
~\ref{fig:injreg}) the fringe field and main field of the AGS main
magnets B19 and C01 become significant and are taken into account
in the calculations that determine: (1) the location of the
stripping foil inside the magnet C01; (2) the beam parameters at
the injection point (see Section ~\ref{sec:bpipmp} below).

\item[e.] The trajectories of the H$^{ - }$ beam that are not
stripped by the stripping foil ( the partially stripped H$^{ - }$
beam and the  electrons emanating from the stripping foil)  must
all be directed  into a ``dump'' downstream of the stripping foil.
\end{description}
\lhead{AGS Super Neutrino Beam Facility} \rhead{Local Beam Bump at
the Injection Region } \rfoot{April 15, 2003}
\subsubsection{The ``Local Beam-Bump'' at the Injection Region }
\label{sec:lbbir}

\begin{figure}[h]
\begin{center}
\resizebox{4.8in}{4.0in}{\includegraphics*[0.5in,0.5in][8.5in,8.5in]{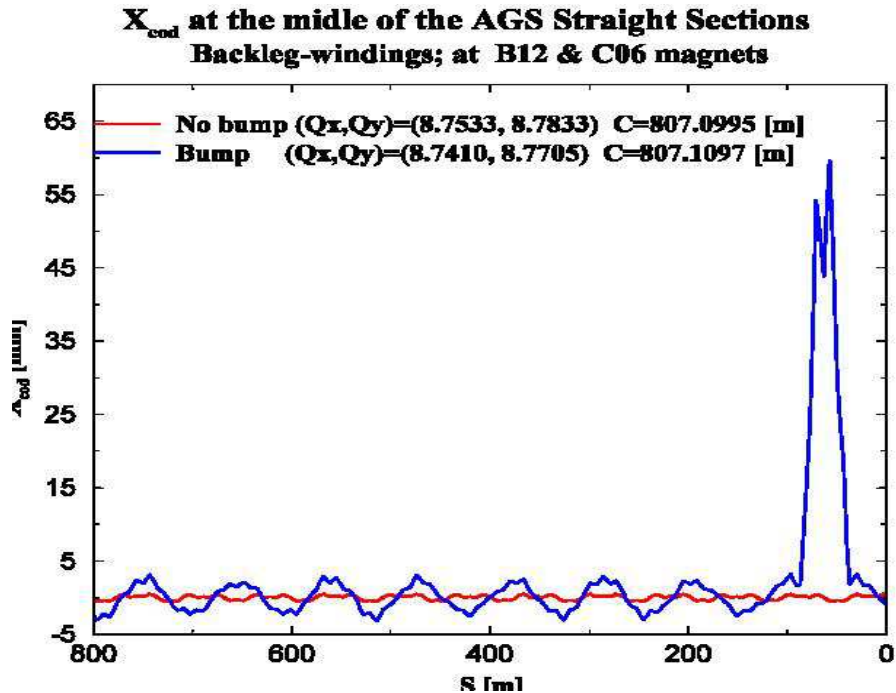}}
\end{center}
\caption{\label{fig:agscox}The transverse displacement of the AGS
closed orbit X$_{cod}$ at the middle of the straight sections of
AGS. The red and the blue lines correspond to the X$_{cod}$
without bumps and with bumps respectively. }
\end{figure}

 In order to
make the circulating proton beam collinear with the H$^{ - }$
injected beam at the injection point (stripping foil), circulating
proton beam is ``locally bumped'' by using two horizontal "bump-
magnets". The first "bump-magnet" is located at $\sim $-90$^{0}$
phase advance, upstream of the stripping foil (location of the
first "bump-magnet" is at the straight section SS{\_}B12) , and
the second "bump-magnet" is located at $\sim $+90$^{0}$ phase
advance downstream of the stripping foil (location of the second
"bump-magnet" is at the straight section SS{\_}C06). Figure
~\ref{fig:agscox} plots the displacement (y-axis) of the closed
orbit at the center of the straight sections (x-axis) of the AGS
for the two cases:
\begin{description}
\item[a.]When the ``bumped-magnets'' are turned off, and
the orbit (red line) is not ``locally bumped'';
\item[b.]when the ``bumped-magnets'' are turned on , and the orbit (blue line) is
``locally bumped''. The strength of the ``bumped-magnets'' whose
location is  systematically shown in Figure ~\ref{fig:agscox}, is
$\sim $3 mrad each. The horizontal and vertical tunes
(Q$_{x}$,Q$_{y})$ and the lengths of the closed orbits for the two
cases is shown in the Figure ~\ref{fig:agscox}.
\end{description}

\input epsf

\begin{figure}[h]
\begin{center}

\resizebox{4.8in}{4.0in}{\includegraphics*[0.5in,0.5in][7.5in,7.5in]{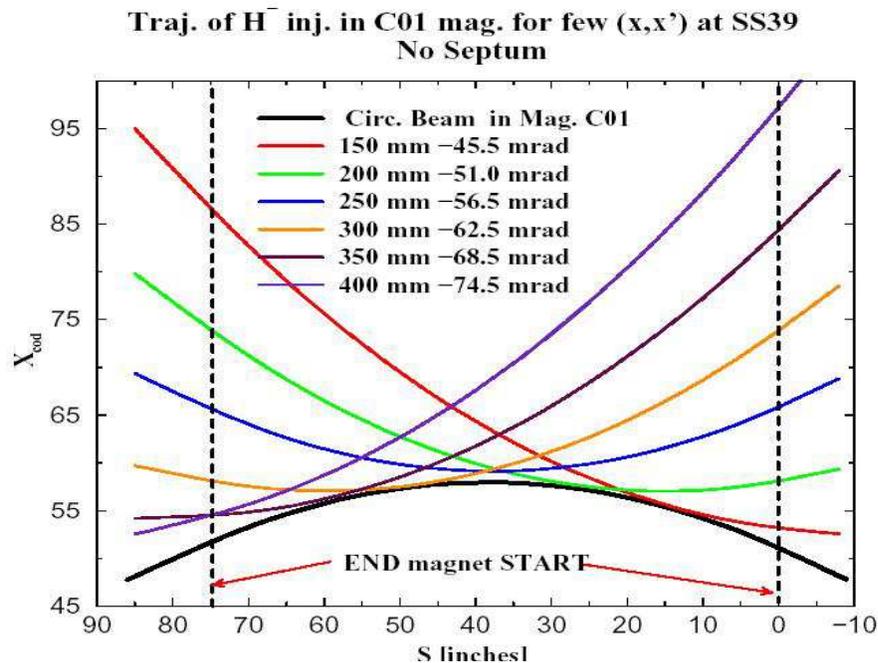}}
\end{center}
\caption{\label{fig:agsbumpcox}The black line corresponds to  the
``bumped closed orbit'' X$_{cod}$ in the AGS magnet CO1. This
orbit has been generated using two magnet located at $\pm
$90$^{0}$ phase advance from the C01 magnet (see text). The C01
magnet is 75 inches long. The rest of the lines are various
trajectories of the H$^{ - }$ injected beam. Each of these
trajectories corresponds to a different initial condition (x,x')
of the injected H$^{ - }$ beam. }
\end{figure}

Figure ~\ref{fig:agsbumpcox} shows the displacement (y-axis) of a
section of the ``locally bumped'' closed orbit (black line) at the
location of the C01 main magnet of the AGS. The maximum
displacement ($\sim $56 mm) of the closed orbit occurs at the
center ($\sim $38 inches) of the 75 inches long C01 magnet. This
displaced orbit has been generated using the two magnet mentioned
earlier which are located at $\pm $90$^{0}$ phase advance from the
C01 magnet (this section).Each of the other lines are trajectories
of the H$^{ - }$ injected beam corresponding to a different (x,x')
direction of the H$^{ - }$ injected beam (see Figure
~\ref{fig:injreg}, Figure ~\ref{fig:agsbumpcox} and Section
~\ref{sec:layoutinjreg} (d) for definition of x,x').

The values of (x x') that were used to generate the trajectories
shown in Figure ~\ref{fig:agsbumpcox} are shown also on the same
Figure ~\ref{fig:agsbumpcox}. Each of the H$^{ - }$ trajectories
is tangent to the ``bumped-closed orbit'' (black line) and this
has been achieved by adjusting the initial conditions (x,x') of
the H$^{ - }$ injected beam.

\input epsf

From all the possible H$^{ - }$ trajectories shown in Figure
~\ref{fig:agsbumpcox}, the H$^{ - }$ trajectory that corresponds
to the (x,x')=(250 mm, -56.5 mrad) has been selected as the
``injection trajectory''. Both, the closed orbit calculations for
the derivation of the X$_{cod}$ as a function of distance, and the
various H$^{ - }$ trajectories shown in Figure
~\ref{fig:agsbumpcox} have been calculated using the code
AGS{\_}BATE\cite{agsbate}. This code is a modification of the code
BEAM. The code uses the experimentally measured fields at the
median plane of the AGS magnets to ray trace the charged
particles. The code also provides  transport matrices at any
location along a trajectory and beam parameters at any location
along a closed orbit. Detailed description of the code and
additional references are given in reference \cite{rhicrd75}. In
addition to the ``local-bump'' of the closed orbit discussed
above, a set of two fast time varying bumps (horizontal and
vertical) of the closed orbit my also be superimposed to help
optimize the ``beam painting'' of the AGS phase space.

\lhead{AGS Super Neutrino Beam Facility} \rhead{Beam Parameters at
Injection Point } \rfoot{April 15, 2003}
\subsubsection{Beam Parameters at the H$^{ - }$
Injection Point } \label{sec:bpipmp}

In order to minimize the emittance blow-up of the injected H$^{ -
}$ beam, the beam parameters of the injected H$^{ - }$ beam must
match those of the circulating beam at the ``injection point'',
which is chosen to be the location of the stripping foil.

The beam parameters of the circulating beam in AGS at the
``injection point'' have been calculated at the injection beam
energy of 1.2 GeV, using the computer code
AGS{\_}BATE\cite{agsbate} and these beam parameters appear in
Table ~\ref{tab:bpatipmp}.

\begin{figure}[htbp]
\begin{center}
\centerline{\includegraphics[width=4.512in,height=4.875in]{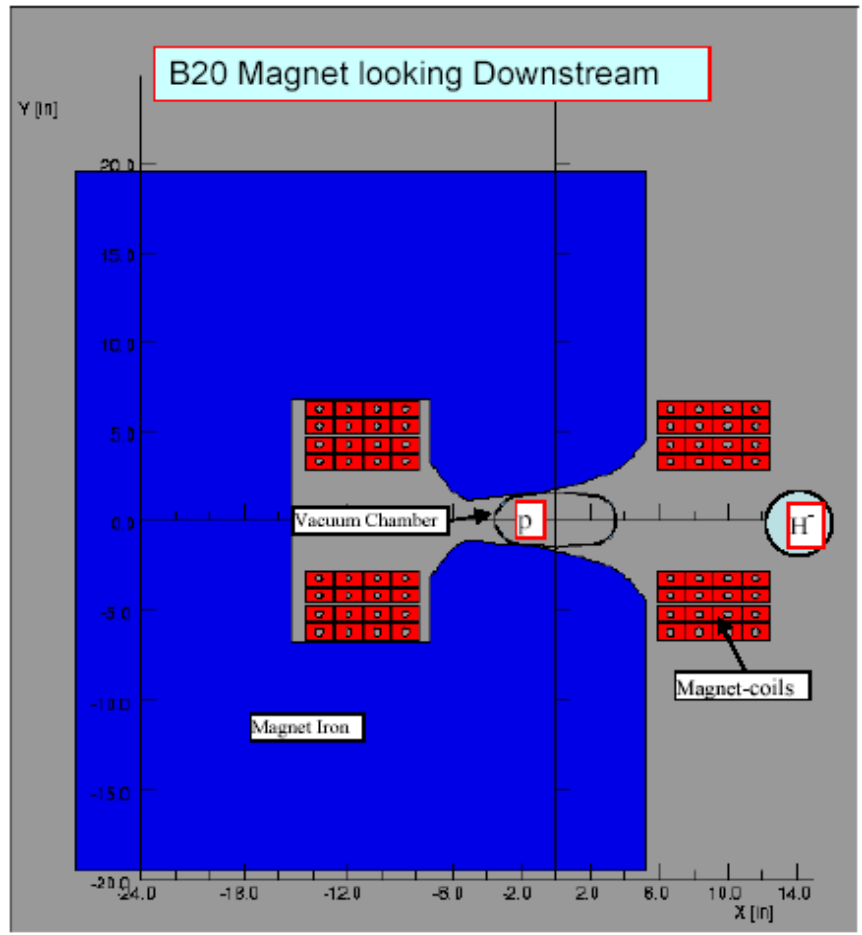}}
\caption{ Cross section of the B20 combined-function main magnet
of the AGS, looking downstream. The oval line between the pole
pieces is the vacuum chamber of the circulating proton beam. The
circle shown between the top and bottom coils (red regions) of the
magnet is an approximate location of the vacuum chamber of the
H$^{ - }$ injected beam.} \label{fig:b20cs}
\end{center}
\end{figure}

The proximity of the H$^{ - }$ injection line to the AGS main
magnet B20, shown in Figure ~\ref{fig:b20cs}, suggests that the
fringe field of the B20 magnet will affect the beam ellipse of the
H$^{ - }$ injected beam. Therefore, the fringe field of the B20
magnet as well as that of the C01 magnet (see Figure
~\ref{fig:injreg}) are included in the calculations of the first
order transfer matrices (R$_{H}$,R$_{V})$ which define the beam
transport between the ``Matching point'' (the point (x,x') shown
in Figure ~\ref{fig:injreg} is defined in this note as the
``matching point'') and the ``Injection point''.

Using the code AGS{\_}BATE\cite{agsbate} we have calculated the
first order horizontal and vertical transport matrices
(R$_{H}$,R$_{V})$ between the ``Matching point'' and ``Injection
point''. These matrices appear below and correspond to the central
trajectory with coordinate at the ``matching point'' (x,x')=(250
mm,-56.5 mrad).

\[
R_{H}= \left|
\begin{array}{cc}
 0.7015 & 6.058 \\
-0.0292 & 1.173
\end{array}
\right|
\]
\[
R_{V}=\left|
\begin{array}{cc}
 0.7015 &  6.058 \\
 0.00497 & 0.796
\end{array}
\right|
\]

In order simplify the matching of the beam parameters of H$^{ - }$
injected beam to those of the circulating beam we have calculated
the beam parameters at the ``Matching point'' (see Table
~\ref{tab:bpatipmp}) by using both, the values of the beam at the
``Injection Point'', and the values of the transfer matrices
R$_{H}$ and R$_{V }$ above.

\begin{table}[htbp]
\caption{ The beam parameters at the ``Injection Point'' and
``Matching Point'' (see text). } \label{tab:bpatipmp}
\begin{tabular}
{|p{117pt}|p{43pt}|p{43pt}|p{43pt}|p{43pt}|p{43pt}|p{43pt}|}
\hline Location& $\beta _{x}$[m]& $\alpha _{x}$& $\eta _{x}$[m]&
$\eta $'$_{x}$& $\beta _{y}$[m]& $\alpha _{y}$ \\ \hline Injection
 Point& 28.0& -1.00& -1.25& -0.05& 8.00& 0.43 \\ \hline
Matching Point& 26.6& -1.32& -1.00& -0.10& 16.10& 1.75 \\ \hline
\end{tabular}
\end{table}
\lhead{AGS Super Neutrino Beam Facility} \rhead{H$^{-}$,H$^{0}$
and Electron Beam Dump} \rfoot{April 15, 2003}
\subsubsection{H$^{ - }$, H$^{0}$ and Electron Beam Dump }
\label{sec:h-h0ebd}
 The H$^{ - }$ injected beam will not be
stripped totally of its electrons by the stripping foil which is
located at the injection point. Therefore the emerging beam from
the foil will consists of:
\begin{description}
\item[a.]protons (H$^{ + })$ which constitutes the circulating beam to be accelerated to 28
GeV.

\item[b.]H$^{ - }$ ions that did not loose any electrons passing
through the foil.

\item[c.]H$^{0}$ neutral particles that lose only one electron.

\item[d.] electrons ($e^{-}$).
\end{description}
Except for the fully stripped H$^{ - }$ injected beam, which is
the circulating proton beam, the rest of the particles (H$^{ - }$,
H$^{0}$ , and the electrons) must be ``dumped'' into a material
which can absorb the deposited energy. The different rigidity of
these three particles species, may require a specially designed
``beam dump''. Similar problems of how to dump the partially
stripped H$^{ - }$beam have been encountered in the 200 MeV H$^{ -
}$injection in AGS\cite{ir32784}, and the 200 MeV H$^{ -
}$injection in the Booster\cite{BTN195A}. A study for the design
of the beam dump of the partially stripped 1.2 GeV H$^{ - }$beam
is required.


 \lhead{AGS Super Neutrino Beam Facility} \rhead{Extraction from
AGS} \rfoot{April 15, 2003}

\subsection{Extraction from AGS at 28.0 GeV}
\label{sec:extags} At present the AGS synchrotron is equipped with
a Fast Beam Extraction (FEB) system which is used to extract the
circulating bunched beam in AGS to the AtR beam transfer line. The
FEB system will be used for the extraction of the 28 GeV high
intensity proton beam which will irradiate the ``neutrino
production'' target. A section (the U-line) of the AtR beam
transfer line which is used to transfer the extracted beam from
AGS into the Relativistic Heavy Ion Collider (RHIC) will also be
used to transport  the 28 GeV proton beam to the ``neutrino
production'' target. Detailed description of the FEB system has
been presented in previous technical notes
\cite{rhicrd75},\cite{caap54} therefore in this report we will
only outline the main features of the (FEB) system.

\lhead{AGS Super Neutrino Beam Facility} \rhead{Layout of the
Extraction Region} \rfoot{April 15, 2003}
\subsubsection{Layout of the Extraction Region}
\label{sec:layoutext}
 The main steps of the procedure for fast
beam extraction are outlined below:
\begin{description}
\item[a.]Beam acceleration from injection energy to 28 GeV.
\item[b.] Generate two ``local beam bumps'', one
(BLWG09 see Figure ~\ref{fig:FEB}) which displaces the circulating
beam into the C-type Fast kicker magnet (FK) which will ``kick''
the beam out of the AGS, and the other (BLWH11 see Figure
~\ref{fig:FEB} ) which displaces the circulating beam close to the
H10 septum magnet (SM) which is located downstream from the G10
kicker at $\sim $270$^{o}$ phase advance. It is advantageous to
use a C-type Fast kicker magnet which is displaced from the ideal
orbit, instead of a full aperture kicker
\item[c.]The circulating beam bunches  pass through the aperture of the G10
Fast kicker, and are kicked by the G10 Fast kicker which displaces
the beam bunches on the other side of the H10 septum, into the
U-line (see Figure ~\ref{fig:FEB} ) which will be used as the
first  section of the beam transport line to the ``Neutrino
production'' target. A schematic diagram of the FEB system in AGS
is shown in Figure ~\ref{fig:FEB}. In this figure shown are:
\end{description}
\noindent (i) the BLW{\_}G09 and BLW{\_}H11 ``local-bumps'';

\noindent (ii) the Fast Extraction kicker (FK);

\noindent (iii) the Beam Extraction Septum Magnet (SM);

\noindent (iv) and the beginning of the U-line.

Detailed description of the FEB system is given in references
\cite{rhicrd75},and \cite{caap54}.

\lhead{AGS Super Neutrino Beam Facility} \rhead{Extraction
Devices} \rfoot{April 15, 2003}
\subsubsection{Extraction Devices}
\label{sec:extdev}
\begin{figure}[htbp]
\begin{center}

\centerline{\includegraphics[width=5.46in,height=4.23in]{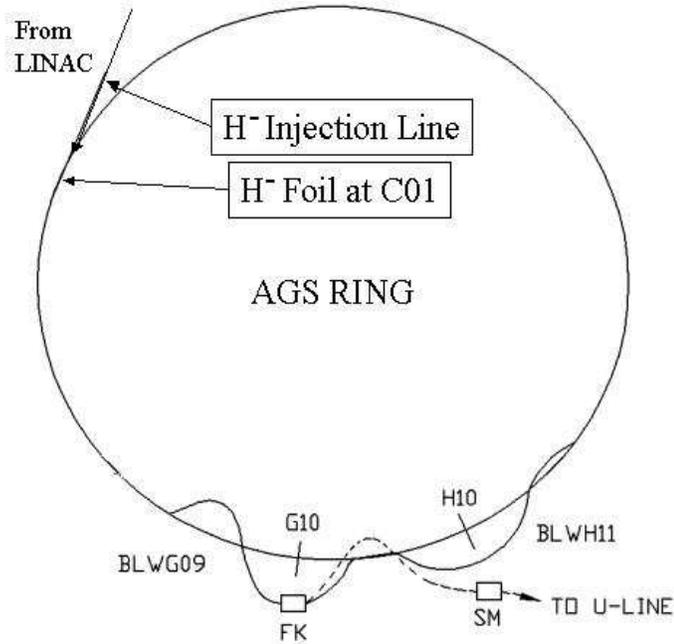}}
\caption{ Schematic diagram of the AGS RING, Fast-Beam-Extraction
(FEB) system. Shown are: a) the BLW{\_}G09 and BLW{\_}H11
``local-bumps'' b) the Fast Extraction kicker (FK) c) the Beam
Extraction Septum Magnet (SM) d) and the beginning of the U-line.}
\label{fig:FEB}
\end{center}
\end{figure}

 In this section we describe  briefly the
function of the devices which are used for the FEB in the time
order the devices are involved in the fast beam extraction.
 \begin{enumerate}
\item The two 3$\lambda $/2 ``local beam bumps''
  BLWG09 and BLWH11 (see Figure ~\ref{fig:FEB} )
 Each of the bumps is generated by back-leg (BLW) windings on the
return iron of specified main magnets of the AGS.
\begin{description}
\item[a.]The BLWG09 ``beam-bump'' which displaces the closed orbit inside the
aperture of the G10 fast kicker that is displaced radially from
the ideal orbit on the outside of the AGS ring, because it has
acceptance large enough to accept the size of the 28 GeV beam but
not large enough to accept the size of the injected beam. A
proposed ``full-aperture'' kicker which can accept the size of the
injected beam, does not need to be displaced radially nor does the
beam need to be bumped by the BLWG09 into the kicker. The large
gap of a ``full aperture'' kicker, however, will increase the
power supply requirements of the fast kicker.
\item[b.] The BLWH11 ``beam-bump'' which displaces the closed orbit of the
circulating beam near the septum of the H10 extraction magnet
(SM). This displacement of the beam closed to the septum magnet
reduces the strength requirements of the kicker to displace the
beam into the other side of the septum and into the aperture of
the magnet.
\end{description}
\item The G10 fast extraction kicker (FK Figure ~\ref{fig:FEB} )
which kicks each circulating beam-bunch into the aperture of the
septum magnet.
\item The H10 extraction septum magnet
(SM Figure ~\ref{fig:FEB} ) which extracts the beam bunch from the
AGS into the U-line.
\end{enumerate}

\lhead{AGS Super Neutrino Beam Facility} \rhead{Beam Parameters at
the Extraction Point } \rfoot{April 15, 2003}
\subsubsection{Beam Parameters at the Extraction Point}
\label{sec:bpatep} The extracted beam to be transported to the
``neutrino production'' target will use part of the AtR transport
line, the U-line which will be extended to the production target.
The beam transport line is discussed in more detail in one of the
chapters of this report. In this section we only report the beam
parameters of the extracted beam (Table ~\ref{tab:exttwiss} at the
extraction point of the transport line. The knowledge of the beam
parameters of the extracted beam at the beginning of the transport
line are required for the design of the transport line. Details of
the method which is used to calculate the beam parameters and the
calculation of the beam parameters are given in references
\cite{rhicrd75} and \cite{caap42}.

\begin{table}[htbp]
\begin{center}
\caption{The beam parameters at the ``Extraction Point'' of the
beam transport line to the ``neutrino production'' target (see
text). The dispersion and angular dispersion ($\eta _{x}\eta
$'$_{x })$ were not measured.}
 \label{tab:exttwiss}
\begin{tabular}
{|p{176pt}|p{36pt}|p{36pt}|p{43pt}|p{39pt}|p{43pt}|p{36pt}|}
\hline Location& $\beta _{x}$[m]& $\alpha _{x}$& $\eta _{x}$[m]&
$\eta $'$_{x}$& $\beta _{y}$[m]& $\alpha _{y}$ \\ \hline
Extraction Point (Theory)& 37.5& -4.1& -1.5& -0.13& 6.5& 0.85 \\
\hline Extraction Point (Measr.)& 36.7& -4.5& & & 7.7& 1.2 \\
\hline
\end{tabular}

\end{center}
\end{table}

\clearpage
\newpage
\lhead{AGS Super Neutrino Beam Facility} \rhead{ Beam Transport to
Target} \rfoot{April 15, 2003}

\section{Beam Transport to Target}
\label{sec:btt}

\setcounter{table}{0}
 \setcounter{figure}{0}
 \setcounter{equation}{0}

\begin{figure}[htbp]
\centerline{\includegraphics[width=5.4594in,height=5.9886in]{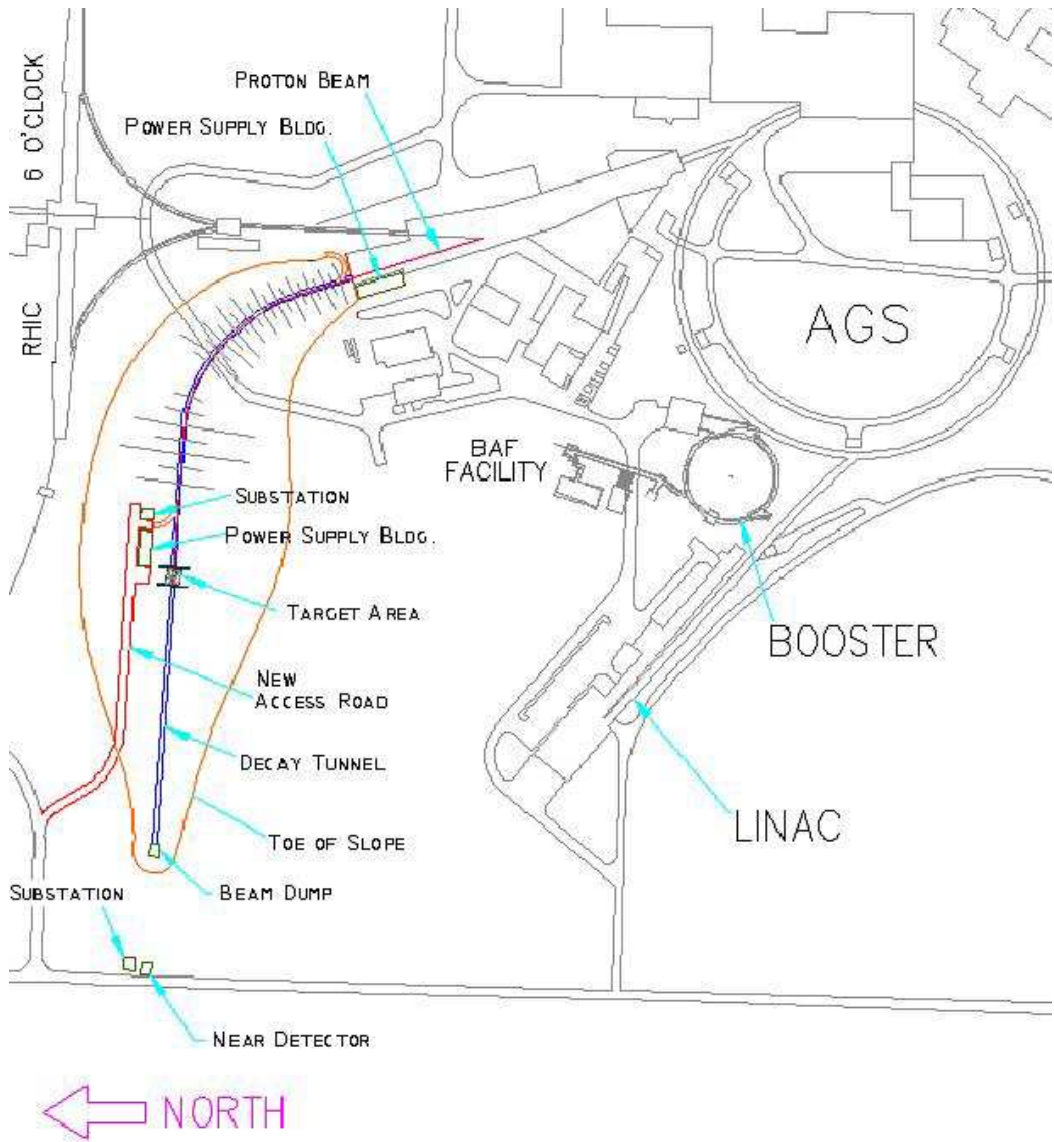}}
\caption{A plan view of the neutrino beam facility at the AGS. The
1 MW proton beam will be taken from the existing fast extraction
line (the U-line) and continued on the up-slope of the hill. The
vertical and horizontal bending magnets are separate in this plan.
The beam will be bent downwards at the top of the hill to aim it
towards Homestake at 11 degrees.} \label{fig:ultbt}
\end{figure}

The extracted beam will enter into the existing U-line of the AGS.
For aiming the neutrino beam to the   Homestake site in South
Dakota, the proton beam has to bend 68 degrees and 4 seconds
horizontally and 11.26 degrees downwards. Figure ~\ref{fig:ultbt}
shows the proposed layout of the beam transport to the neutrino
target at the Homestake site.

\begin{figure}[htbp]
\centerline{\includegraphics[width=6.05in,height=3.872in]{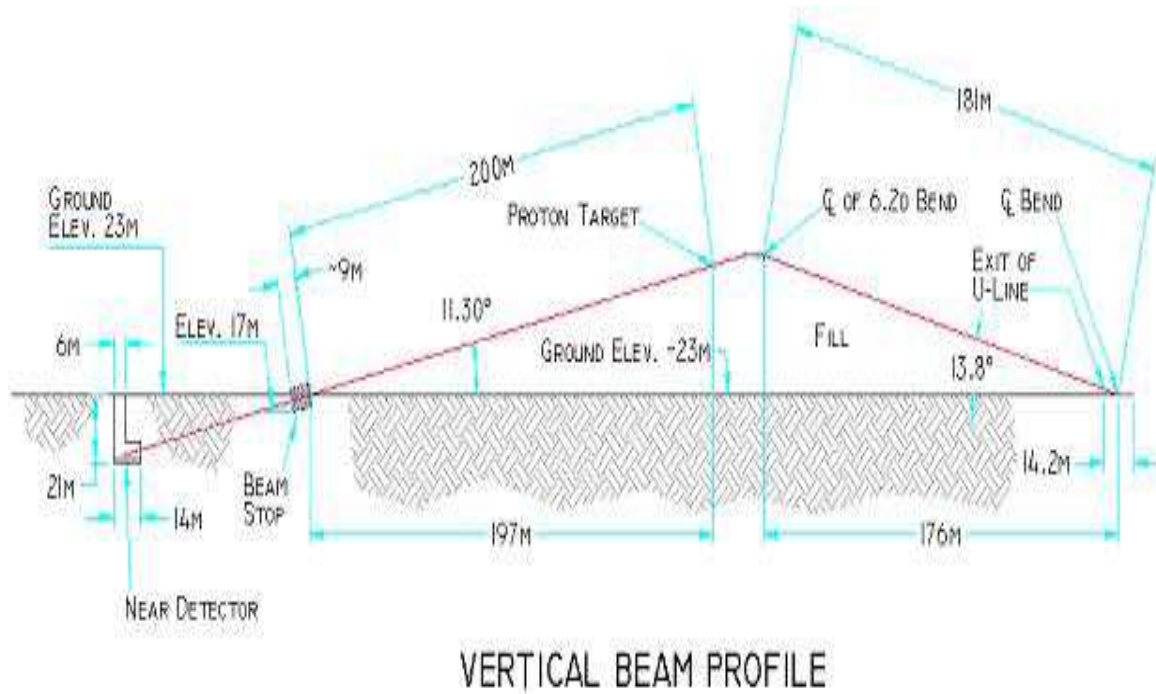}}
 \caption{Elevation view of the neutrino beam
line to Homestake, South Dakota.} \label{fig:homsd}
\end{figure}

\begin{figure}[h]
\begin{center}
\epsfxsize=5.0in \epsfysize=1.5in
\centerline{\includegraphics[width=6.03in,height=2.82in]{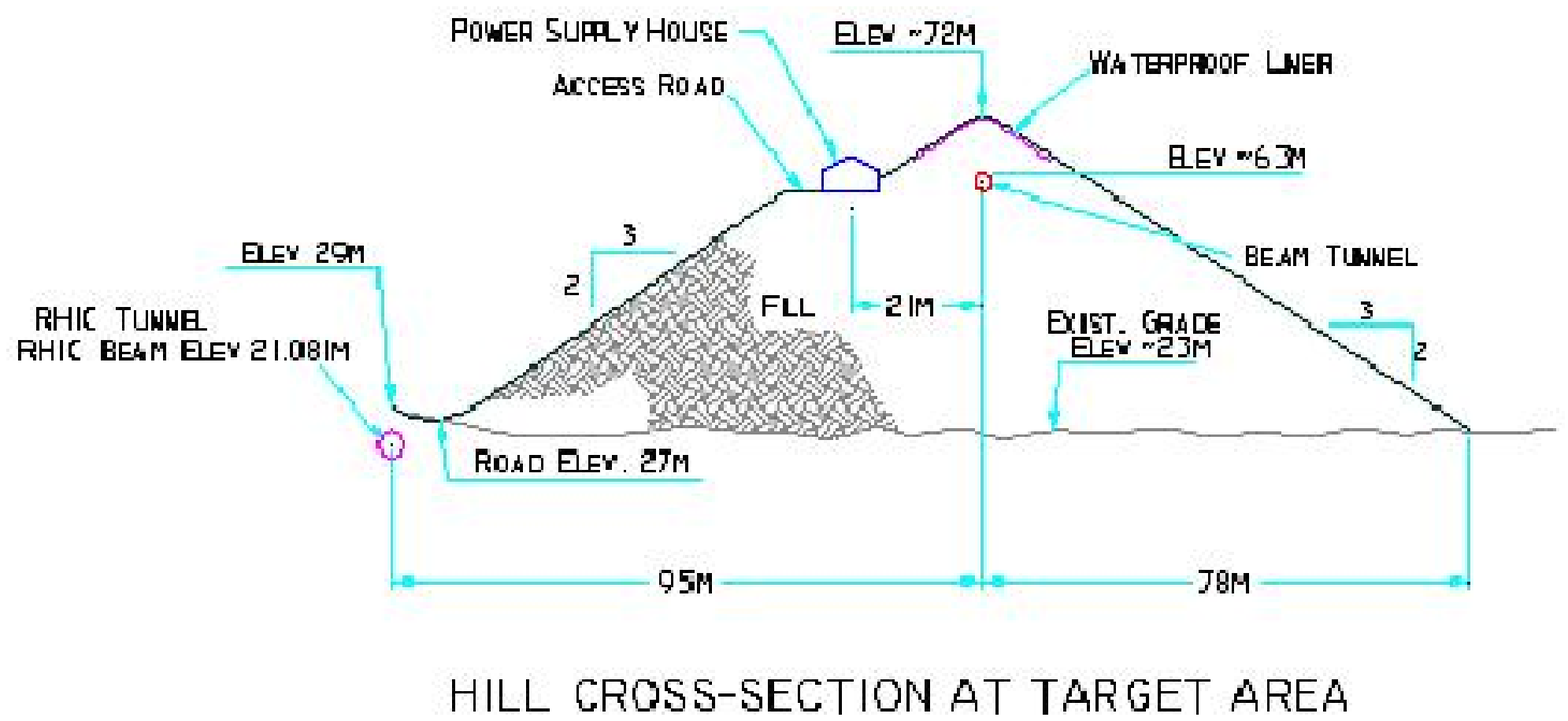}}
\end{center}
\caption{Hill cross section at target area.}
 \label{fig:nb1}
\end{figure}

To aim 11.26 degrees downwards, the proton beam has to climb up a
high hill to the target and decay channel. The hill arrangement
keeps the target and hadronic decay channel  well above the water
table of Long Island. Figure ~\ref{fig:homsd} shows the sketch of
the hill. Figure ~\ref{fig:nb1} shows the cross section of the
hill.

\begin{figure}[htbp]
\centerline{\includegraphics[width=5.98in,height=4.25in]{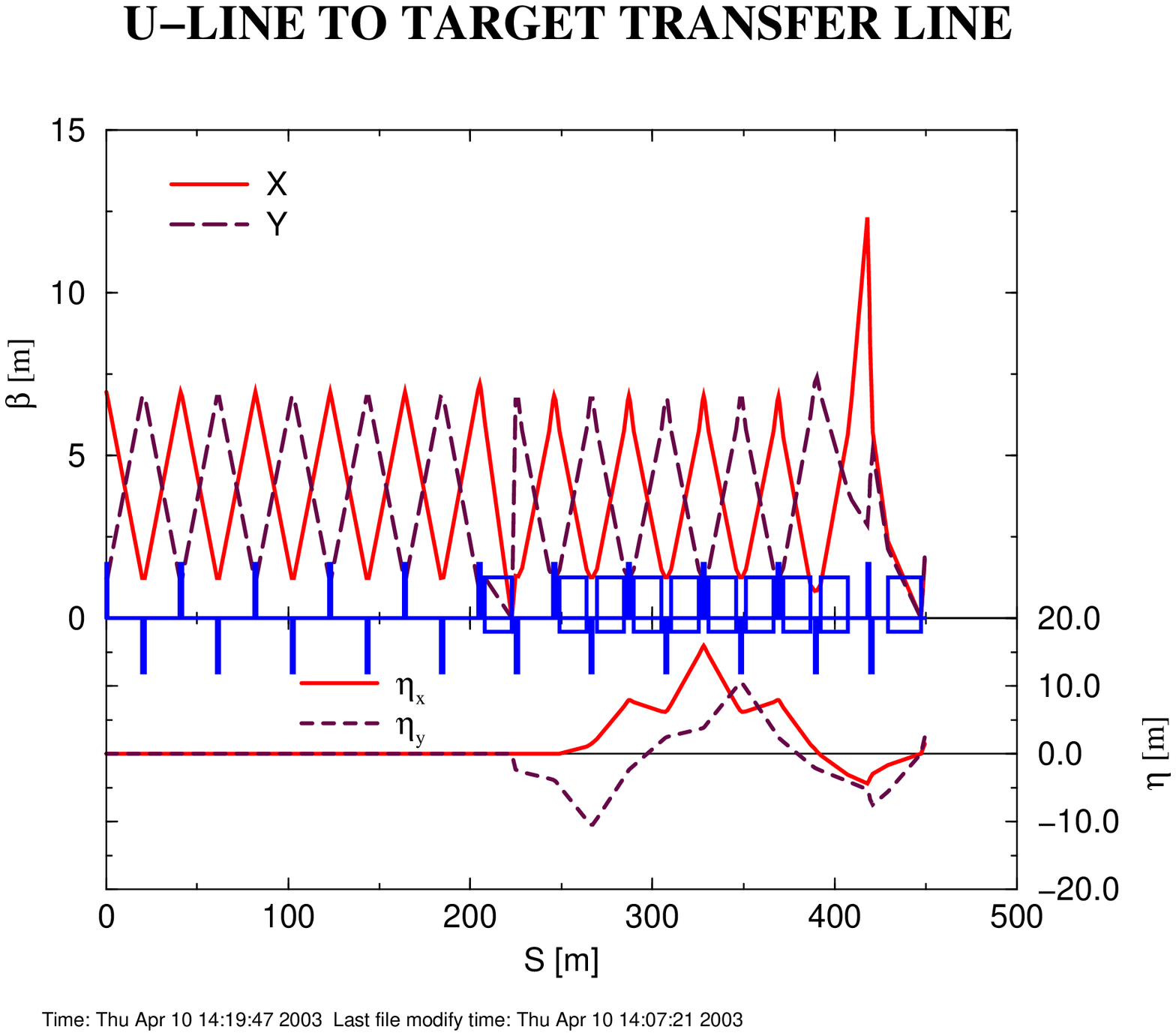}}
 \caption{$\beta$ and $\eta$ functions along
the ULTBT.} \label{fig:betabtt}
\end{figure}

The beam is bent vertically upward by 11 degrees with help of one
cell with phase advance of 90 degrees. In the vertical incline
plane, beam is then bent 68 degrees and 4 minutes horizontally in
four cells with a phase advance of 90 degrees. Finally, the beam
is  bent downwards by 11.26 degrees; aiming it towards Homestake,
South Dakota. Before the final bend, the line uses one doublet to
focus the beam with 2 mm radius (rms) on target. This arrangement
results in an  achromatic system in both horizontal and vertical
planes. The transport system uses a 4Q16 existing quadrupole and
dipole for the vertical bend at 1 degree/meter and for the
horizontal bend at 0.5 degree/meter. Figure ~\ref{fig:betabtt}
shows the $\beta$ and $\eta$ functions for the beam transport to
the target from the U-line (ULTBT).

\clearpage
\newpage
\lhead{AGS Super Neutrino Beam Facility} \rhead{The Target and
Horn System} \rfoot{April 15, 2003}

\label{sec:tarhorns}

\section{Target/Horn Conceptual Design}
\setcounter{table}{0}
 \setcounter{figure}{0}
 \setcounter{equation}{0}

\lhead{AGS Super Neutrino Beam Facility} \rhead{Background and
Overall Issues} \rfoot{April 15, 2003}
\subsection{Background and Overall Issues}
\label{sec:bgandoi}
\begin{figure}[htbp]
\centerline{\includegraphics[width=7.43in,height=2.92in]{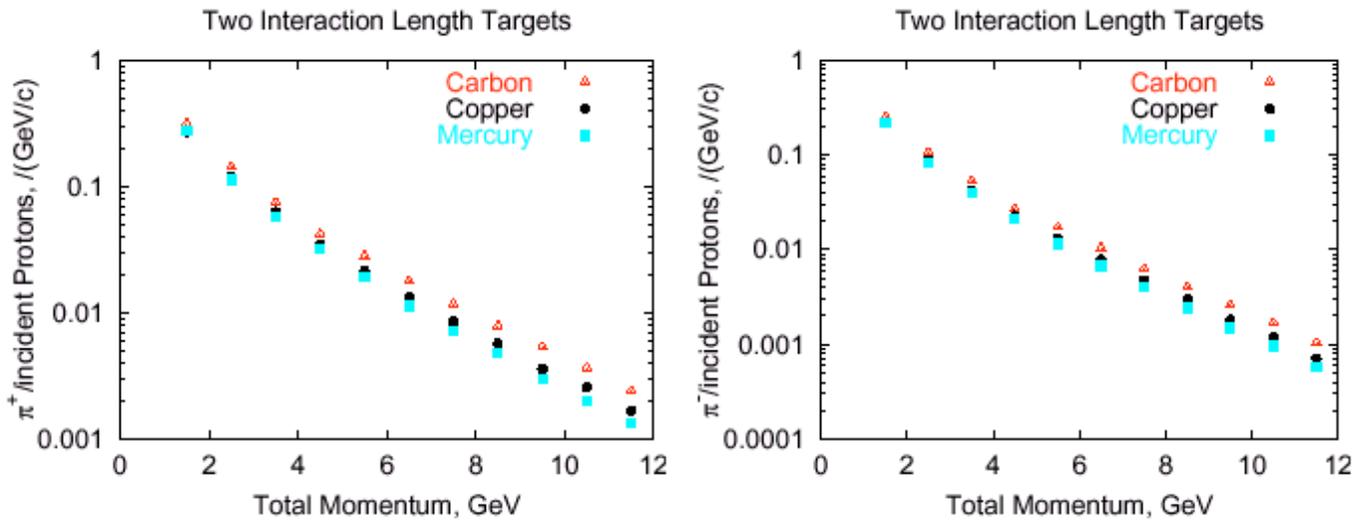}}
\caption{Pion production as function of proton momentum for
different materials. }
 \label{fig:pionprod}
\end{figure}

 To achieve the 1 MW
upgrade option of the proton driver at BNL serious consideration
must be given to both the target material and selection and horn
configuration.A solid target is a viable choice for a 1 MW beam.
Low and high Z materials have been investigated both in terms of
the material endurance as well as the feasibility of target/horn
configuration. Results of the parametric studies on material
choices regarding pion production are shown in
~\ref{fig:pionprod}.

A graphite-based carbon-carbon composite is selected as a target
material both for its low-Z and thermo-mechanical properties.
Various aluminum-based materials are being considered for horn
conductors, such as 3000-series and 6061-T6 aluminum, that will
combine low resistivity, high strength and resistance to corrosion
and micro-cracking. According to the baseline design the following
beam parameters are guiding the material choice, horn and target
design, cooling schemes, etc.

\begin{table}[htbp]
\begin{center}
\caption{\label{horntreq}  Horn/Target design parameters.}
\begin{tabular}{|@{}l|l|}
\hline

Proton Beam Energy & 28 GeV \\

\hline
 Protons per Pulse & 8.9x10$^{ 13 }$\\
\hline

Average Beam Current & 35.7 $\mu $A \\ \hline

Repetition Rate & 2.5 Hz\\ \hline

Pulse Length & 2.58 $\mu $sec \\ \hline

Number of Bunches & 23\\
\hline Number Protons per Bunch  &
3.87x10$^{12}$ \\ \hline
 AGS Circumference  & 807.1 m \\
\hline
 Bunch Length & 40 ns \\
\hline
 Bunch Spacing & 60 ns \\
\hline
 Normalized Emittance-X & 100 $\pi$ mm-mrad \\
\hline
 Normalized Emittance-Y  & 100 $\pi$ mm-mrad \\
\hline
 Longitudinal Emittance  & 5.0 eV-sec \\
\hline
 Target Material & carbon-carbon composite \\
\hline
 Target Diameter  & 1.2 cm \\
\hline
 Target Length  & 80 cm \\
\hline
 Horn Small Radius & 7 mm \\
\hline
 Beam Size (Radius) on Target & 2 mm (rms)\\
\hline
 Horn Smallest Radius & 6 mm \\
\hline
 Horn Large Radius & 61 mm \\
\hline
 Horn Inner Conductor Thickness & 2.5 mm \\
\hline
 Horn Minimum Thickness & 1 mm \\
\hline
 Horn Length & 217 cm \\
\hline
 Horn Peak Current & 250 kA \\
\hline
 Current Repetition Rate  & 2.5 Hz \\
\hline
 Power Supply Wave Form & Sinusoidal, Base Width 1.20 ms
\\

\hline
\end{tabular}
\end{center}
\end{table}

In the process of  designing  the horn/target system the following
key elements were closely addressed:

\begin{itemize}

\item {Heat generation and removal from the target/horn system}

\item {Target thermo-mechanical response from intercepting energetic,
high intensity protons}

\item {Irradiation and corrosion effects on materials}

\item {Horn/target integration issues}

\item {Horn mechanical response and long term integrity (irradiation,
corrosion and thermal fatigue)}

\item Beam windows integrated in the system to (a) separate the
vacuum space in the transfer line from the final beam run to the
target and (b) to maintain the coolant around the target in a
close-system loop.

\end{itemize}


\lhead{AGS Super Neutrino Beam Facility} \rhead{Description of the
Integrated System } \rfoot{April 15, 2003}
\subsection{Description of the Integrated System}
\label{sec:desints}

\begin{figure}[htbp]
\begin{center}
\centerline{\includegraphics[width=4.536in,height=3.456in]{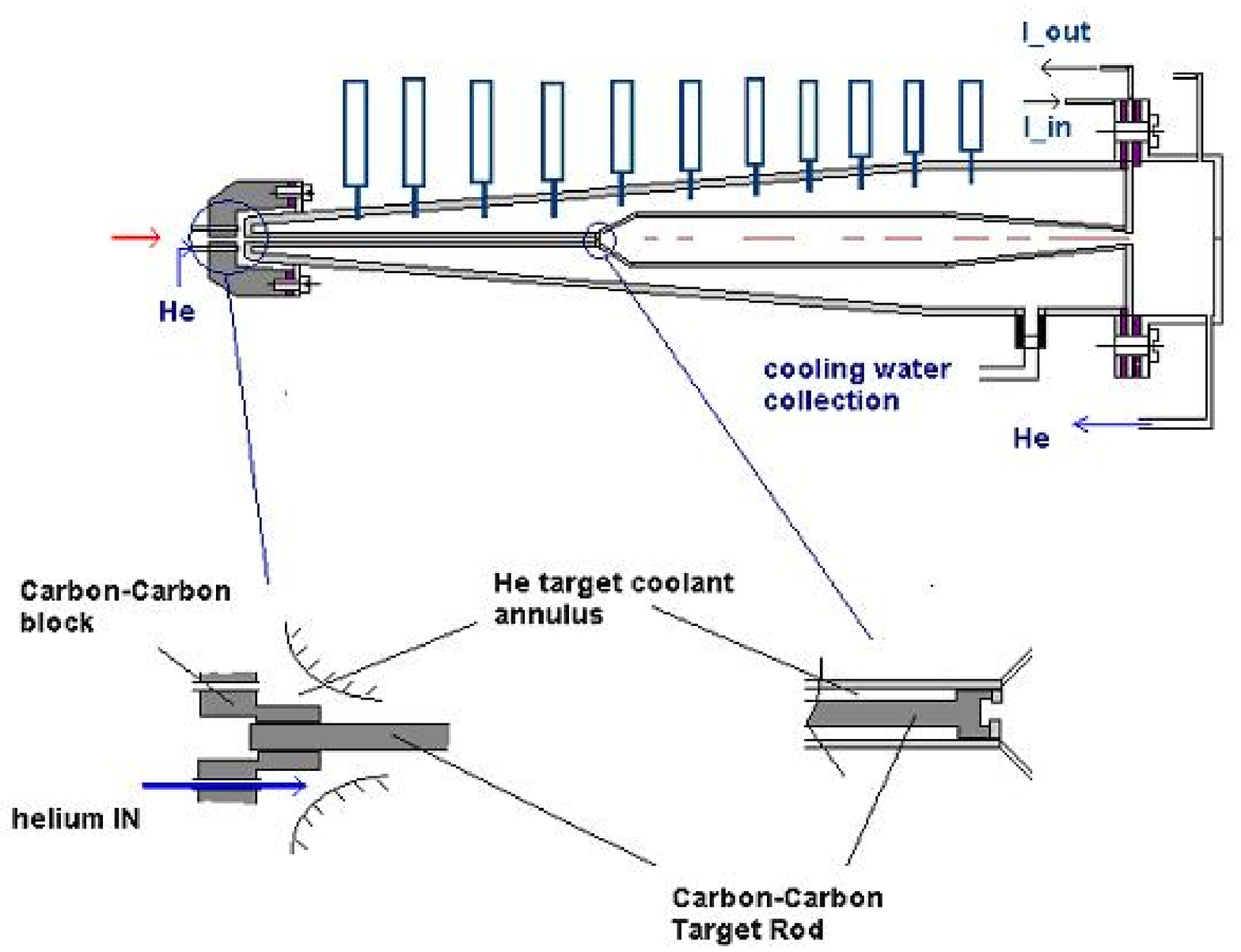}}
\caption{Conceptual arrangement of horn and target.}
 \label{fig:caht}
 \end{center}
\end{figure}

Figure ~\ref{fig:caht}  is a conceptual description of the target
and horn integrated system being considered in this study. The 1.2
cm diameter carbon-carbon composite target is fully inserted into
the inner horn conductor allowing a 1 mm annular gap between the
target and horn surfaces for the helium coolant to flow and remove
heat. Shown in the front of the target is a carbon-carbon beam
``collimator'' or baffle that has dual role. Specifically, it
provides the upstream target support and includes the channels
that will force the coolant flow into the annular space. Further,
it plays the role of the beam diffuser in the event the proton
beam strays off beam axis and thus protects the horn. The size of
this baffle needs to be optimized to play that role effectively.
In addition to the above two functions, the front end of the
target will be maintained at a low temperature which will help in
removing heat deposited on the target by conducting into the mass
of the baffle.

It is envisioned that the baffle/target arrangement will be a
monolithic structure for system optimization. At the downstream
end of the target, a special design allows for the forced coolant
to leave the horn's neck-down section and also provides a simple
support for the target in the current design. The option of
maintaining the target as a cantilever beam (only supported from
the upstream end is being considered.

The horn, made out of an aluminum alloy, has a diameter in its
narrowest section of 1.4 cm while the thickness over that section
is 2.5 mm. The thickness of the inner conductor reduces to $\sim $
1 mm downstream of the neck-down section. As shown, the current
supply and return take place at the downstream end of the horn.
The baseline design calls for a 250 kA peak current to be achieved
with a half-sine wave shape that has a base of 20 $\mu $s with a
repetition rate of 2.5 Hz. The magnetic pressures and joule
heating generated in the conductor are the driving forces in the
optimization of the horn conductor. While heat generated in the
narrowest section of the horn will be partly removed by the fluid
flowing in the annular space, heat generated in the skin of the
current surface will be being removed by spraying  a coolant
through a set of optimally positioned jets. Two options are being
considered, namely, the spraying of water and cold helium gas. The
schematic of Figure ~\ref{fig:caht} depicts the arrangement with
water-cooling. The choice of cooling, as discussed later in
detail, is significant in maximizing the useful life of the
aluminum material in that environment.

Under consideration, and shown in Figure ~\ref{fig:caht}, is a
downstream thin window whose role is to hold the target coolant in
a closed system. This way the coolant is collected, cooled and
returned to the target upstream to be re-ejected into the annular
space. The key issue with such window is the fact that it will see
a significant portion of the incoming beam power ($\sim $ 80{\%})
and will be subjected to high thermo-mechanical stress conditions.
Further, the presence of additional material in the flight path of
pions generated and focused by the horn is a further impediment.
However, since the only role of such window is the prevention of
coolant to escape from the closed envelope, the materials
considered are low-Z (for minimal interaction with secondary
particles or generation from intercepting the beam protons) and
minimal heating, which combined with the small thickness will
minimize the problem further.

\lhead{AGS Super Neutrino Beam Facility} \rhead{Evaluation of Heat
Removal Principles } \rfoot{April 15, 2003}
\subsection{Evaluation of Heat
Removal Principles }

\label{sec:hrp}

As part of the design process, a number of cooling schemes have
been considered for the integrated system of horn and target.
Utilizing past experience on horn/target system performance we
arrived at the current baseline scheme which calls for a
de-coupling of the target from the horn with regards to their
cooling and the use of helium gas as the fluid to remove the heat
from the target.

Specifically, as shown in Figure ~\ref{fig:caht}, cold helium is
forced past the surface of the CC target at sufficient velocity to
ensure that the heat generated per beam pulse is removed before
the next pulse arrives. The helium flowing in the annular space is
also responsible for removing some of the heat generated in the
horn inner conductor that surrounds the target. This heat, as will
be discussed in detail, originates in three separate mechanisms,
namely, joule heating from the flowing current, energy deposition
due to the proton beam interaction with the target, and heat by
radiation from the surface of the CC target. Further, heat
generated in horn conductors (inner {\&} outer) due to current and
beam is removed with spraying of cooling water over the inner horn
surface. Since different mechanisms, namely heat convection,
conduction and radiation between target and horn, will be at work,
the heat balance of the overall system as it reaches an operating
temperature will be addressed by the use of a sophisticated finite
element model which is described in a later section. As a first
step, however, and in order to assess the heat removal capacity of
surfaces in the system, a set of basic calculations are performed
that will indicate whether the adopted schemes are feasible.
Furthermore, the results of this preliminary analysis will serve
as a benchmark and check for the finite element simulations. The
following analysis addresses the joule heating in the horn, the
radiation heat exchange between the target and the horn conductor,
the heat removal capacity of the flowing He in the annular space
and,  the heat removal capacity/requirement of the water spray
inside the horn.


\lhead{AGS Super Neutrino Beam Facility} \rhead{Joule Heating
Estimate in the Horn } \rfoot{April 15, 2003}
\subsubsection{Joule Heating Estimate in the Horn}

\label{sec:joulhh}

\begin{figure}[htbp]
\centerline{\includegraphics[width=2.78in,height=2.095in]{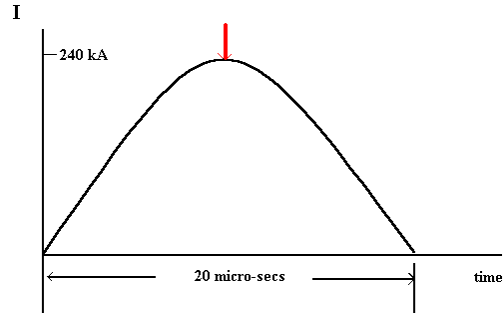}}
\caption{Current pulse structure. } \label{fig:cps}
\end{figure}

Given the current pulse structure shown in Figure ~\ref{fig:cps}
with an effective frequency of 0.025 MHz, the current is expected
to flow over a skin depth of the inner surface of the inner
conductor. Key to the evaluation of joule heating is the
estimation of the skin depth for the parameters of this study.

The skin depth $\delta $ for a horn conductor made out of
3000-series aluminum, for example, with   resistivity $\rho $ =
4.2 mohm-cm, is calculated based on the following relations:

\begin{equation}
 \delta  =( 6.61/ f^{1/2})k_l,
 \label{7eq1}
\end{equation}

\noindent where, f = 0.025 MHz, {k$_{1 }$= [$\rho $/$\rho
_{c}$]$^{1 / 2}$ } and $\rho _{c = }$1.724 mhoms-cm is the
resistivity of copper.

From these relations one obtains,

\begin{equation}
\underline {\delta_{Al}= 0.06525 \textrm{cm}}. \label{7eq2}
\end{equation}

By focusing attention on the neck-down section of the inner
conductor where conditions are most severe, the joule heating
generated by the current pulse and  the anticipated temperature
rise is estimated. The current density is a function of the
penetration into the conductor according to the relation:

\begin{equation}
J(z) = J_{o} e^{ - z / \delta } \label{7eq3}
\end{equation}

\noindent such that,

\begin{equation}
I(z) = \int\limits_A J(z)dA, \label{7eq3a}
\end{equation}

\noindent where, \textbf{J}$_{o}$\textbf{ = 689 kA/cm}$^{2}$ is
the current density at the conductor surface and z is the
penetration and \textbf{dA = 2$\pi $ (R}$_{out }$\textbf{-- z) dz,
}where R$_{out}$ = 0.95 cm\textbf{. }The joule heating generated
by the pulse per cm-length of the inner conductor is

\begin{equation}
JH_{pulse - cm}=\int\limits_0^{20\mu s}  \int\limits_A J^{2}(z,t)
\rho  dA dt. \label{7eq4}
\end{equation}
\noindent where, J(z,t) = J(z) sin(\textit{$\pi $t/20$\mu $s}). By
carrying out the integration above one obtains,

\begin{equation}
 JH/\textrm{(pulse-cm)} = 3.88 ~~\textrm{Joules}.
 \label{7eq5}
\end{equation}

\noindent The total power generated in the neck-down section of
the horn alone is estimated to be:

\begin{equation}
P_{neck} = 2.5 (Hz) \cdot  3.88 ~~J/\textrm{(cm-pulse)} \cdot 86~
\textrm{cm} = 835 ~~\textrm{Watts}. \label{7eq6}
\end{equation}

The peak temperature rise in the conductor, experienced at the
inner surface where the current density is at maximum during each
current pulse, is estimated to be

\bigskip

\begin{center}
\textbf{$\Delta $T = 7.4}$^{o }$\textbf{C}.
\end{center}


\lhead{AGS Super Neutrino Beam Facility} \rhead{Estimate of
Radiation Heat Transfer } \rfoot{April 15, 2003}
\subsubsection{Estimate of Radiation Heat Transfer}
\label{sec:erht}

Given that the CC target will be operating at much higher
temperatures than those of the horn ($\sim $ 840 $^{o}$C) some
heat will radiate from the target surface to the surface of the
horn over the neck-down section. Assuming that the surface
temperature of the aluminum is maintained at $\sim $ 90 $^{o}$C,
then the radiating heat flux can be estimated from the relation:

\begin{equation}
\frac{q_{CC - > Al} }{A_{CC} } = \frac{\sigma \left[ {T_{CC}^4 -
T_{Al}^4 } \right]}{\frac{1}{\varepsilon _{CC} } + \frac{A_{CC}
}{A_{Al} }\left( {\frac{1}{\varepsilon _{Al} } - 1} \right)},
\label{7eq9}
\end{equation}

\noindent where, $\sigma  =  5.669$ x 10$^{ - 8}$ W/m$^{2}$ --
K$^{4}$, $\varepsilon _{cc}$ = 0.98, $\varepsilon _{Al }$= 0.19

\noindent Based on the above parameters, the heat flux from the
target surface is approximately $q$/A =11916 W/cm$^{2}$ and the
total heat transfer from the target to the horn inner conductor $q
$= 11916. A$_{target}$ = 360 W. As seen the amount of heat
exchanged by heat transfer is quite low. However, if the surface
of the horn seeing the target is treated in order to increase its
emissivity to $\varepsilon _{Al }$= 0.19, then the heat
transferred to the horn amounts to $\sim $ 1.363 kW.


\lhead{AGS Super Neutrino Beam Facility} \rhead{Target Cooling
Calculations } \rfoot{April 15, 2003}
\subsubsection{Target Cooling Calculations}
\label{sec:tcc}

Given that the amount of heat transferred by radiation alone from
the target to the horn is quite small, the primary mechanism is
heat removal by forced convection of cold He in the annular space
between the target and the horn. The key elements in assessing how
feasible such a mechanism is are: the operating temperature of the
target; the temperature of the cooling fluid; and the heat
generation rate in the volume of the target that needs to be
removed. These working parameters will establish the velocity of
the coolant necessary to remove the generated heat. Using as a
baseline, the energy deposition in the volume of the target for a
2 mm RMS beam spot, (which amounts to 7.1 kJ/pulse or 17.75 kW) as
an acceptable average operating temperature of surface of the
target T$_{s}$ = 800$^{o }$C and a temperature of the helium of
$\sim $ 144 K, the flow velocity is estimated as follows,
\begin{equation}
\text{Heat flux from target surface} = q/A_{target}=h_{f} (T_{s}-
T_{He}). \label{7eq10}
\end{equation}

\noindent One can estimate the necessary film coefficient $h_{f}$
from the above relation to be 618 W/m$^{2}$-C. By relating the
film coefficient to the Nusselt number Nu$_{De}$

\begin{equation}
Nu_{De}=h_{f} D_{e}/ k = 13.326, \label{7eq11}
\end{equation}

\noindent where, $k$ = 0.14 W/m-K and D$_{e}$ = 0.014-0.012 =
0.002 m (equivalent annular diameter). By relating the Nusselt
number to the Reynold's number N$_{Re}$ through the relation
\begin{equation}
Nu_{De} = 0.023 (N_{Re})^{0.8} (P_{r})^{0.3},
 \label{7eq12}
 \end{equation}

\noindent where P$_{r}$ is the Prandtl number, the Reynolds number
is estimated to be $\sim $3400. Reynolds number on the other hand
is related to the velocity of flow U$_{f}$ through,
\begin{equation}
N_{Re} = U_{f} D_{e}/\nu .
 \label{7eq13}
\end{equation}

\noindent where $\nu $ is the dynamic viscosity of He ($\nu $ =
37.1 x 1 0$^{ - 6}$ m$^{2}$/s). From the above relations it is
estimated that a flow of He with velocity U$_{f}$ = 65 m/s is
required.

These calculations, however, do not account for the heat moved
away from the target by radiation and conduction. The later takes
place near the front of the CC target by maintaining the block of
the baffle (or collimator) that  holds the target in place at a
low temperature.


\lhead{AGS Super Neutrino Beam Facility} \rhead{Heat Removal from
Horn Neck-down ... } \rfoot{April 15, 2003}

\subsection{Heat Removal from Horn Neck-Down Through Forced Convection}

\label{sec:hrhn}

In this section estimates are made of the required convective heat
transfer capacity of He in the annular space to enable the removal
of all the heat generated in the volume of the 86cm-long neck-down
section. The best estimates, thus far, of the heat deposited over
that volume are as follows:

Joule Heat = 1.335 kW

Heat from beam interaction with target = 36{\%} of 17.75 kW = 6.39
kW

Radiation from target = 1.363 kW

TOTAL = 9.088 kW.

\noindent The high value of energy deposited in that section alone
from beam-target interaction needs further verification.
Independent analysis is currently being performed to verify this
energy deposition. The available area of the horn surface with
radius R$_{i}$ = 0.7 cm is A$_{i}$ = 0.0352 m$^{2}$. This leads to
a surface flux of
\begin{equation}
q/A_{i} = 258,181 W/m^{2}. \label{7eq14}
\end{equation}

\noindent With forced helium temperature of 144 K and surface
temperature of the horn $\sim $ 90$^{o}$ C, the required film
coefficient is h$_{f}$ = 1179 W/m$^{2}$-C. Following the
procedures of the previous section that relate the convective film
coefficient with the Nusselt and Reynolds numbers, one arrives at
a flow velocity required to remove the heat in the neck-down
section equal to U$_{He}$ = 135 m/s.

\lhead{AGS Super Neutrino Beam Facility} \rhead{Heat Removal from
Horn Inner Surface ...} \rfoot{April 15, 2003}
\subsubsection{Heat Removal from Horn
Inner Surface  by Coolant Spray} \label{sec:hrhis}

\vspace {0.25in}

\underline{\textbf{Option I -- RAW Spray}}

\vspace{ 0.25in}
 \noindent The baseline method utilized to remove the generated heat over the entire
inner surface (current side) as well as the energy deposited
throughout the horn from beam-target interaction is by spraying
water through special nozzles that penetrate the outer conductor.
While this approach has been used extensively to cool horns in the
past (and in many instances to cool the target as well through
conduction across the horn inner conductor), there are some issues
associated with such approach, namely, the potential acceleration
of corrosion of the aluminum surface coming in contact with water
and the fatigue failure through propagation of surface cracks plus
the fact that water within the horn will be in the path of pions.
The attempt to maintain the aluminum surface temperature, when in
contact with water, below 100$^{o}$ C (based on experience from
the reactor industry) in an attempt to extend the life of the
horn, results in larger quantities of water being sprayed against
the conductor inner surface.

To estimate the required spraying capacity, attention is again
focused on the inner conductor which will experience higher joule
heating and gamma-ray heating. Assuming that the spraying jets are
positioned in such a way that the entire inner surface experiences
forced flow sweeping by (resembling a cylinder in cross-flow), the
following forced convection relations apply,

\begin{equation}
Nu_D = 0.3 + \frac{0.62Re_D^{1 / 2} \Pr ^{1 / 3}}{\left[ {1 + (0.4
/ \Pr )^{2 / 3}} \right]^{1 / 4}}\left[ {1 + (\frac{Re_D
}{282000})^{5 / 8}} \right]^{4 / 5} ,\label{7eq14a}
\end{equation}

\noindent where, \textbf{\textit{Nu}}$_{D }$\textbf{\textit{ =
hD/k}} is the Nusselt number and \textbf{\textit{h}} is the
effective film coefficient.

Assuming that the water jets force a flow with a free velocity of
U = 2.5 m/s at a temperature of T$_{water}$ = 20$^{o}$ C and that
the conductor surface is maintained at a temperature of T$_{wall}$
= 80$^{o}$ C, the following fluid properties apply: Reynolds
number Re$_{D }$ = U D/$\nu $ = 87963, Pr = 3.6, $\nu $ = 0.0054
cm$^{2}$/sec and k = 0.645 W/m-K.

Substitution into the above relation leads to,

\begin{equation}
Nu_{D} = hD/k = 367. \label{7eq14b}
\end{equation}

And to an effective film transfer coefficient of:
\textbf{\textit{h = 12485 W/m}}$^{2}$\textbf{\textit{ --K}}.

By relating the heat flux from the conductor surface to the
convective heat transfer,

\begin{equation}
q/A = h (T_{wall} - T_{water}) W/m^{2} , \label{7eq14c}
\end{equation}

\noindent the heat transferred through the surface area A is
\textbf{q = 38.5 kW}. Clearly this cooling scenario can satisfy
the needs for removing the heat from the inner surface of the
conductor even for operations that will increase the heat
deposition in the conductor.

\vspace {0.25in}
 \underline{\textbf{OPTION II -- Helium Spray}}

 \vspace {0.25in}

\noindent In an effort to eliminate the RAW altogether, an attempt
is made to remove excess heat by spraying He onto the surfaces
instead of water. To estimate the capacity of such system, we
again focus on the neck-down section that represents a horizontal
cylinder of radius R$_{o}$ = 9.5 mm and length L = 86 cm. The
relations that apply are those of gases flowing normal to a single
cylinder. The Reynolds number for this configuration is N$_{Re}$ =
U$_{m}$ D$_{o}$/$\nu $ while the Nusselt number is related to it
by the empirical relation

\begin{equation}
Nu_{Do} = C(N_{Re})^{m} .
 \label{7eq15}
\end{equation}

Where C and m are parameters that are based on the range of the
Reynolds number. For a free velocity of He (after leaving the
nozzle and before impinging on the surface) of 30 m/s and
temperature T$_{He}$ = 144 K, while maintaining the aluminum
surface temperature to 90$^{o}$ C, one can estimate N$_{Re}$ =
1536 and,
\begin{equation}
Nu_{Do} = 54.9 = h_{f} D_{o}/ k .
 \label{7eq16}
\end{equation}
Which leads to h$_{f}$ = 270 W/m$^{2}$-C.

Using the relation: q/A$_{o}$ = h$_{f}$ (T$_{o}$ -- T$_{He})$
where A$_{o}$ = 0.0513 m$^{2}$, the heat transfer capacity of the
inner surface is approximately 3000 W. Any increase of the
velocity of He coming out of the nozzle or a potential reduction
in the heating of the horn from beam-target interaction, will make
the cooling scheme that depends entirely on He sufficient to
remove the heat deposited on the horn. Further experimental
studies are needed to assess the potential of this option.


\lhead{AGS Super Neutrino Beam Facility} \rhead{Target
Considerations} \rfoot{April 15, 2003}
\subsection{Target Consideration }
\label{sec:tcfea}

\vspace{ 0.25in}

\underline {\textbf{Material Selection -- Issues}}

\vspace{ 0.25in}

The baseline material for the superbeam experiment is a special
Carbon-Carbon composite. It is a 3-D woven material that exhibits
extremely low thermal expansion for temperatures up to 1000
$^{o}$C. For higher temperatures it behaves like typical graphite.
Its thermal expansion behavior is significant in the sense that
the thermo-elastic stresses induced by intercepting the beam will
be quite small,  thus extending the life of the target. The Table
~\ref{tab:teccc} below lists the variation of expansion as a
function of temperature.

\begin{table}[htbp]
\begin{center}
\caption{Thermal expansion data for a 3-D fine weave Carbon-Carbon
composite. }
\begin{tabular}
{|p{77pt}|p{85pt}|} \hline \textbf{Temp.}& \textbf{{\%}
Elongation}
\\ \hline 23 $^{o }$C& 0{\%} \\ \hline 200 $^{o }$C& -0.023{\%} \\
\hline 400$^{o}$ C& -0.028{\%} \\ \hline 600$^{o}$ C& -0.020{\%}
\\ \hline 800$^{o}$ C& 0{\%} \\ \hline 1000$^{o}$ C& 0.040{\%} \\
\hline 1200$^{o }$ C& 0.084{\%} \\ \hline 1600$^{o}$ C& 0.190{\%}
\\ \hline 2000$^{o}$ C& 0.310{\%} \\ \hline 2300$^{o}$ C&
0.405{\%} \\ \hline
\end{tabular}
\label{tab:teccc}
\end{center}
\end{table}

In addition to the low thermal expansion, the material is stronger
than typical graphite, reaching compressive strengths $ \ge $ 120
MPa.

Experience on this material has been acquired as a result of the
BNL E951 experiment and its use in the SNS foil scrapers. Some of
the findings, relevant to this study, are discussed in a later
section.

The long term behavior of this special material in an irradiation
environment is not known. There was an initiative, as part of the
E951 experiment, to irradiate this material to dpa levels that
could be comparable to running times that are approaching
anticipated lifetimes for this experiment. While this particular
material was not  irradiated, it is anticipated, as part of the
R{\&}D effort, to be tested in the future to enable the assessment
of its integrity after the accumulation of several dpa.

 \vspace{ 0.25in}

\underline {\textbf{Beam Energy Deposition and Heat Removal}}

\vspace{ 0.25in}

In the current option, the target is an 80 cm long cylindrical rod
with two competing diameter sizes (6 mm and 12 mm). The 6 mm
diameter target is chosen to intercept the 100-TP 1 mm rms proton
beam and the 12 mm for the 2 mm rms beam. There are advantages and
dis-advantages in both choices. When intercepting the 1 mm beam,
the total energy deposition in the target per pulse amounts to
about 5.1 kJ. This corresponds to a peak temperature rise of about
1000 $^{o}$C in the center of the target. With the larger beam
spot (2 mm rms) the total energy deposited and converted to heat
in the target is 7.3 kJ per pulse with peak temperature rise of
about 270 $^{o}$C. Heat will be removed from the target through
forced convection of helium past the outside surface. While the 2
mm beam/12 mm target diameter option appears to be the optimum,
the fact that the heat within the target will have to travel a
longer distance to reach the surface and be removed, may make the
1 mm beam option attractive.

Engineering calculations and detailed finite element analysis were
employed to qualify the two options while taking into
consideration the special target/horn configuration. Figure
~\ref{fig:caht} depicts the current thinking of system layout.
Specifically, the target rod is placed within the neck-down
section of the inner horn conductor forming an annular space with
the inner surface of the conductor. The long target is supported
at both ends. In the front, a specially machined carbon-carbon
block holds the target while at the same time it allows for the
beam to go through without intercepting it. In addition the this
special block allows for special nozzles that will force helium
into the annulus for target cooling and it provides a conduction
path in the vicinity of the target section that is heated the
most. Further, the block will play a diffusing role in the event
of a beam straying from center. At the far end, the target
includes four special fins that allow it to rest on the inner
surface while allowing the helium to pass through. A special notch
on the inner conductor will prevent the target from moving further
into the horn.

Calculations of heat removal of independent heat transfer
mechanisms while using the basic principles were performed and are
presented in an earlier section. The engineering calculations for
removing the deposited energy revealed the following: Since the
target is to be decoupled from the horn, the forced helium (and
the conduction near the front of the target) must remove all the
heat that amounts to 18.25 kJ for the 2 mm rms beam/12 mm diameter
target. The energy deposition varies along the beam and so the
engineering calculations based on the total energy in the volume
can only provide a rough estimate. Analytic calculations are
expected to refine the engineering numbers. Specifically, to
remove the 18.25 kJ from the surface of the target through forced
convection, a helium temperature of $\sim $5 $^{o}$C and a target
operating temperature of 840 $^{o}$C were assumed. The heat
transfer calculations (using detailed finite element analysis)
reveal that helium flowing continuously in the annulus with
velocity near 40 m/s will, in conjunction with radiation to the
horn surface and conduction near the front of the target, enable
the removal of all the heat deposited in the target.

The detailed finite element model incorporated both the target and
the horn in one single system. The only connection between the
horn and the target was provided by the radiation exchange of the
target surface and the surface of the horn inner conductor.
Further, the two surfaces shared similar values of convective heat
transfer due to the flowing helium in the annulus. In order for
the overall system to achieve a ``steady-state'' (or base
operating temperatures) throughout a transient analysis with all
the heat transfer mechanisms in place was performed.

\begin{figure}[htbp]
\centerline{\includegraphics[width=6.12in,height=3.348in]{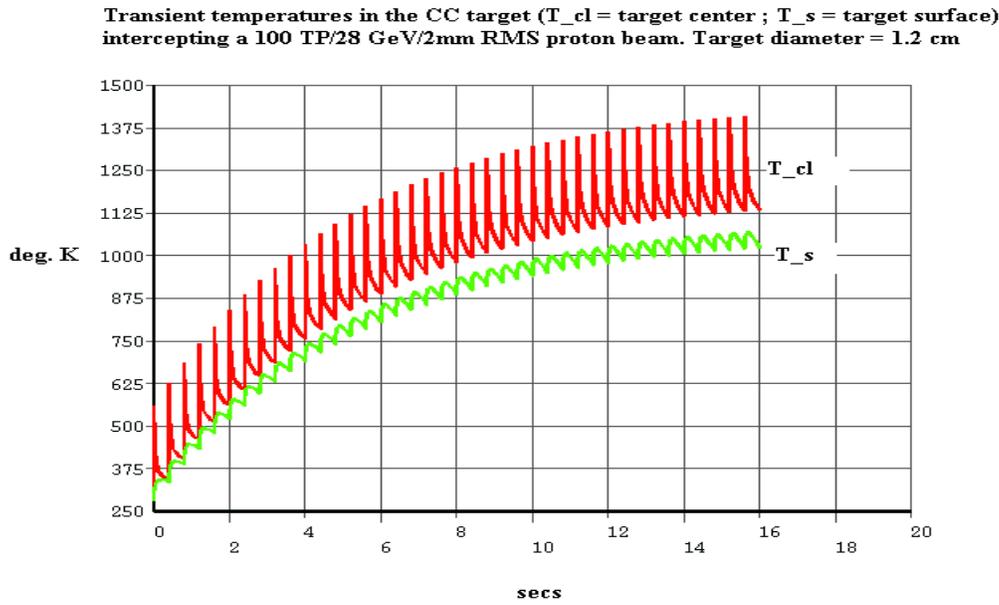}}
\caption{Transient temperatures in the CC target subjected to a
100 TP/28 GeV/2 mm RMS beam. Coolant temperature in the annular
space T$_{He}$ = 5 $^{o}$C.} \label{fig:ttincc}
\end{figure}

\begin{figure}[htbp]
\centerline{\includegraphics[width=4.1664in,height=3.2192in]{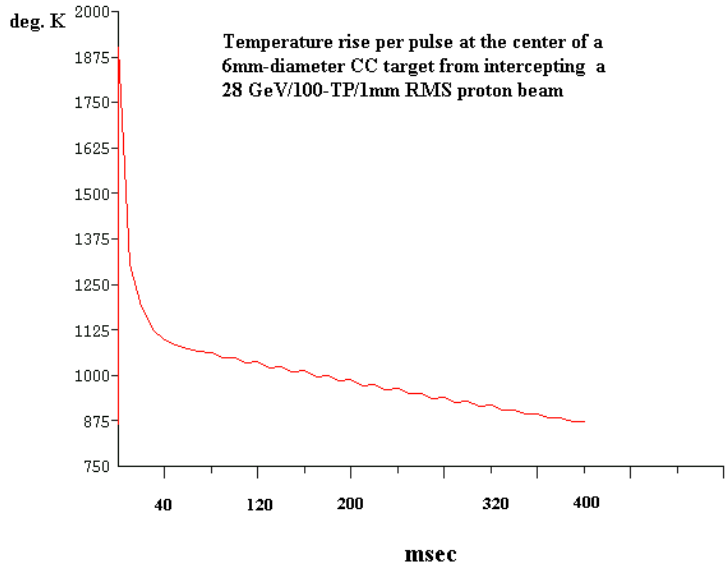}}
\caption{ Temperature rise per pulse in the 6 mm-diameter CC
target from a 100 TP/28 GeV/1 mm RMS beam.} \label{fig:trpp}
\end{figure}

Shown in Figure ~\ref{fig:ttincc} are the transient temperatures
at the hottest point in the center of the CC target and at the
surface that result from a 100 TP/28 GeV/ 2 mm RMS proton beam on
target. As mentioned previously, the temperature rise per pulse in
the target is of the order of 270$^{o}$ C. For a 1 mm RMS beam
spot on a 6 mm diameter target the temperature rise per pulse is
$\sim $ 1000$^{o}$ C, as shown in Figure ~\ref{fig:trpp}. The
total energy deposited on the target, however, is only 5.1 kJ per
pulse as compared to the 7.3 kJ for the 2 mm RMS beam. While the
benefit in smaller energy deposition would be welcomed, the
temperatures in the CC target will be in the regime of higher
strains and the resulting stresses will exceed the material
strength.

\vspace{ 0.5in}

\underline{\textbf{Evaluation of the  Option with the Target {\&}
Horn in Contact}}

\vspace{ 0.25in}

A scheme  used in past experiments  involves solid targets and
horns and allows for the removal of heat from both the target and
the horn by cooling the outer face of the horn conductor (current
side) through massive water spray. In this arrangement, the solid
target is housed within the narrowest inner conductor and through
surface contact it passes the heat generated by the beam/target
interaction onto the conductor. The heat travels by conduction to
the opposite face (current side) where it is removed by the forced
cooling. To assess its potential application in this experiment,
the system was analyzed by considering a perfect contact between
the two systems (CC target and aluminum horn) and the required
heat removal capacity was estimated.

In performing this analysis, the following key points are kept in
mind, namely:
\begin{itemize}
\item Energy deposition and thus heat generated in the target in
this study are almost an order of magnitude higher than the
previous exercises;

\item The repetition rate in this study is also higher;

\item The goal in this study is to hold the aluminum temperatures as
low as possible for  surfaces that are in contact with water.
\end{itemize}

 An additional concern is the fact that, at the
operating temperatures of the target, it will be impossible to
maintain perfect contact between the surfaces. This will
inevitably lead to the formation of hot spots and the subsequent
melting of the aluminum.

\begin{figure}[htbp]
\centerline{\includegraphics[width=3.645in,height=3.00in]{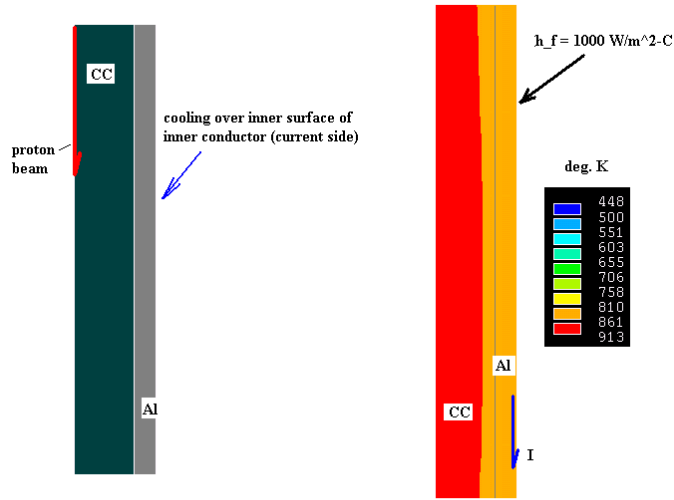}}
\caption{Evaluation of cooling scenario with elimination of
cooling in the annulus and heat removal by conduction through
target/horn contact.} \label{fig:ecs}
\end{figure}

Figure ~\ref{fig:ecs} depicts the two materials in contact as they
were modeled in the finite element analysis) and the temperature
distribution when the aluminum surface is being cooled with an
effective film coefficient of 1000 W/m$^{2}$-C. As seen in Figure
~\ref{fig:ecs}, a large section of the aluminum conductor (only a
segment is shown in the figure) operates at temperatures near the
melting point of aluminum (660$^{o}$ C). A parametric study
revealed that in order to maintain temperatures in the aluminum
that are below 100 C  would require a film coefficient of 3000
W/m$^{2}$-C.   Such a heat transfer coefficient will be almost
impossible to achieve in such setting. However, even if that was
achievable, the issue of hot spot formation cannot be eliminated
since the two materials expand differently. Thus this cooling
method is not suitable here.

\vspace{ 0.25in}

\underline{\textbf{Thermo-elastic target Stresses}}

\vspace{ 0.25in}

\begin{figure}[htbp]
\centerline{\includegraphics[width=3.99in,height=2.796in]{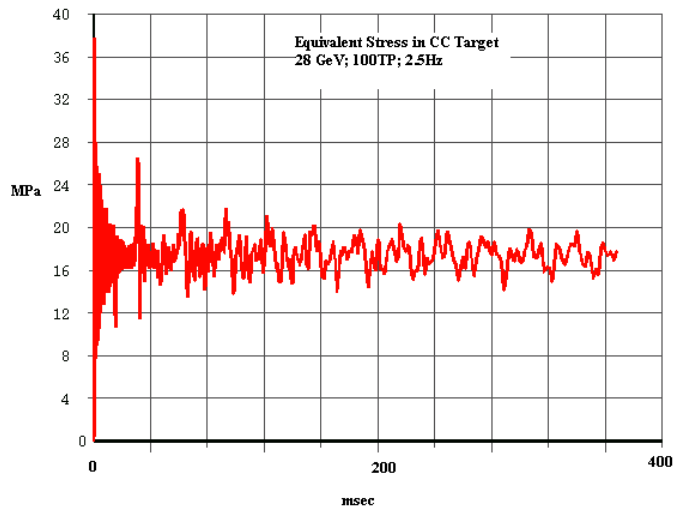}}
\caption{ vonMises dynamic stresses generated in the 2.2 cm
diameter carbon-carbon composite and their attenuation before the
arrival of the subsequent beam micro-bunch train. Results depicted
for a 6-microbunch pulse train totaling 100 TP.}
 \label{fig:vonmises}
\end{figure}

\begin{figure}[htbp]
\centerline{\includegraphics[width=4.056in,height=2.832in]{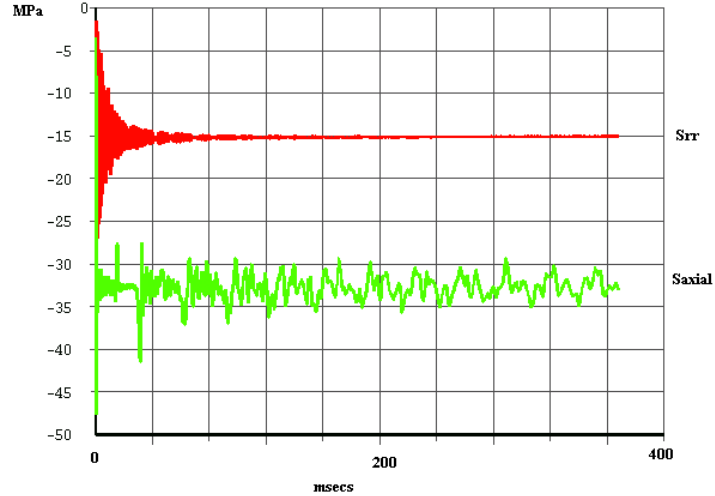}}
\caption{Radial and axial stresses in the 1.2 cm diameter
carbon-carbon target intercepting a 6-microbunch train 28 GeV
proton beam Totaling 100 TP.} \label{fig:rastress}
\end{figure}

 Figures ~\ref{fig:vonmises} and ~\ref{fig:rastress}
depict the dynamic stresses in the target intercepting the train
of six micro-bunches that comprise each 100 TP pulse. The
resulting equivalent stresses, estimated for a pulse which arrives
when the target has achieved a ``steady-state'' operating
temperature ($\sim $840$^{o}$ C) peak at $\sim $ 38 MPa. Such
stresses are well below the strength limits of $\sim $ 120 MPa.
However, as it will be discussed in the following section,
experimental measurements have shown that the material exhibits
some dynamic inertia during fast beam energy deposition that is
expected to slightly increase the anticipated stress levels that
are based on the thermal expansion properties which, in turn, are
estimated with slow heating of the material.

As a result of beam optimization, the new parameters call for a
beam spill of 23 40 ns-long micro-bunches spaced by 60 ns. While
calculations for such case have not been performed, it is expected
that the thermal shock stresses will experience some relaxation
between micro-bunches thus  driving the stresses even lower.

\vspace{ 0.25in}

\underline {\textbf{Assessment of Long-term Target Survival}}

\vspace{ 0.25in}

Given that little is known about the behavior of this target
material when it is highly irradiated, it is envisioned that, as
part of the R{\&}D, representative target material will be exposed
to levels of dpa anticipated to be seen during the lifetime of the
target and the its long-term integrity be assessed.

\vspace{ 0.25in}
 \underline {\textbf{Verification of CC Mechanical
Properties }}

\vspace{ 0.25in}

\begin{figure}[htbp]
\centerline{\includegraphics[width=4.146in,height=2.478in]{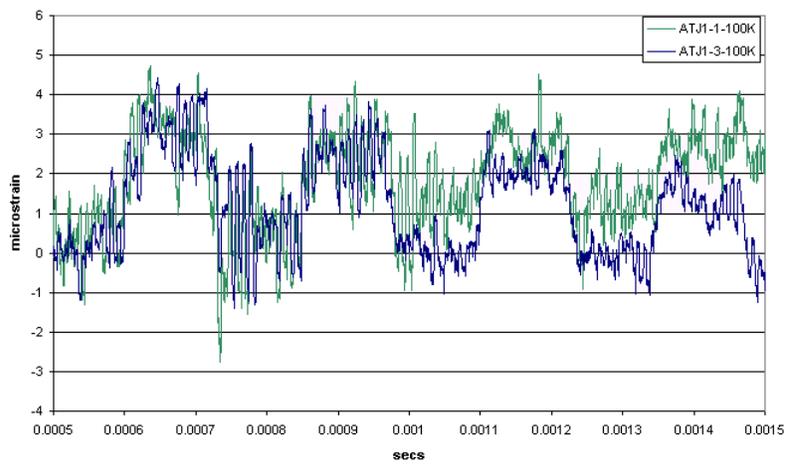}}
\caption{Axial strains measured in the ATJ graphite target during
E951 experiment. }
 \label{fig:asme951}
\end{figure}

While the chosen operating temperatures (including the temperature
rise) are still within the very low thermal expansion regime and
the expected thermal stresses and strains are within the safety
envelope, some questions remain regarding the response of the
target material during fast deposition of heat. In the course of
experiment E951, 1 cm diameter carbon-carbon targets (along with
ATJ graphite) were exposed to the 24 GeV AGS beam and their
response was recorded on the outer surface in the form of axial
strains. Figure ~\ref{fig:ccce951} below shows the target
arrangement. Figure ~\ref{fig:assmcc} depicts strains recorded on
the surface of the CC target along the axial dimension. While
these strains are, by comparison, much smaller than those recorded
on the ATJ graphite target with same beam parameters (Figure
~\ref{fig:asme951}), the generated temperatures from the beam
energy deposition were very small to cause these strain amplitudes
(based on the material property data sheet). It appears that the
material responds dynamically during very rapid energy deposition.
For the 2 mm rms beam the anticipated thermo-mechanical stresses
are well below the material strength limits even with a potential
increase in stresses due to inertia in the material (as shown in
Figure ~\ref{fig:assmcc}). It is anticipated that further tests of
intense proton beams on such targets will shed light in verifying
the true thermo-mechanical response of this special material.

\begin{figure}[htbp]
\centerline{\includegraphics[width=4.17in,height=2.868in]{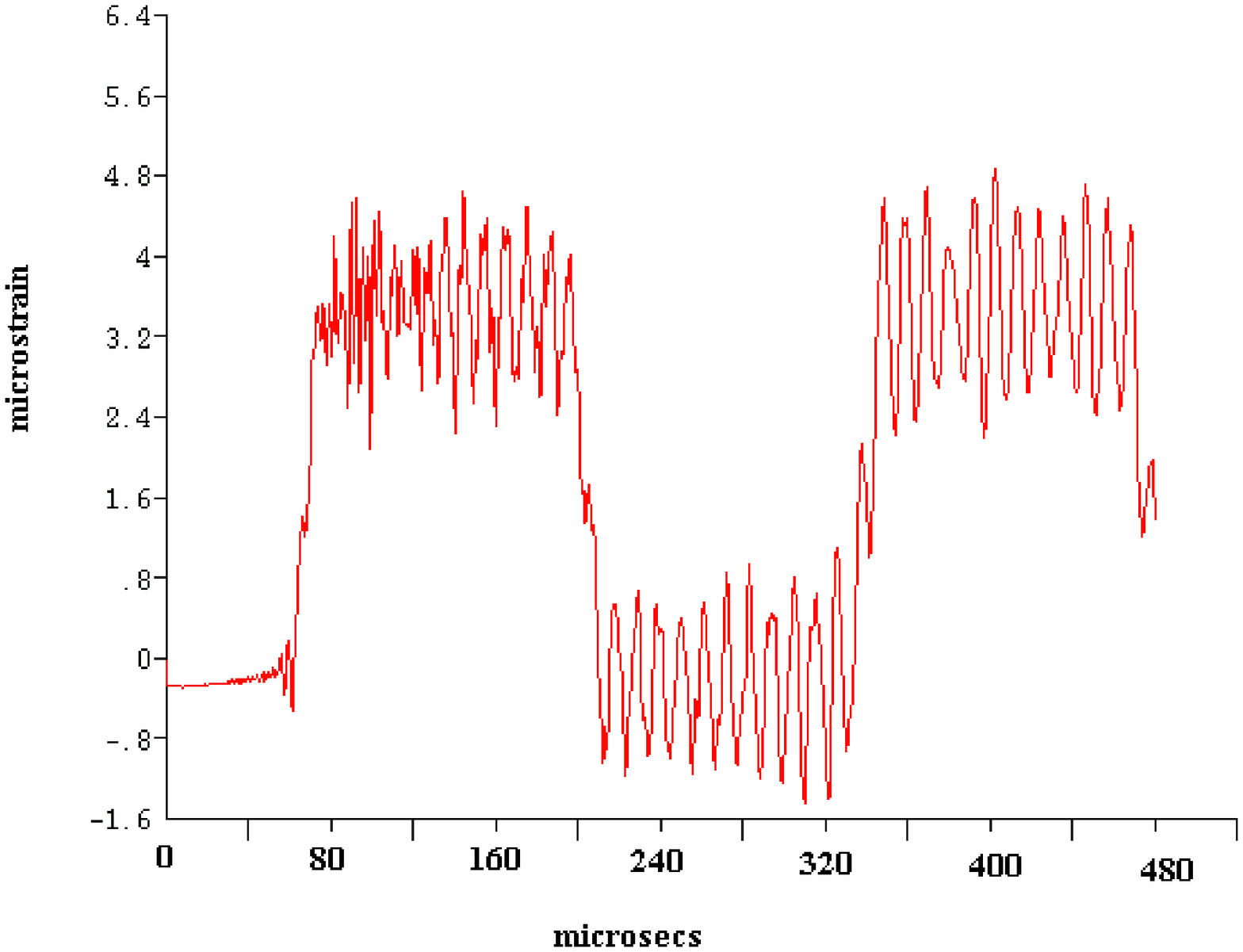}}
\caption{Predicted dynamic strains at approximately same locations
as strain gauges. } \label{fig:pds}
\end{figure}

Figure ~\ref{fig:pds} depicts the predictions in axial strain on
the ATJ graphite target at a location similar to the location of
the two fiber-optic strain gauges of Figure ~\ref{fig:asme951}.
The predictions were made using similar transient finite element
methodologies as the one utilized in the analysis of the CC target
and horn. It is clear that these computational techniques can
provide an excellent estimate of the actual system response.

\begin{figure}[htbp]
\centerline{\includegraphics[width=4.05in,height=3.04in]{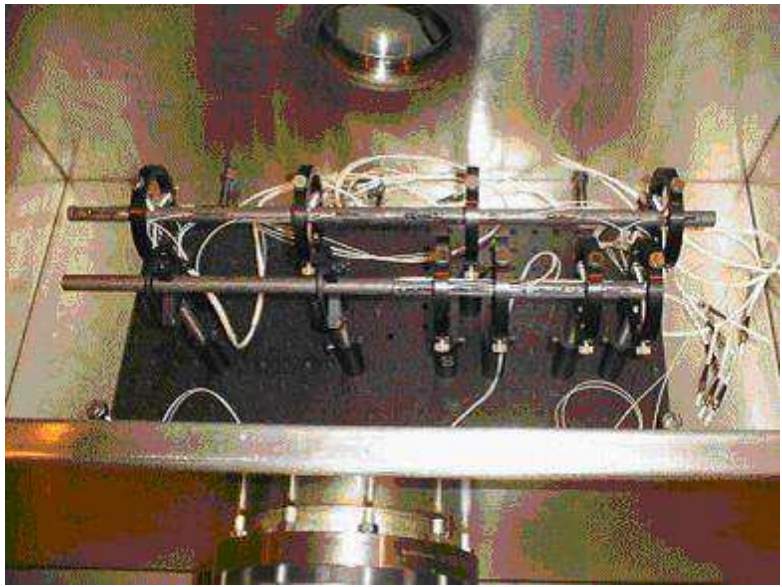}}
\caption{Carbon-Carbon composite and ATJ graphite target
arrangement for the E951 experiment.}
 \label{fig:ccce951}
\end{figure}

\begin{figure}[htbp]
\centerline{\includegraphics[width=4.753in,height=2.816in]{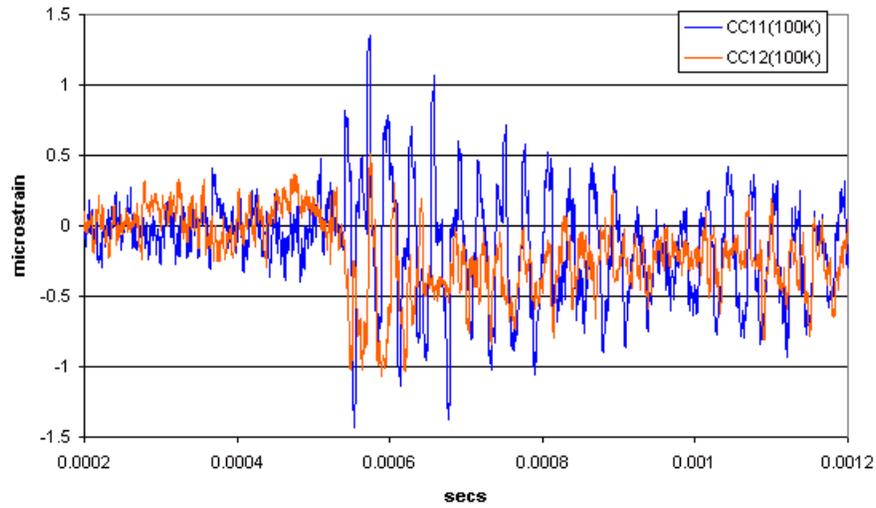}}
\caption{Axial surface strain measurements in the CC target. }
 \label{fig:assmcc}
\end{figure}

\lhead{AGS Super Neutrino Beam Facility} \rhead{Horn Design
Issues} \rfoot{April 15, 2003}
\subsection{Horn Design Issues}
\label{sec:hdi} Figure ~\ref{fig:caht} depicts the first horn made
of aluminum and with a thickness in the neck-down section of the
conductor is 2.5 mm. The radius of the inner conductor for the
following calculations was assumed to be 7 mm. The issues
associated with the horn are the following:
\begin{itemize}
\item Joule heating in the inner conductor

\item Energy deposited on the horn due to beam-target interaction

\item Heat received from target by radiation exchange

\item Heat removal for the 2.5 Hz operation

\item Thermal stresses (including highly transient loads)

\item Magnetic field pressures

\item Vibration of thin horn conductors (from pulsed load and
possibly from He coolant)

\item Differential expansion between inner and outer conductor

\end{itemize}

The most serious issue is the amount of heat generated and
deposited in a small volume of the inner conductor where current
is flowing mostly in the skin depth. For the engineering
calculations 240 kA will flow into the horn during each pulse in a
half-sine form with duration of 20 $\mu $s.

\vspace{ 0.25in}

\underline{\textbf{Skin Depth -- Effective Resistance -- Joule
Heating Calculations}}

\vspace{ 0.25in}

In a previous section the parameters governing the current flow in
the horn and the resulting joule heat were estimated. The primary
goal, given the structure of the current pulse is to estimate the
skin depth and subsequently the joule heating throughout the horn.
For a horn made of aluminum 3000-series and the 20 $\mu $s
half-sine 240 kA peak current pulse the following relation lead to
the estimate of skin depth and joule heating:
\begin{itemize}
\item Resistivity $\rho $ of Al-3000 = 4.2 mohm -- cm

\item Skin depth $\delta $ = [6.61/f$^{1 / 2}$] k$_{1}$

\item  k$_{1}$ = [$\rho $/$\rho _{c}$]$^{1 / 2}$

\item $\rho _{c}$ = 1.724 mohms-cm

\end{itemize}

Utilizing only the dominant frequency (f = 0.025 MHz) the skin
depth has been estimated to be:
 \begin{center}
 $\delta _{al}$ =0.06525 cm.
 \end{center}

Using the parameters above (and a 240 kA current in the horn),
calculations were made with the simplified assumption that all the
current flows uniformly over one skin depth. The purpose of this
simplification was to generate joule heat and temperature rise
estimates that can be used to validate and benchmark the finite
element simulation model being employed to estimate the heat
generation and transfer in the integrated system. By assuming that
the current only flows over one skin depth, an effective
resistance can be estimated,

\begin{equation}
R_{eff}=\rho /A_{s} = 4.2 e-6 \textrm{(Ohm-cm)}/(2\pi  \cdot  0.95
\cdot 0.06525) = 0.107836 \cdot 10^{-4} \textrm{Ohm/cm}.
\label{7eq16a}
\end{equation}

For this simplified case the joule heating per current pulse can
be estimated from

\begin{equation}
JH_{pulse - cm}=\int\limits_0^{20\mu s} I^{2} R_{eff} dt ,
\label{7eq16b}
\end{equation}

\noindent where,$ I $= 250 kA sin(\textit{$\pi $t/20$\mu $s}).
From the above relation one arrives at

\begin{equation}
JH/(pulse-cm) = 6.211 ~\textrm{Joules}.\label{7eq16c}
\end{equation}

The total power generated in the neck-down section of the horn
alone is estimated to be:

\begin{equation}
P_{neck} = 2.5 (Hz) \cdot  6.211 J/\textrm{(cm-pulse)} \cdot  86
\textrm{ cm} = 1.335 ~\textrm{kW}. \label{7eq16d}
\end{equation}

The temperature rise in the horn conductor per pulse is estimated
by calculating the volume per cm-length of the conductor,

\begin{equation}
V = 2 \pi  r_{o}  \delta  = 2 \pi  \cdot  0.95 \cdot  0.06525 =
0.38947 ~cm^{3}. \label{7eq16e}
\end{equation}

The mass of aluminum in the neck-down volume (with density $\rho
_{AL}$ = 2.77 g/cc) is 1.0788 g. The temperature rise in the horn
conductor is estimated from,

\begin{equation}
Q = m c_{p}  \Delta T ,
 \label{7eq16f}
\end{equation}

\noindent where, c$_{p}$ = 0.967 J/g-C. This leads to an estimated
temperature rise per pulse \textbf{$\Delta $T = 5.95}$^{o
}$\textbf{C}. As will be shown in the next section, the calculated
temperature rise per pulse using a detailed finite element
analysis is of the order of $\sim $ 6$^{o }$C.

 \vspace{ 0.25in}

\underline{\textbf{HORN Thermal Finite Element Analysis}}

\vspace{ 0.25in}

\begin{figure}[htbp]
\begin{center}
\centerline{\includegraphics[width=2.4048in,height=2.697in]{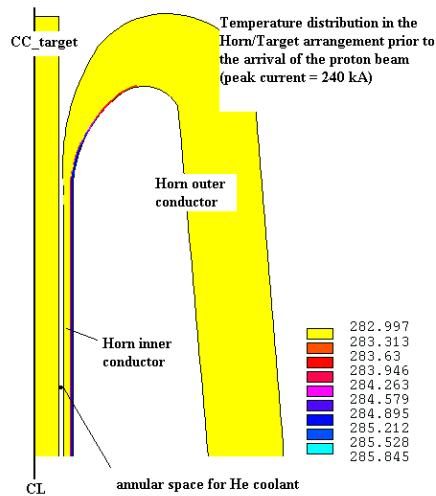}}
\caption{ Geometry of the upstream end of the horn/target
arrangement and temperature distribution in the narrow inner horn
conductor prior to the arrival of the proton beam due to 240 kA
current pulse.} \label{fig:ght}
\end{center}
\end{figure}

\begin{figure}[htbp]
\begin{center}
\centerline{\includegraphics[width=4.302in,height=2.982in]{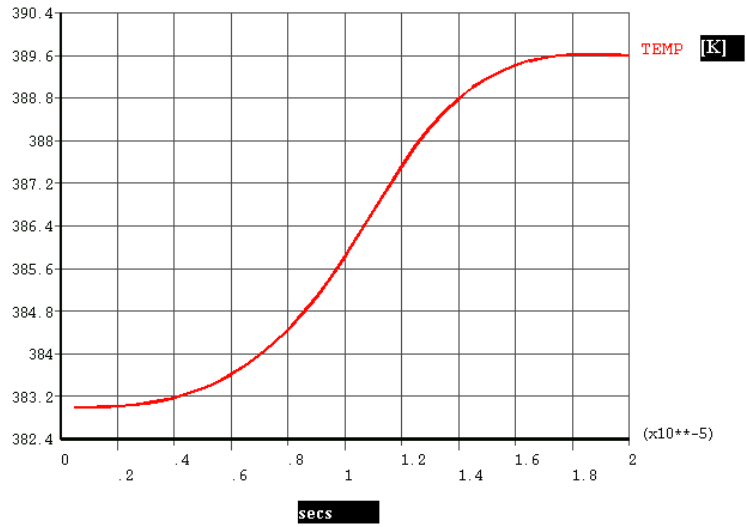}}
\caption{ Calculated temperature rise in the conductor skin depth
(narrowest section) due to the half-sine 240 kA peak current pulse
using finite element analysis.} \label{fig:ctr}
\end{center}
\end{figure}

\begin{figure}[htbp]
\begin{center}
\centerline{\includegraphics[width=4.332in,height=3.198in]{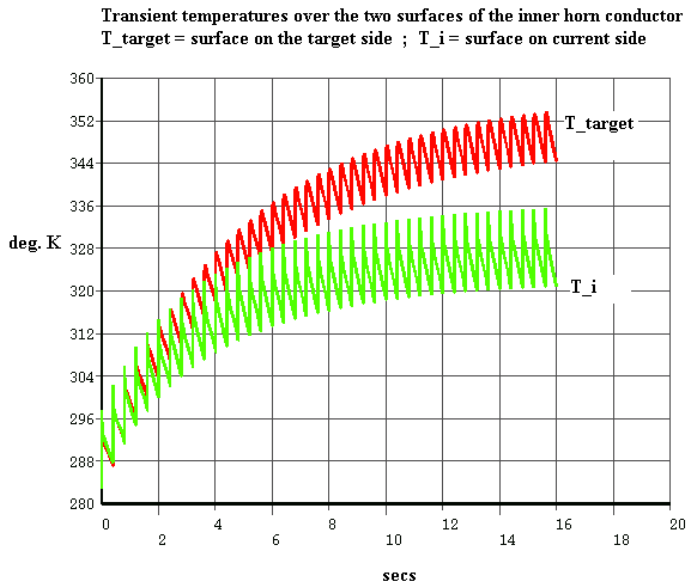}}
\caption{ Transient temperatures over the two surfaces of the
narrowest section of the inner horn conductor.}
 \label{fig:ttihc}
 \end{center}
\end{figure}

\begin{figure}[htbp]
\begin{center}
\centerline{\includegraphics[width=3.985in,height=2.855in]{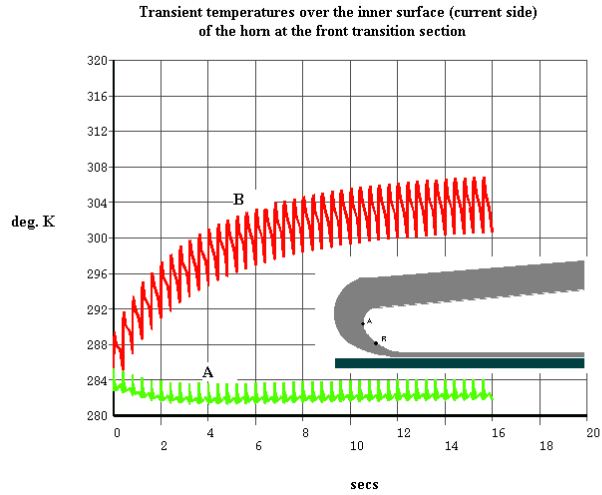}}
\caption{ Transient temperatures over the inner surface of the
upstream transition section of the horn.}
 \label{fig:tttsh}
 \end{center}

\end{figure}

\begin{figure}[htbp]
\begin{center}
\centerline{\includegraphics[width=3.615in,height=2.705in]{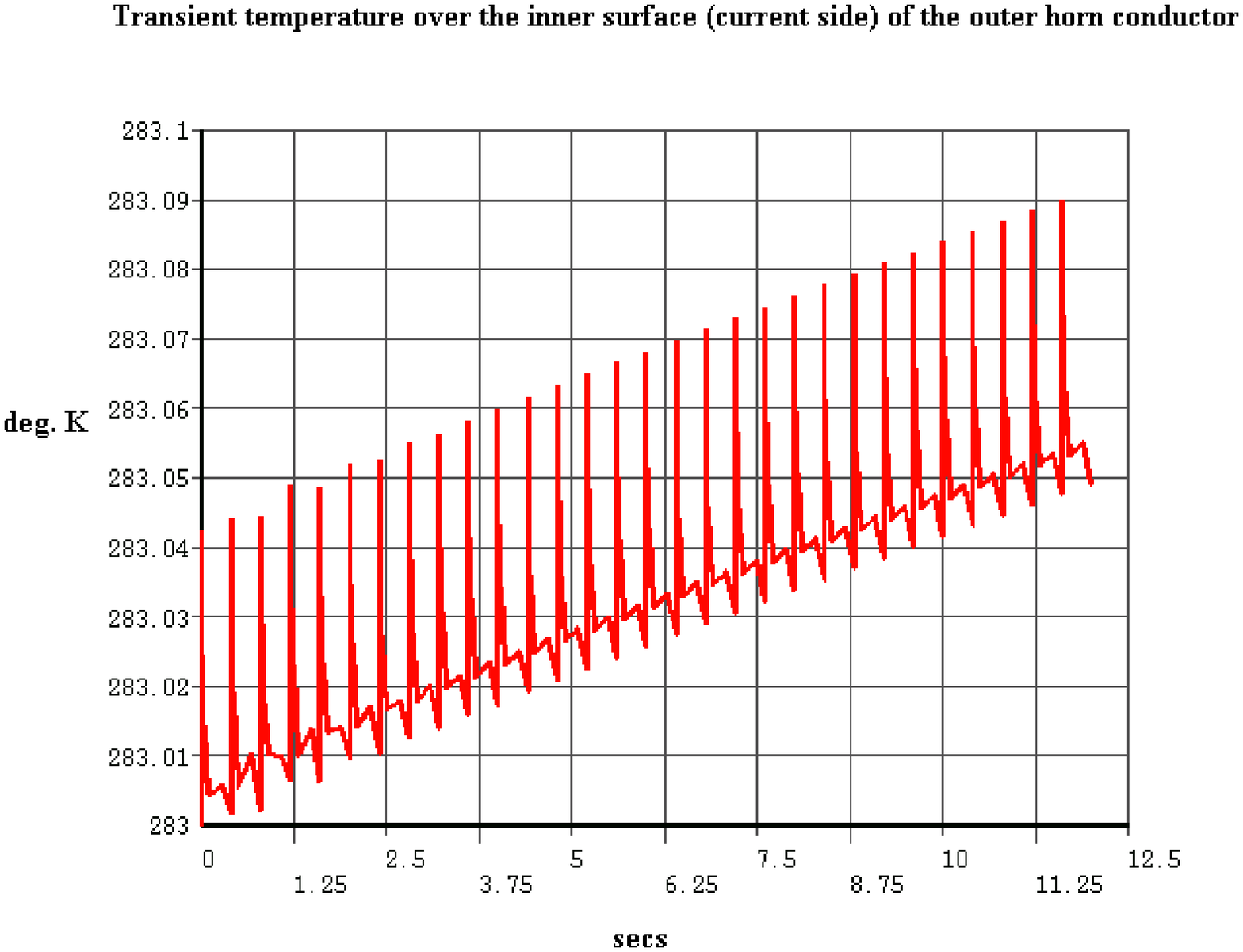}}
\caption{ Transient temperatures estimated over the inner surface
of the outer horn conductor.}
 \label{fig:ttohc}
 \end{center}
\end{figure}

By focusing on the neck-down section, it is estimated that
approximately 1.4 kW are generated due to joule heating. The
temperature rise per pulse at the inner surface of the conductor
(based on aluminum density of 2.77 g/cc and c$_{p}$ = 0.967 J/g-C)
is about $\sim $6$^{ o}$C per pulse while the melting temperature
of aluminum is 660 C. Numerical calculations based on a
sophisticated finite element analysis verified the temperature
rise in the inner horn conductor. Figures ~\ref{fig:ght} -
~\ref{fig:ttohc} depict temperature distributions and transients
throughout the horn. The transient analysis of the integrated
system is performed while considering all the participating
mechanisms of heat transfer in order to find out the operating or
``steady-state'' temperatures of the system. As seen from the
following results, the goal of maintaining the aluminum
temperature below 100$^{o}$ C is achievable. It should be noted
that with the replacement of RAW for the inner surface horn
cooling with sub-cooled He the threshold of 100$^{o}$ C can be
increased to several hundred degrees and thus reducing the demand
for massive cooling since the heat transfer effectiveness will
increase with higher allowable surface temperatures in the
aluminum.

\vspace{ 0.25in}

\underline {\textbf{Exploration  of Alternative  Horn Materials}}

\vspace{ 0.25in}

While aluminum is the baseline material for this study, attention
has been focused to a new material called AlBeMet which is
considered to be the highest performance aluminum alloy in the
industry. The Figures ~\ref{fig:cAlBeMet} - ~\ref{fig:pAlBeMet}
provide its composition and properties relative to a number of
other materials. In this on-going study this material will be
furthered assessed for potential use in the horn structure.

\begin{center}
\begin{figure}[htbp]
\centerline{\includegraphics[width=3.918in,height=2.088in]{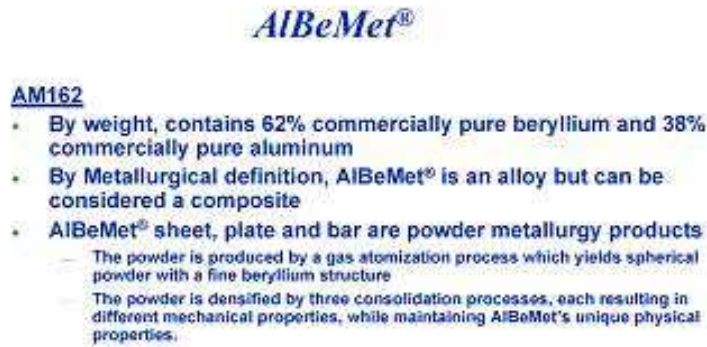}}
\caption{Composition of AlBeMet alloy.}
 \label{fig:cAlBeMet}
\end{figure}

\begin{figure}[htbp]
\centerline{\includegraphics[width=5.66in,height=3.97in]{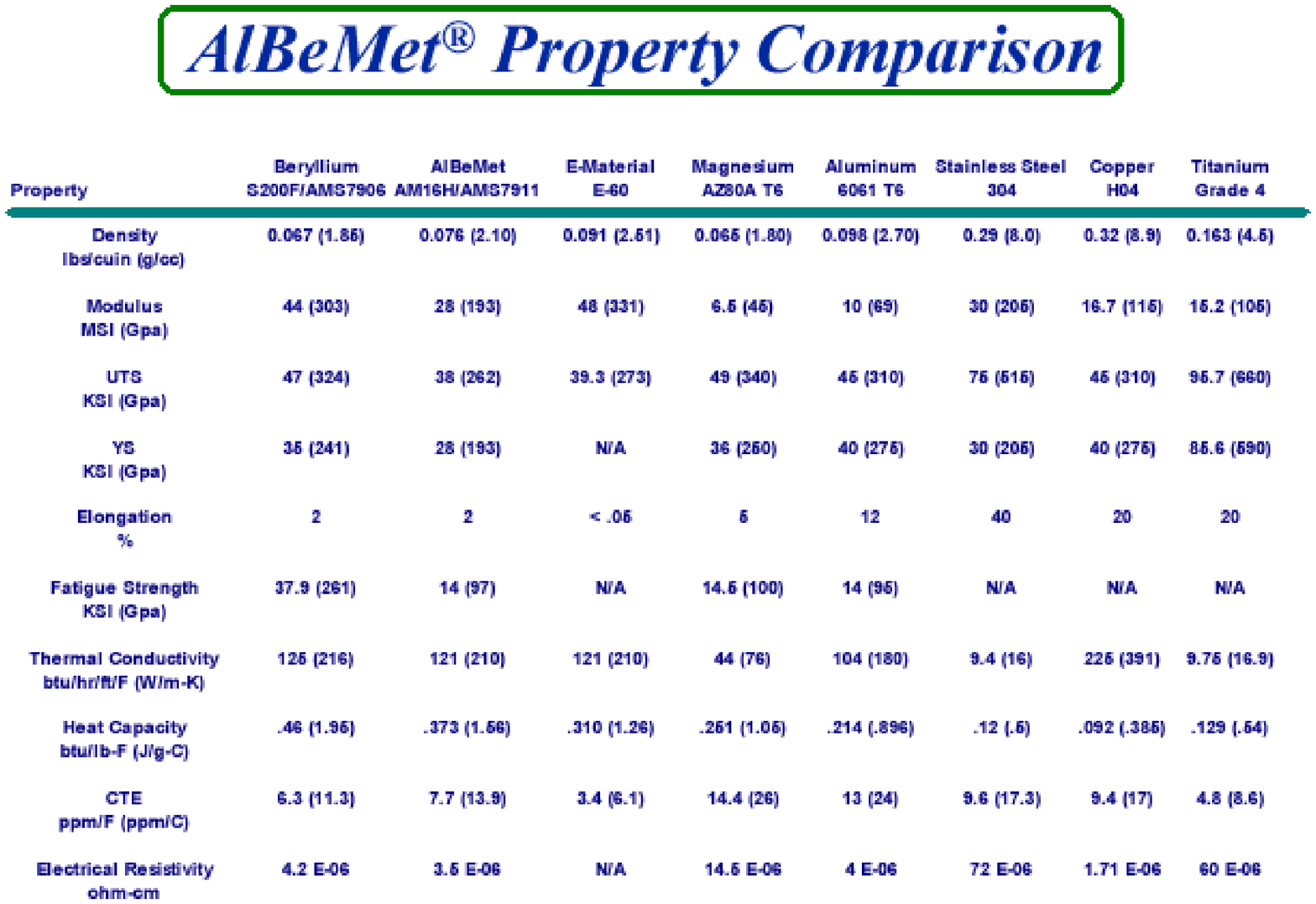}}
\caption{ Properties of AlBeMet Alloy.} \label{fig:pAlBeMet}
\end{figure}
\end{center}

\vspace{ 0.25in}

\textbf{Work in Progress}

\vspace{ 0.25in}

In the course of this on-going study a number of issues still
remain to be addressed in the effort to optimize the horn/target
system and maximize its useful life. The ``open'' issues that are
currently being addressed through material selection, detailed
modeling and analysis and envisioned R{\&}D efforts are summarized
below.
\begin{itemize}
\item \textbf{Gamma-Ray Heating of the Horn }(especially of the
inner conductor) Independent hadronic analysis are being performed
to confirm the energy deposited in the inner conductor of the horn
closest to the target.

\item \textbf{Horn Thermal Stresses}

\item \textbf{Horn Magnetic Forces and Induced Stresses}

\item \textbf{Horn Vibration - Modal Analysis}

\end{itemize}

Detailed finite element analysis are under way to estimate the
level of stresses generated in the horn, its modes of excitation
and the means to optimize the horn structurally so it can meet all
the physics criteria.

\begin{itemize}
\item \textbf{Horn Material (Aluminum) Irradiation {\&} Fatigue}
It is envisioned that as part of the R{\&}D and fatigue estimation
effort, experience data will be reviewed further and, possibly,
exposure of the material to dpa values will take place. Further,
tapping on the expertise of collaboration members, the effects of
the water environment on the changes induced on the aluminum
surface and the issue of crack formation will be examined.

\item \textbf{Discussion of Issues Related to Choice of Coolant}
In connection with the previous bullet, the effects and heat
removal capacity of the choice of coolant fluids will be further
assessed.

\end{itemize}

\clearpage
\newpage
\lhead{AGS Super Neutrino Beam Facility} \rhead{Conceptual Design
of  the Horn Power Supply} \rfoot{April 15, 2003}

\subsection{Conceptual Design  of the Horn Power Supply}

\label{sec:hornps} The main objectives of the horn \cite{ref:hps1}
power supply design is to achieve an output pulse amplitude of
250KA, a pulse flat top of 2.5 $\mu $S or longer with a 1{\%}
flatness and pulse to pulse repeatability, and a repetition rate
of 2.5 pulses per second.

Most commonly used scheme is the capacitor discharge type
\cite{ref:hps2}\cite{ref:hps3}\cite{ref:hps4}\cite{ref:hps5}. In
this type of circuit, a capacitor bank stores the energy, and a
main discharge switch release the energy to the load through
transmission lines. For very long distant transmission, pulsed
transformers have been added into the KEK design and CNGS horn
system. The horn as an electrical load is usually being described
as an inductor in series with a resistor. Discrete parameters of
inductance and resistance are also used to formulate the short
length, low impedance transmission lines when associated with low
bandwidth pulse. Hence, the circuit can be simplified as a RLC
discharge circuit.

The pulse rise time, T$_{r}$, is usually approximated by the
quarter period of sine wave. For a lossless LC oscillation
circuit, this can be determined by the equation $T_r \; =
\;\frac{\pi }{2}\;\sqrt {LC} \;$, and the load current is given by

\begin{equation}
I(t) = V_o \sqrt {\frac{C}{L}}\sin (\frac{1}{\sqrt {LC}}t ).
\label{eq:hornps1}
\end{equation}

The maximum current amplitude is $I_{\max } = V_o \sqrt
{\frac{C}{L}} $for lossless LC circuit.

In the case of critical damped RLC circuit, $R\; = \;2\,\sqrt {L /
C} $ and

\begin{equation}
I(t) = \frac{V_o }{L}t\;e^{ - \,\;\frac{R}{2L}t}.
\label{eq:hornps2}
 \end{equation}

 The maximum output current can
be determined by

\begin{equation}
I_{\max } (t = \frac{2L}{R}) = V_o \sqrt {\frac{C}{L}} \;e^{ - 1},
\label{eq:hornps3}
\end{equation}

\noindent and the current rise time is $T_r = \frac{2L}{R} = \sqrt
{LC} $.

Compare two cases, for a given L, C and charging voltage V$_{o,}$
the output current of the critical damped one has peak amplitude
of 0.3679 of the lossless one. To reach the same maximum output
current, the initial voltage of the critical damped circuit has to
be 2.7183 times of the lossless one. It implies that, for the
critical damped one, the energy stored in the capacitor bank as
well as the charging power has to be 7.389 times larger.

In the high current path, the resistance caused voltage drop and
energy dissipation are critical factors to be considered. The
cooling system for heat removal from the effective resistor, and
the additional charging power required to make up the resistive
dissipation can be very costly. Hence the low circuit resistance
is preferred.
\subsubsection{Effective Resistance {\&} Skin
Effect} \label{hornpsskineff}

The load and transmission line resistance varies with frequency
due to skin effect. For any given material, the skin depth $\delta
_{s}$ is inverse proportional to the square root of the frequency
f, and the effective resistance R$_{eff}$ is proportional to the
oscillating frequency. Here

\begin{equation}
\delta _s = \frac{1}{\sqrt {\pi \,f\,\mu _R \,\mu _o \sigma } },
\label{eq:horns4}
\end{equation}

\noindent and

\begin{equation}
R_{eff} = \frac{l}{\sigma \,b\,\delta _s }, \label{eq:hornps5}
\end{equation}

\noindent where l  is the conductor length and b is the conductor
width. The material's conductivity, relative permeability, and the
free space permeability are described by $\sigma $, $\mu _{R}$,
and $\mu _{o }$, respectively.

One can see that slower frequency leads to lower effective
resistance. The other factors associated with effective resistance
are the length, width, and permeability and material conductivity.
For non-magnetic material, the relative permeability is close to
unit. The switching device on-state resistance and hardware
connection joints resistance also contribute to the total
resistance.

In summary, the lower effective resistance can be achieved by
using:
\begin{enumerate}

\item Lower frequency, higher skin depth;

\item Wider conductor width;

\item Shorter conductor length;

\item Higher conductivity material;

\item Lower switch on-state resistance; and

\item Lower connection joint resistance.
 \end{enumerate}
\lhead{AGS Super Neutrino Beam Facility} \rhead{Inductance}
\rfoot{April 15, 2003}
\subsubsection{Inductance
Issues} \label{sec:hpsinduct}

The total inductance includes the horn inductance, transmission
line inductance, series inductance of the capacitor, switch
inductance, and circuit loop stray inductance. The external
inductance depends on the inductor geometry and material
permeability. The internal inductance has frequency dependence.
The voltage across the inductor is defined as $V_L (t) =
L\frac{dI(t)}{dt}$. Hence, the larger inductor and faster current
rate of change requires higher voltage.

The current going through the inductor is the inductively stored
energy. For a lossless circuit, the total inductive energy shall
be equal to the total capacitive energy storage, i.e.
\begin{equation}
\frac{1}{2}\,L\,I^2 = \;\frac{1}{2}\,C\,V^2. \label{eq:hornps6}
\end{equation}

Therefore, we have that the higher the inductance and current, the
higher the capacitance and its initial voltage.

For a reasonable design, the total inductance shall be kept as low
as possible, and the current rise time shall be chosen to
accommodate the device operating voltage. \lhead{AGS Super
Neutrino Beam Facility} \rhead{Inductance} \rfoot{April 15, 2003}
\subsubsection{Principle Design Example} \label{sec:pde}

Let us consider a system with overall inductance of 2.5 $\mu $H, a
total resistance of 2 m$\Omega $, and a capacitor bank of 16 mF.
If the initial capacitor voltage is 4000 Volts, the peak output
current amplitude is about 288 kA with 314 $\mu $s rise time.

\begin{figure}[htbp]
\centerline{\includegraphics[width=5.40in,height=3.60in]{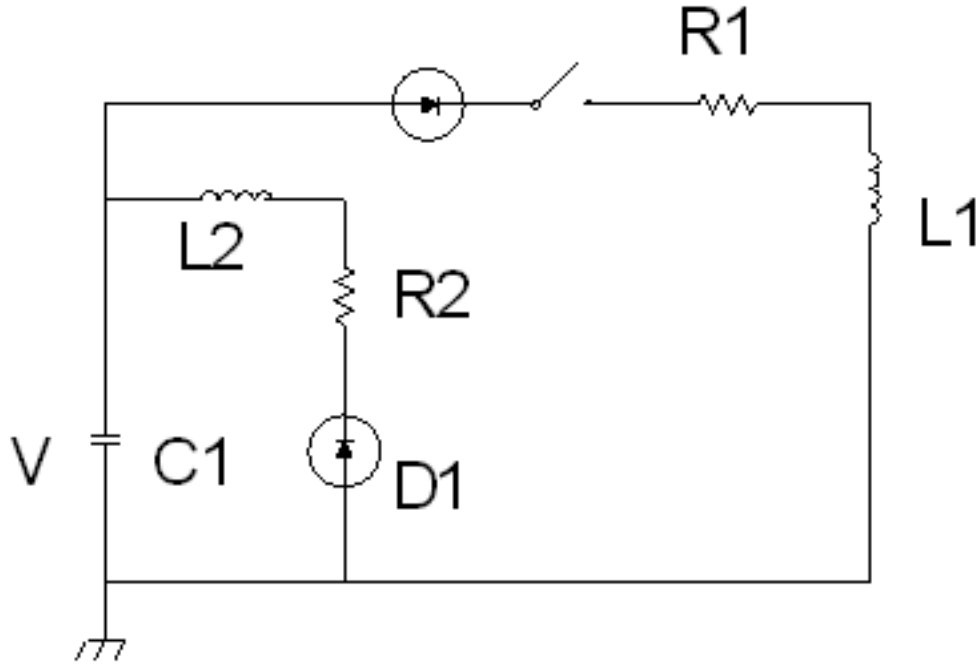}}
\caption{Simplified circuit diagram.} \label{fig:hornps11}
\end{figure}
\begin{figure}[htbp]
\centerline{\includegraphics[width=4.658in,height=3.0in]{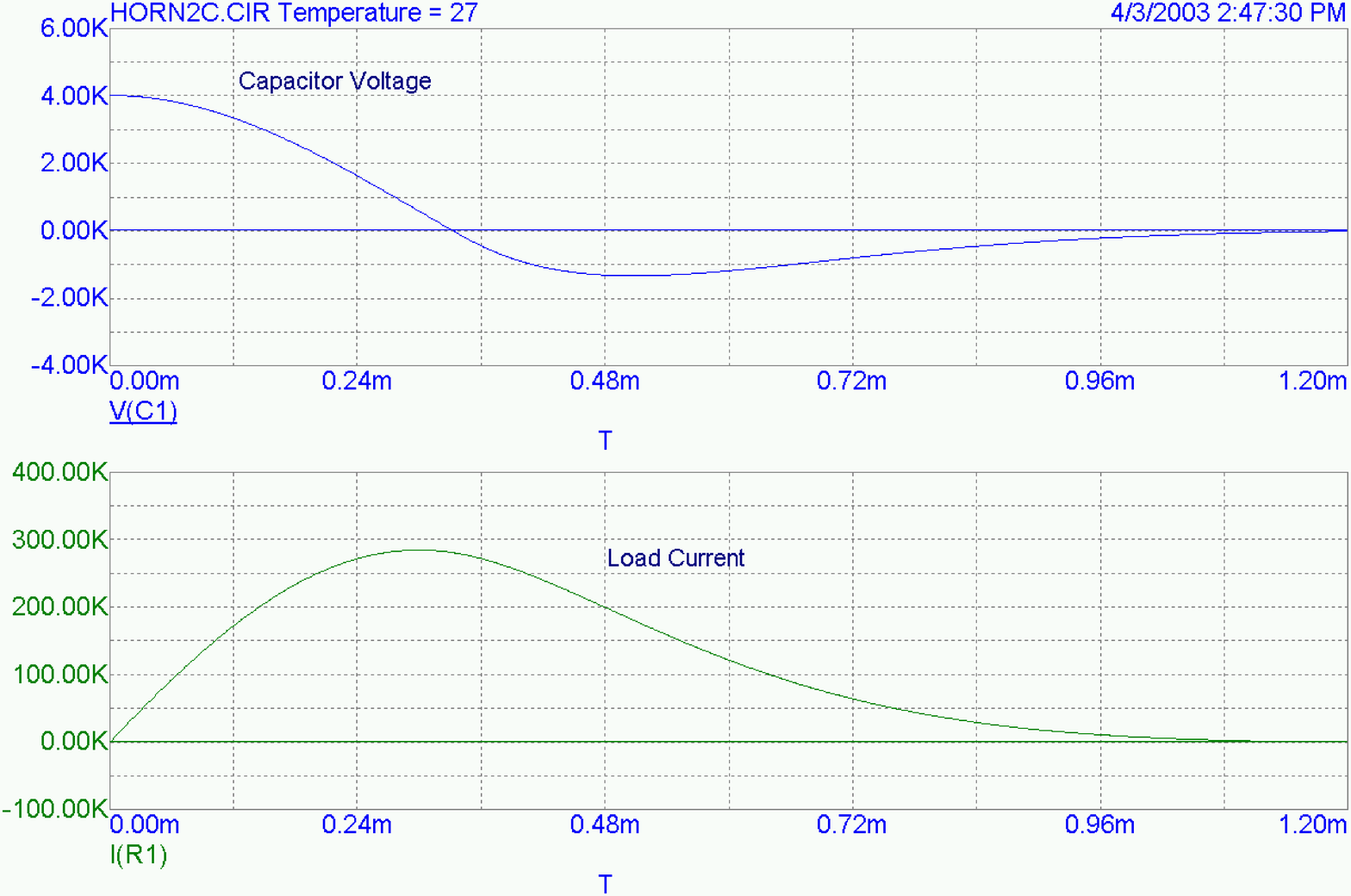}}
\caption{Circuit simulation with V=4000, C1=16 mF, R1= 2 m
$\Omega$, L1=2.5 $\mu$H, L2=0.1 $\mu$H, R2=7.5 m$\Omega$.}
\label{fig:hornps12}
\end{figure}
The stored energy in the capacitor is 128 kJ. For 2.5 Hz pulse
repetition rate, the minimum charging power supply is 320 kW.

For the 3000-series aluminum being considered in the Horn
mechanical design, the material skin depth is 0.06525 cm at 25
kHz. Derive from it, we have the skin depth is 0.3574 cm at 833
Hz.

The parameters used in this example are rough estimates based on
similar systems, as listed on Table ~\ref{tab:hps1}. The
resistance used in the example is tight for the chosen frequency.
If higher resistance has to be used, then the voltage and
capacitance have to be increased.

\begin{sidewaystable}[htbp]
\begin{center}
\caption{The parameter used here based on similar  systems.}
\begin{tabular}
{|p{32pt}|p{25pt}|p{25pt}|p{25pt}|p{25pt}|p{25pt}|p{25pt}|p{25pt}|p{25pt}|p{25pt}|p{25pt}|p{25pt}|p{25pt}|p{25pt}|p{25pt}|p{35pt}|}
\hline
 &
\multicolumn{4}{|c|}{Pulse} & {Cap.}&
\multicolumn{4}{|c|}{Induct.} & \multicolumn{5}{|c|}{Resist.} &
{Rep.}
\\ \hline
 &
Curr.& Volt.& Tr& 10{\%} Width& & Horn {\#}1& Horn {\#}2& Stray&
\textbf{Tot.}& Horn {\#}1& Horn {\#}2& Stray& \textbf{Tot.}& Rev.
Dump& Rate \\ \hline
 & kA& kV& $\mu$S& $\mu$S&
mF& nH& nH& nH& nH& $\mu \Omega$ & $\mu \Omega$& $\mu \Omega$&
$\mu \Omega$& $\mu \Omega$& Hz \\ \hline \textbf{AGS Narr. Band}&
240& 12.44& 58& 350& 0.850& 462& 567& 635& \textbf{1664}& & & &
\textbf{9600}& 5000&
 \\
\hline \textbf{AGS Wide Band}& 285& 10.98& 58& 350& 1.116& & & &
\textbf{1198}& & & & \textbf{7500}& 5000&
 \\
\hline \textbf{Mini Boone}& 170& 5.35& 143& & 1.500& 680& & 660&
\textbf{1340}& 230& & 770& \textbf{1000}& & 5 \\ \hline
\textbf{NuMi}& 205& 0.97& 2600& & 900.& 689& 510& 1208&
\textbf{2407}& 270& 71& 690& \textbf{1031}& & 0.53 \\ \hline
\textbf{CERN}& 300& 7.275& 81& & 1.075& 420& & 200& \textbf{620}&
128& & 200& \textbf{328}& & 75 \\ \hline
\multicolumn{16}{|c|}{\textbf{Pulser with Output Transformer }}
\\ \hline \textbf{KEK}& 250& 4.702& 3000& & 6.000& 1030& & &
 &
210& & &
 &
& 0.5 \\ \hline \textbf{CNGS}& 150& 7.21& 4300& & 4.008& 2150& & &
 &
405& & &
 &
& 2 pul./6s, 50ms apart \\ \hline
\end{tabular}
\label{tab:hps1}
\end{center}
\end{sidewaystable}

\normalsize
\subsubsection{Major Components}
 \label{sec:majorcom}

The major components of this system include:

\begin{enumerate}
\item Charging Power Supply;

\item Capacitor Bank;

\item Discharge Switch;

\item Reverse Diode;

\item Transmission Line.
\end{enumerate}

In this design, Self-Healing capacitors are being considered for
the fault tolerance, and increased reliability. This type of
capacitor is usually rated under a few kilo-volts. Multiple
capacitor cells have to be used to divide capacitor bank into
smaller units with lower stored energy per cell for safety
concerns.

The discharge switch in favor is the light triggered thyristor.
The newest in this category is the EUPEC T2563N80. The advantage
of this device is its high voltage rating of 8000 V, high forward
current rating of 5600 A rms, and 63 kA surge current. Multiple
thyristors have to be used in parallel to carry the 250 kA pulse
current.

This thyristor features light triggered gate structure. It
eliminates the high voltage isolation trigger transformers
normally used in conventional thyristors, and improves the high
voltage hold-off and the noise immunity. Large reverse diodes are
available from the same company.

The trend of new designs is to use solid-state switch, which has
much longer lifetime compared to gas discharge switches. The very
high current capability required in this system limits the
selection to thyristor types for its high power rating per single
unit and cost effectiveness.

The traditionally used ignitron is mercury vapor filled device.
With rapid advancement of solid state devices, it is being
replaced by thyristors. Other solid state devices, such as IGBT
and MOSFET are limited by their current capability.

The transmission line being considered is similar to FERMI NuMI
\cite{ref:hps6} design for its ultra low resistance.

The design options of high voltage, high current pulsed system are
often limited by the industry development and available
components. In this case, the preferred operating voltage is under
5 kV. The total resistance shall be kept to less than or around 2
m$\Omega $.

\clearpage
\newpage
\lhead{AGS Super Neutrino Beam Facility} \rhead{The Conventional
Facility and Target Hill} \rfoot{April 15, 2003}

\section{Conventional Facilities and Target Hill}

\label{sec:cfth}

\setcounter{table}{0}
 \setcounter{figure}{0}
 \setcounter{equation}{0}

\begin{figure}[htbp]
\begin{center}
\centerline{\includegraphics[width=3.6in,height=3.0in]{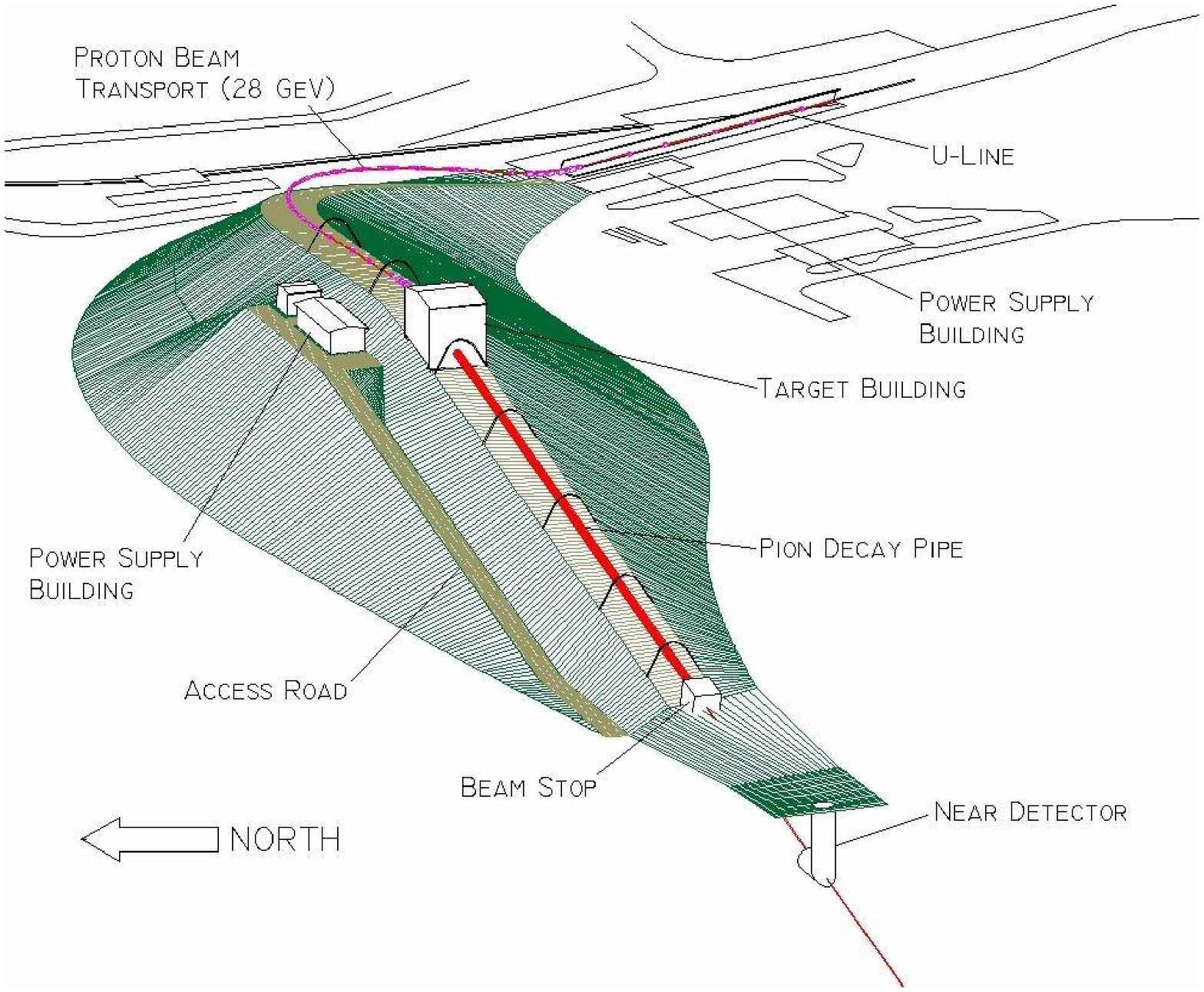}}
\caption{The hill, truncated $\sim $2 meters below the beam and
the new beam transport elements are illustrated.}
\label{fig:rtbtandth}
\end{center}
\end{figure}

 The proton beam will be extracted from the AGS and will utilize part of the
RHIC beam transport before exiting the decommissioned neutrino
beam line tunnel in a northerly direction and at an upward angle
of approximately 13.8 degrees. The beam will bend towards the west
by approximately 68.5 degrees, then down a total of 25.1 degrees
to the proton target. A 200 m decay region follows with the
neutrino beam entering the hadron beam stop at 11.26 degrees to
the horizontal. The Near Detector Facility is located 285 m from
the target, 21 m below ground level. A plan view of this
arrangement at BNL is illustrated in the Figure~\ref{fig:ultbt}and
the overall beam facility is shown in Figure ~\ref{fig:rtbtandth}.

The vertical beam geometry results in a hill 50 meters high at the
apex of the proton beam. This geometry provides for the neutrino
beam's 11.3-degree entry into the earth and avoids potential
irradiation of soil close to the Long Island water table. This,
together with impermeable rain water barriers to prevent rain
water penetration of potentially irradiated soil, is consistent
with present ground water protection practices at BNL. A
simplified vertical beam profile and a cross-section of the hill
is shown in Figures ~\ref{fig:homsd} and ~\ref{fig:nb1}.

\begin{figure}[htbp]
\centerline{\includegraphics[width=2.61in,height=1.926in]{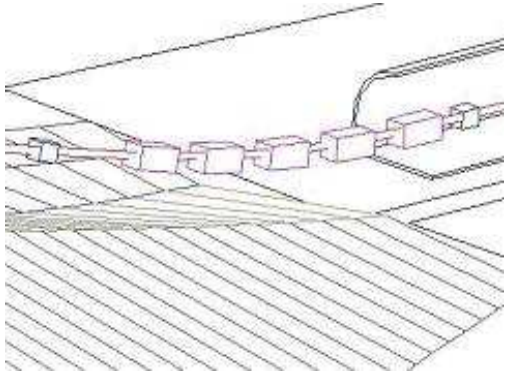}}
\caption{13.8 Degrees vertical bend.} \label{fig:166vertbend}
\end{figure}

\begin{figure}[htbp]
\centerline{\includegraphics[width=1.944in,height=1.656in]{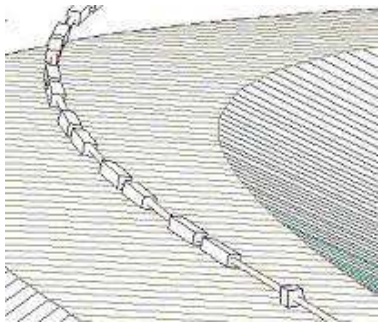}}
\caption{68.2 Degrees horizontal bend.} \label{fig:682horzbend}
\end{figure}

\begin{figure}[htbp]
\centerline{\includegraphics[width=2.415in,height=2.145in]{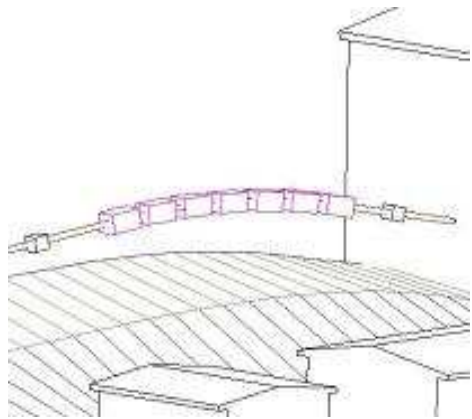}}
\caption{16.6 Degrees vertical bend.} \label{fig:138vertbend}
\end{figure}

The new conventional facilities can be subdivided into the
following areas:

\begin{itemize}

\item Proton beam transport

\item Target area

\item Decay tunnel

\item Hadron beam stop

\item Near Detector Facility
\end{itemize}

\lhead{AGS Super Neutrino Beam Facility} \rhead{Proton Beam
Transport} \rfoot{April 15, 2003}
\subsection{Proton Beam Transport}
\label{sec:pbt}

Existing utilities and roads will be relocated and approximately
726,350 cubic meters of sand fill will be placed forming the hill.
The sand fill will be placed in 0.3 m lifts and compacted to
98{\%} of its maximum density. The fill will be placed early in
the project allowing it to settle for several years before
re-excavation for placement of the tunnel. Approximately 330 m of
3-m diameter tunnel, overburdened with 6.0 m of fill will be
required for proton transport. There will be a waterproof liner
installed .6 m below the surface of the overburden. Portions of
beam transport are depicted in Figures ~\ref{fig:166vertbend},
~\ref{fig:682horzbend}, and ~\ref{fig:138vertbend}.

\begin{figure}[htbp]
\centerline{\includegraphics[width=3.003in,height=2.26in]{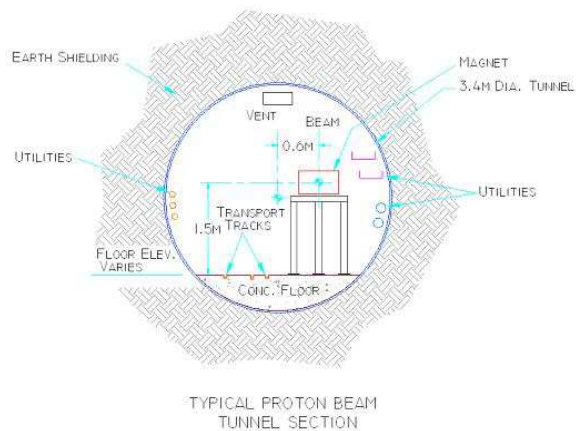}}
\caption{Typical proton beam tunnel section.} \label{fig:pbts}
\end{figure}

Access to the tunnel will be provided at the beginning of the
vertical rise and near the target area. Equipment transport up
{\&} down the slope will be facilitated by use of a motorized
railway on the aisle side of the transport elements in the tunnel.
A typical section of the proton tunnel is shown Figure
~\ref{fig:pbts}.

Two power supply/utility buildings will be provided, one located
low near the existing beam line, the other located high near the
target area. These buildings will house power distribution
systems, power supplies, water pumping systems, instrumentation
and controls for the beam line.

The cooling water system will use a 3.5 MW cooling tower for
primary heat rejection with four isolated, closed loop cooling
systems for:
\begin{itemize}
\item All transport magnets ($\sim $2.8 MW)

\item Two power supply areas ($\sim $. 5 MW)

\item Horn cooling ($\sim $. 2 MW)

\end{itemize}

Each system will contain redundant pumps, a heat exchanger, a full
flow filter and a side stream deionizer capable of maintaining the
system at 2-5 Megohm-cm. The system controls will be PLC based and
be capable of monitoring and reacting to water leaks if they
occur. In general, electrical power will be distributed around the
site at 13.8 kV. Unit substations will transform the power into
convenient voltages, typically 480 and 208/120 volts. Electrical
power is divided into two major categories: conventional and
experimental. Conventional power encompasses building power for
lighting and convenience power, HVAC, and miscellaneous equipment;
 and site distribution. Emergency power will be provided as
required from existing circuits.

Experimental power feeds all the power supplies and associated
equipment, such as, cooling water pumps, cooling towers, etc.
Designs will follow the requirements of the National Electrical
Code and industry standards.

\begin{table}[htbp]
\begin{center}
\caption{Preliminary power requirements.}
\begin{tabular}
{|p{115pt}|p{75pt}|p{75pt}|p{65pt}|} \hline Location& Conventional
Power (KVA)& Experimental Power (KVA)& Total \par Power (KVA) \\
\hline Lower Power House& 150& 2750& 2900 \\ \hline Upper Power
House& 150& 2050& 2200 \\ \hline Near Detector& 100& 900& 1000 \\
\hline Total Power (KVA)& 400& 5700& 6100 \\ \hline
\end{tabular}
\label{tab:powerforcf}
\end{center}
\end{table}

The present radiation security system in the RHIC transport line
will be extended and three new radiation security access gates
will be installed. Access to the new tunnel will be allowed during
RHIC operations by implementing the necessary features such as
shield walls and beam plugs in the existing tunnel. A fence will
be provided to limit access over the beam tunnel while the beam is
in operation.

Fire detection and protection will be provided where appropriate
throughout the facility and connected to the BNL site fire alarm
system. High sensitivity smoke detection systems will be utilized
in increased risk or high value areas.

The proton beam tunnel will be provided with dehumidification,
ventilation and a compressed air system for device controllers.
Heat and air-conditioning will be provided in the power supply
buildings. All buildings will be connected to the BNL site
sanitary system where the appropriate monitoring of effluent is
provided.

Ground water monitoring wells will be provided to insure
compliance with all Local, State and Federal ground water
protection requirements.

\lhead{AGS Super Neutrino Beam Facility} \rhead{Target Area}
\rfoot{April 15, 2003}
\subsection{Target Area}
\label{targetarea}

\begin{figure}[htbp]
\centerline{\includegraphics[width=6.018in,height=3.72in]{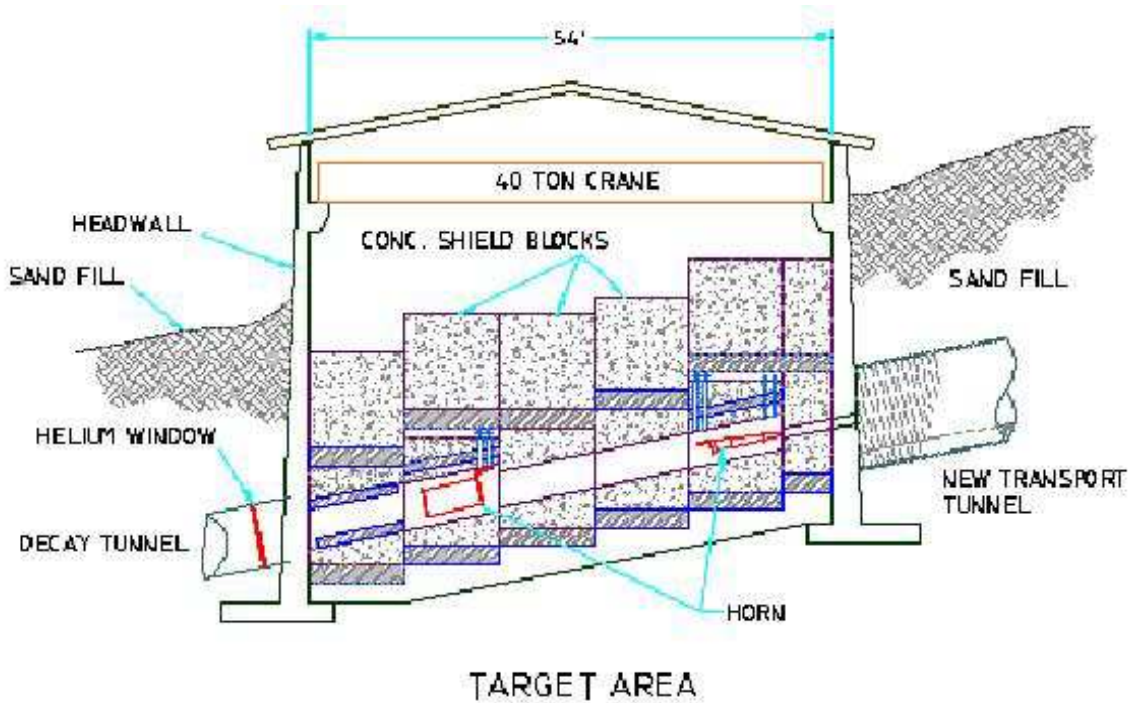}}
\caption{Target area.} \label{fig:targetarea}
\end{figure}

\begin{figure}[htbp]
\centerline{\includegraphics[width=3.35in,height=2.31in]{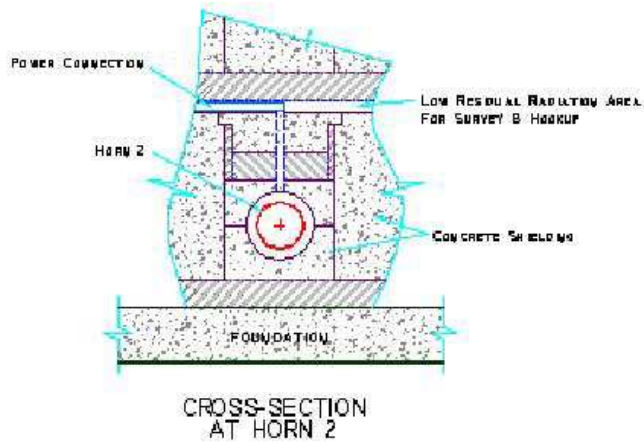}}
\caption{ Cross section at horn{\#}2.} \label{fig:csathorn2}
\end{figure}

The target area will include the proton target, horns, horn power
supply and water cooling system.The facility will be a high bay
building with a stepped foundation and an overhead crane of
sufficient capacity to handle the largest load of approximately 40
tons, see Figure ~\ref{fig:targetarea}. A shielded storage area
will be provided for radioactive component storage and repair.
Modular concrete and steel shielding will provide radiation
shielding. Access to the horn vault for installation and removal
of the horns is accomplished by removing the modular shielding.
Present plans call for hanging the horns from shielded supports
with the ability to survey and connect or disconnect the horns
from above the shield in a relatively low residual radiation
environment as shown in Figure ~\ref{fig:csathorn2}. The design of
this area, as well as all areas, will incorporate ALARA radiation
 principles.  A collimator downstream of Horn {\#}2 will be
installed to intercept a portion of the off-axis beam that would
otherwise interact in the soil along the decay tunnel.


Survey tolerances, both absolute and relative, are driven by the
physics requirements which are less stringent than for the
``MINOS'' experiment in Soudan, MN.. The geodetic position of BNL
is well known and tied to the National Geodetic Survey system. The
geodetic position of the beam front-end elements and the Detector
site will be taken in 2 to 3 simultaneous epochs giving the
relative positions of the two sites within a few centimeters. A
more difficult issue will be in determining the position of the
subterranean detector relative to the surface. Gyro-theodolite
observations coupled with traditional line-of-site observations
(if possible) will provide the data necessary. If line-of-site
observations are not possible, an Inertial Navigation System may
be used giving the relative position within a few meters.

\lhead{AGS Super Neutrino Beam Facility} \rhead{Decay Tunnel}
\rfoot{April 15, 2003}
\subsection{Decay Tunnel}

\label{decaytunnel}

The decay tunnel will be a 2-meter diameter steel tube; 185 meters
long, with seal welded joints. There will be a thin helium window
at the downstream end of Horn {\#}2 and a helium window at the
upstream end of the hadron beam stop. Simulations will be done to
predict the temperatures and stresses on the windows and the
appropriate measures taken to insure they withstand the
radioactive environment. The contained volume will be purged with
helium gas. Access to the upstream window will be provided through
the target area. There will be no utilities or access to the decay
tunnel between helium windows. The tunnel will be overburdened
with 9 meters of earth fill with a waterproof liner installed
approximately 0.6 meters below the surface, extending 9 meters on
each side of the tunnel.
\begin{figure}[htbp]
\centerline{\includegraphics[width=6.042in,height=3.21in]{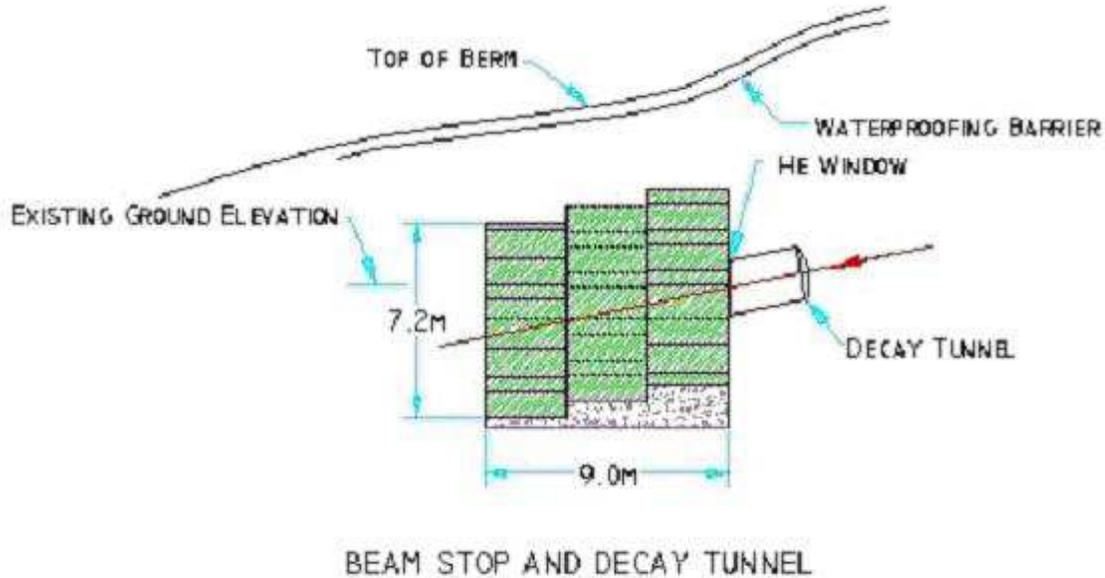}}
\label{fig:bsatdt} \caption{Beam Stop and decay tunnel.}
\end{figure}
\lhead{AGS Super Neutrino Beam Facility} \rhead{Hadron Beam Stop}
\rfoot{April 15, 2003}
\subsection{Hadron Beam Stop}

The hadron beam stop will be approximately 7 m wide x 7 m high x 9
m long. Elevation view is shown if Figure ~\ref{fig:bsatdt}. It
will have a stepped foundation to approximate the 11.3-degree
downward angle. The lowest portion of the beam stop will be
approximately 3m above the ground water table. The stop will
consist mainly of existing steel plate from the decommissioned
neutrino beam line at BNL, overburdened with 4m of soil. A
waterproof liner will be placed .6m below the surface of the soil
to insure no rainwater can penetrate. Simulations will be carried
out to predict the temperatures and stresses on the front face of
the stop and the appropriate materials chosen. It is presently
assumed that no cooling of the material is necessary and only
temperature interlocks are required.

\lhead{AGS Super Neutrino Beam Facility} \rhead{Near Detector
Facility} \rfoot{April 15, 2003}
\subsection{Near Detector Facility}

\begin{figure}[htbp]
\centerline{\includegraphics[width=4.5in,height=3.9in]{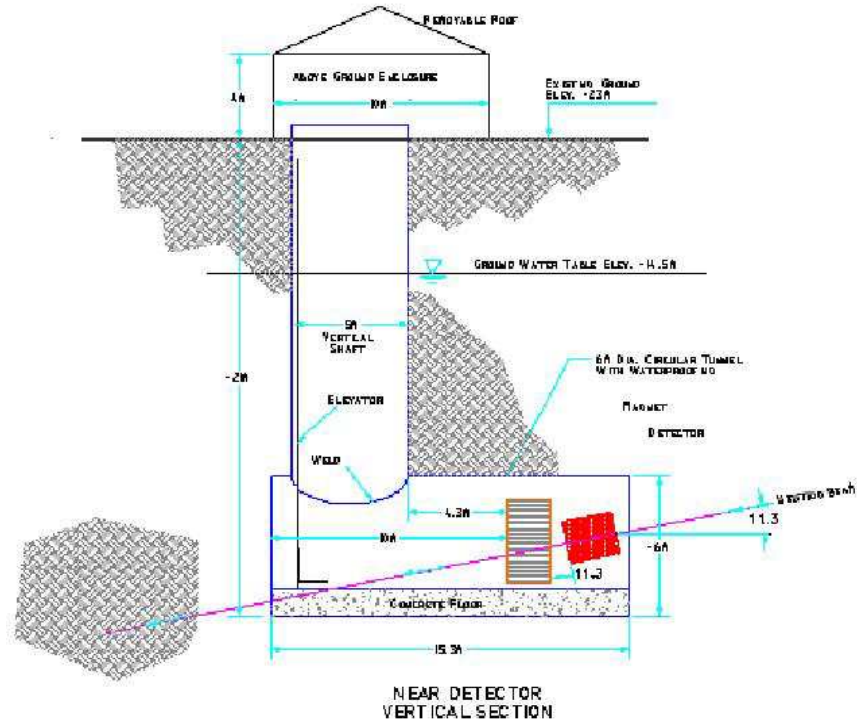}}
\caption{Near detector vertical section.} \label{fig:neardect}
\end{figure}

The near detector facility will be located 285 m from the target,
21 m below ground level. The near detector facility is depicted in
elevation view in Figure ~\ref{fig:neardect} The facility will
consist of welded steel tunnel sections 6 m in diameter, with a 5
m access shaft to the surface. A 9 m wide x 15 m long service
building will be constructed over the access shaft with a
removable roof to access equipment below with a mobile crane. This
building will contain electrical distribution, HVAC units, water
cooling systems, power supplies and experimental equipment
associated with the detector. An elevator will be provided for
accessing the detector from the service building. A unit
substation and cooling tower will be provided for utilities in
this area. Estimated requirements are:

Power -------- 1000 KVA

Cooling ------ 500 kW

Since the facility is located below the water table, installation
of the tunnel sections will involve excavation of soil to the
water table, installation of sheet piling, excavation and
de-watering of the sheet pile site, and installation of the tunnel
sections. The tunnel is then seal welded and back-filled with
soil.

\lhead{AGS Super Neutrino Beam Facility} \rhead{Radiation
Shielding} \rfoot{April 15, 2003}

\subsection{Radiation Shielding for the Super Neutrino Beam}
\label{sec:shield}

The conceptual shielding design for the Super Neutrino Beam meets
the necessary standards for radiation protection and provides a
basis for the initial cost estimates for the project. The design
has been developed using existing reports, calculations, and
simple analytic techniques to scale to the proposed facilities for
the proton transport, target station, decay tunnel, and beam stop.
More detailed modeling will be conducted when the design of the
various components are more mature. The shielding has been
designed to:
\begin{itemize}
\item Meet standards for chronic exposure to people at adjacent
areas and off-site.

\item Prevent contamination of the ground water.

\item Reduce direct exposure sufficiently so that personnel can
service equipment at the facility during beam operations.
\end{itemize}

A brief description of the methods and assumptions used will be
given in this section.

\lhead{AGS Super Neutrino Beam Facility} \rhead{Source Terms}
\rfoot{April 15, 2003}

\subsubsection{Source Terms}

The shielding design requires estimations of the source terms and
their locations. Logically, the Super Neutrino Beam breaks into
four simple areas, proton transport from the AGS to the target
station, target station, decay tunnel, and the beam stop. The
proton beam transport is expected to have an integrated loss of
0.1{\%} with less than 0.03{\%} lost in any one location. The
target station has been designed assuming 100{\%} of the beam
interacts in the target. It is assumed that no beam interacts with
the helium in the decay tunnel. In reality, approximately 1{\%} of
the incident beam will survive the target and interact in the
helium. These neglected interactions are compensated for by the
assumption that  100{\%} of the beam interacts in the target.
20{\%} of the beam will survive the target and strike the beam
dump.

An MCNPX calculation by A. Stevens \cite{ref:sh1} has been used to
estimate the initial radiation pattern in the near shield. The
calculation assumed that the 25 GeV proton beam struck an iron
target 1-meter long and 3 cm in radius. This target was placed in
a tunnel with a 1.5 meter radius and surrounded by soil. The
initial radiation pattern has then been extrapolated using the
analytic techniques of Tesch\cite{ref:sh2} and discussed by A.H.
Sullivan\cite{ref:sh3}. This method was used for the beam
transport, target station, and decay tunnel.

The beam stop was designed by using an existing
design\cite{ref:sh4} of a 29 GeV beam dump in the AGS slow beam
area. The analytic techniques were then used to extrapolated were
necessary.

\lhead{AGS Super Neutrino Beam Facility} \rhead{Ground Water
Protection} \rfoot{April 15, 2003}

\subsubsection{Ground  Water  Protection }

The interaction of high-energy protons will create radioactive
products in the surrounding shield. It is important that these
radioactive products are not transported to the water table. The
BNL standard\cite{ref:sh5} requires that the design of this
facility prevent radioactive products in the groundwater from
exceeding 5{\%} of the drinking water standard. The use of modular
concrete and steel shielding for the entire project is cost
prohibitive. Most of the facility will have an earthen shield.
Modular shielding will be used at the locations of highest loss
and when possible to reduce the amount of radioactive products
created in the surrounding soil. The design standard allows for
the use of approved capping materials to prevent created
radioactive products from being transported by rainwater to the
water table. Caps constructed of geomembranes or concrete will be
used to meet the BNL standard for groundwater protection.

\bigskip

$^{22}$Na is the radioactive isotope that will determine the
design of the cap which protects the groundwater. Based on the
transport model used in the BNL standard, the cap must prevent
rainwater from leaching the soil where the created $^{22}$Na in a
year exceeds 1.05*10$^{6} \quad ^{22}$Na atoms/cc. This
corresponds to a neutron flux (E$\geq$20 MeV) of 9.4*10$^{ - 13}$
neutrons/cm$^{2}$ per incident proton.

The loss assumption for the proton transport has a maximum local
chronic loss of 0.03{\%} of the proton beam. The cap must have at
least 4.6 meters of soil between it and the tunnel wall. For the
present conceptual design the membrane has been placed at 5.5
meters from the tunnel wall giving a factor of 5 more reduction
then required by the standard. The membrane is covered with 0.6
meters of earth.

It is assumed that 100{\%} of the beam interacts at the target
station. If it were sited inside a tunnel shielded by dirt the
liner would need to be greater than 9 meters from the target. The
total amount of activation in the soil would be substantial. It
was decided that the target area should have the shield
constructed from modular steel and concrete blocks. The target
area shield will have a thickness equivalent to 0.6 m of steel and
4.3 m of heavy concrete. The target area will be inside a
building. The concrete floor will prevent rainwater from leaching
activation products from the soil underneath the target station.
The roof of the target building will prevent rainwater from
entering the concrete and steel shielding. Additional shielding
will be placed underneath the target and horns to reduce the
activation of soil underneath the concrete floor as shown in
Figure ~\ref{fig:targetarea}.

The decay tunnel has been designed with the membrane 8.5 meters
from the tunnel wall. The membrane will have 0.6 meters of soil
overburden. At this stage of the design it was decide to be
conservative in the estimate for the decay region. It is expected
when the final design of the target area is complete that the
membrane can be closer to the tunnel and the berm thickness
decreased.

The beam dump is constructed from plates of steel. The dimensions
of the steel were chosen to reduce the neutron flux sufficiently
such that the activation of the soil in contact with the edges of
the steel satisfies the BNL standard. This requires the iron beam
dump to have a radius of 3.4 meters and a length of 9.2 meters.
The water cap for the decay tunnel will be extended over the beam
dump, as shown in Figure ~\ref{fig:bsatdt}. This is necessary to
prevent water from percolating through the beam dump.

There are four monitoring wells planned for the Super Neutrino
Beam area. The beam dump, decay tunnel, target station, and proton
transport will each have an appropriately sited monitoring well.
In addition, existing wells will monitor the portion of the proton
transport, which is close to the AGS.

\lhead{AGS Super Neutrino Beam Facility} \rhead{Sky Shine}
\rfoot{April 15, 2003}

\subsubsection{Sky Shine}

Dose to adjacent facilities occurs due to leakage of neutrons
through the shielding. The dose to people at adjacent facilities
is calculated using the sky shine formulation discussed by A.H.
Sullivan\cite{ref:sh3}. The exposure at a location more than 100
meters from the source is given by:
\begin{equation}
H = 70 (H_{0}A)\exp(-R/600)/(R^{2}),
\end{equation}
where H is the dose rate in mrem/hr, H$_{0}$A is the integrated
dose through the berm in rem-m$^{2}$ per hour, and R is the
distance from the source to the occupied location in meters.

The limits are established by integrating the exposure over a year
of operation. In addition occupancy factors can be used were
appropriate. Five locations were examined for the yearly exposure
from the super neutrino beam. The assumptions on occupancy and
exposure limits are given in the Table ~\ref{tab:sheilding}

\begin{table}[htbp]
\begin{center}
\caption{Occupancy and exposure limits.}
\begin{tabular}
{|c|c|c|c|} \hline Location& Occupancy &Occupancy per& Facility
limit\\ & hours/week&  work hour&   mrem/yr \\ \hline Off-Site&
168& 1& 5 \\ \hline Linac& 40& 1& 100
\\ \hline Blip& 40& 0.25& 25 \\ \hline BAF& 40& 0.25& 25 \\ \hline
Bldg. 1005& 40& 1& 25 \\ \hline
\end{tabular}
\label{tab:sheilding}
\end{center}
\end{table}

Based on these occupancy assumptions the exposure limit for
off-site is the most stringent for all loss locations. The target
station dominates the contribution to off-site dose and is less
than 10{\%} of the allowed limit.

\lhead{AGS Super Neutrino Beam Facility} \rhead{Direct Exposure}
\rfoot{April 15, 2003}

\subsubsection{Direct  Exposure }

Dose rates external to the shielding will typically be less than a
few mrem/hr under normal operating conditions. The highest
localized dose rates occur at the target building and are less
than 60 mrem/hr . Barriers, postings and procedures will control
access to the shielding. Abnormal operating conditions can cause
substantial increases in the dose rates. Elements, which can cause
abnormal conditions, will be monitored and a fast beam interrupt
will limit the duration of the abnormal condition. Radiation
detectors will be distributed along the facility to prevent
personnel exposure and limit fault conditions. A dual PLC based
access control system will prevent access to the transport tunnel,
except under allowed conditions.

\clearpage
\newpage
\lhead{AGS Super Neutrino Beam Facility} \rhead{Cost Estimate and
Schedule} \rfoot{April 15, 2003}

\section{Cost Estimate and Schedule}

\label{sec:costsched}
 \setcounter{table}{0}
 \setcounter{figure}{0}
 \setcounter{equation}{0}

\lhead{AGS Super Neutrino Beam Facility} \rhead{Cost Estimate and
Schedule} \rfoot{April 15, 2003}

The cost estimate of the AGS-Based super Neutrino beam Facility
has been performed following the "bottoms up" approach. After the
performance goals and physics design of each technical systems are
completed, cognizant engineers who have built similar systems were
assigned to prepare a cost estimate of sufficient detail and based
on previous experience. Each estimate typically includes the cost
of the detailed engineering design, procurement, manufacturing,
testing and installation. Material and labor costs are captured as
separate entries for each process.

Most of the cost numbers are based on BNL's recent experience in
building RHIC, SNS ring, and LHC magnets. They also draw on
extensive searches of price given in catalogs and on vendor's
quotations. Although, a formal Work Breakdown Structure (WBS) has
not been implemented yet, all systems have been estimated by going
down to the component and assembly level. From our past
experience, both the manufacturing approaches and the cost
estimates are consistent with good engineering practices and are
credible.

The costs given reflect the 'direct cost" of the item. In other
words, estimates do not include G {\&} A (lab overhead),
contingency and ED {\&}I. These cost elements are added for the
direct cost following guidelines using standard estimating.

All the figures are in FY 2003 US Dollars. No inflation is
included since the construction period is not known at this time.

A preliminary estimate of the direct cost of the accelerator
system is given Table ~\ref{tab:costone}.

\begin{table}[htbp]
\begin{center}
\caption{Accelerator system cost.}
\begin{tabular}
{|p{121pt}|p{58pt}|} \hline \textbf{1.2 GeV SC Linac:}&
 \\
\hline Front End& {\$}2.5 M \\ \hline LE SC Linac& {\$}38.3 M \\
\hline ME SC Linac& {\$}30.7 M \\ \hline HE SC Linac& {\$}28.1 M
\\ \hline \textbf{AGS Upgrades:}&
 \\
\hline AGS Power Supply& {\$}32.0 M \\ \hline AGS RF Upgrade&
{\$}8.6 M \\ \hline AGS Injection Channel& {\$}3.7 M \\ \hline
Full Turn Extraction& {\$}5.5 M \\ \hline Shielding& {\$}3.2 M
\\ \hline Installation& {\$}4.2 M \\ \hline \textbf{Total Direct
Cost}& \textbf{{\$}156.8 M} \\ \hline
\end{tabular}
\label{tab:costone}
\end{center}
\end{table}

The direct cost of the neutrino beam systems is summarized in
Table ~\ref{tab:costtwo}.

\begin{table}[htbp]
\begin{center}
\caption{ Estimated neutrino beam cost.}
\begin{tabular}
{|p{103pt}|p{76pt}|p{58pt}|} \hline \textbf{Item}& \textbf{Basis}&
\textbf{Cost} \\ \hline Proton Transport& RHIC Injector& {\$}14.8
M
\\ \hline Target/Horn& E889& {\$}5.5 M \\ \hline Shielding/Dump&
New& {\$}5.8 M \\ \hline Decay Tunnel& E889& {\$}0.4 M \\ \hline
Hill. Const.& New& {\$}13.0 M \\ \hline Near Detector Vault& RHIC&
{\$}7.3 M \\ \hline Conventional Facil.& RHIC& {\$}7.5 M \\ \hline
Other Const.& New& {\$}2.2 M \\ \hline Installation& RHIC& {\$}5.2
M
\\ \hline \textbf{Total Direct Cost}&
 &
\textbf{{\$}61.7 M} \\ \hline
\end{tabular}
\label{tab:costtwo}
\end{center}
\end{table}

The resultant total direct cost of the 1 MW AGS super Neutrino
Beam facility, not including both near and main detectors, is
about \$ 218.5 M. More detailed cost estimate for each system are
given in Appendix B. The preliminary total estimated cost (TEC) is
{\$}369 M in FY03 dollars, including EDIA {\@} 15 {\%};
contingency {\@} 30 {\%}; BNL project overhead {\@} 13 {\%}.
Escalation cannot be estimated without a project start year.

It is estimated that three years of R \& D are needed to build
prototypes and perform  detail engineering designs to reduce cost
 and improve operation reliability. This will be followed by 4.5 years
of construction and 0.5 year of commissioning to get this facility
ready for physics research. The planned schedule is  represented
in Table ~\ref{tab:costthree}. Where FY1 represents the first year
of project approval.

\begin{table}[htbp]
\begin{center}
\caption{ Schedule.}
\begin{tabular}
{|c|c|c|c|c|c|c|c|c|c|} \hline
 & \multicolumn{3}{|c|}{R\&D} & \multicolumn{5}{|c|}{Construction}& Research \\
 \hline
 & FY1 &2 & 3 & 4& 5& 6& 7& 8& 9- 15 \\
 \hline
 R\&D & XX & XX & XX & & & & & &  \\
 \hline
 Construction & & & & XX & XX & XX & XX & X~~ &  \\
 \hline
 Commissioning & & & & & & & &~~X&  \\
 \hline
 Research & & & & & & & & &  XXXX \\
 \hline
\end{tabular}
\label{tab:costthree}
\end{center}
\end{table}

\clearpage
\newpage
\lhead{AGS Super Neutrino Beam Facility} \rhead{Appendix A}
\rfoot{April 15, 2003}

\appendix
\section{Appendix A: Design Parameters}
 \label{sec:appendixA}

\lhead{AGS Super Neutrino Beam Facility} \rhead{Design Parameters}
\rfoot{April 15, 2003}

\subsection{Facility Level Parameters}
\begin{table}[htbp]
\begin{center}
\caption{\label{tab:flp} Facility level parameters. }
\begin{tabular}{|p{220pt}|p{100pt}|}
 \hline ~~~ Proton Beam Energy & 28 GeV \\ \hline ~~~ Protons per
Pulse  & 8.9$\times 10^{ 13 }$ \\ \hline ~~~ Average Beam Current
&  37.5 $\mu$A \\ \hline ~~~ Repetition Rate & 2.5 Hz
\\ \hline ~~~ Pulse Length & 2.58 $\mu$sec \\ \hline ~~~ Number of
Bunches & 23 \\ \hline ~~~ Protons per Bunch & 3.87$\times 10^{ 12
}$
\\ \hline ~~~ AGS Circumference & 807.1 m \\ \hline
 ~~~ Bunch Length & 40 nsec \\  \hline~~~ Bunch Spacing & 60
nsec \\ \hline ~~~ Extraction Gap & 160 nsec \\ \hline ~~~ Average
Beam Power & 1 MW \\ \hline ~~~ Normalized Emittance-X & 100 $\pi$
mm-mrad
\\ \hline ~~~ Normalized Emittance-Y &100 $\pi$ mm-mrad \\ \hline ~~~ Longitudinal
Emittance & 5.0 eV-sec\\ \hline ~~~ Energy Spread ($\Delta$E/E)&
0.001
\\ \hline ~~~ Target Material & carbon-carbon \\ \hline ~~~ Target
Radius &6.0 mm \\ \hline ~~~ Target Length & 80 cm \\ \hline ~~~
Beam Size on Target & 2 mm (rms)\\  \hline~~~ Beam Elevation  at
Target & 43 m \\ \hline ~~~ Decay Tunnel Length  &200 m
\\  \hline~~~ Beam Dump Length  & 20 m \\
\hline ~~~ Distance of Near Detector from Target & 275 m \\ \hline
 ~~~ Physics Time per Year & 10$^7$ sec/year
\\ \hline ~~~ Number of Protons on Target  & 2.2$\times 10^{21}$
p/year
\\

\hline
\end{tabular}
\end{center}
\end{table}
 \clearpage
\newpage
\lhead{AGS Super Neutrino Beam Facility} \rhead{Front End and Warm
Linac Parameters} \rfoot{April 15, 2003}

\subsection{Front End and Warm Linac Parameters}
\begin{table}[htbp]
\begin{center}
\caption{\label{tab:fewl} Front end and warm linac parameters.}
\begin{tabular}{|p{220pt}|p{100pt}|}
 \hline ~~ Energy of Warm Linac & 200 MeV
\\\hline
~~ Normalized Horizontal Emittance (rms) & 2.0 $\pi$ mm-mrad  \\\hline
~~ Normalized Vertical Emittance (rms) & 2.0 $\pi$ mm-mrad\\
\hline ~~ Longitudinal Emittance (rms)& 0.125 MeV-deg \\ \hline
 \hline~~ Macro-Pulse Average Current & 21 mA
\\ \hline~~ Macro-Pulse Peak Current  & 28 mA \\
 \hline~~ Repetition  Rate & 2.5 Hz
\\ \hline~~ Pulse Length & 720 $\mu$sec \\
\hline~~ Chopping Rate & 0.75 \%\\
 \hline ~~ Energy Spread ($\Delta$E/E)& 0.002
\\ \hline~~ Energy Jitter($\delta$E/E) & 0.001 \\ \hline
\end{tabular}
\end{center}
\end{table}
\clearpage
\newpage
\lhead{AGS Super Neutrino Beam Facility} \rhead{SCL Parameters}
\rfoot{April 15, 2003} \subsection{Superconducting Linac
Parameters}
\begin{table}[htbp]
\begin{center}
\caption{\label{tab:sclp} Superconducting linac parameters. }
\begin{tabular}
{|p{199pt}|p{54pt}|p{58pt}|p{55pt}|} \hline Linac Section& LE& ME&
HE \\ \hline

Kinetic Energy Initial/Final, MeV& 200/400& 400/800& 800/1200 \\
\hline

Frequency, MHz& 805 & 1610& 1610 \\ \hline No. of Protons / Bunch
x 10$^{8}$& 8.70 & 8.70& 8.70 \\ \hline Temperature, $^{o}$K& 2.1&
2.1& 2.1 \\ \hline Cells / Cavity& 8& 8& 8 \\ \hline Cavities /
Cryo-Module& 4& 4& 4
\\ \hline Cavity Separation, cm& 32.0 & 16.0& 16.0 \\ \hline
Cold-Warm Transition, cm& 30 & 30& 30 \\ \hline Cavity Internal
Diameter, cm& 10& 5& 5 \\ \hline Cell Reference $\beta _{0}$&
0.615& 0.755& 0.851 \\ \hline Cell Length, cm& 11.45& 7.03& 7.92
\\ \hline Length of Warm Insertion, m& 1.079& 1.379& 1.379 \\
\hline Total No. of Periods& 6& 9& 8 \\ \hline Length of a period,
m& 6.304& 4.708& 4.994 \\ \hline Total Length, m& 37.82& 42.38&
39.96 \\ \hline Cavities / Klystron& 1& 1& 1 \\ \hline No. of rf
Couplers / Cavity& 1& 1& 1 \\ \hline Coupler rf Power, kW (*)&
263& 351& 395
\\ \hline Total No. of Klystrons& 24& 36& 32 \\ \hline
Z$_{0}$T$_{0}^{2}$, $^{ }$ohm/m& 378.2& 570.0& 724.2 \\ \hline
Q$_{0 }$ x 10$^{10}$& 0.97 & 0.57& 0.64 \\ \hline Ave. Axial
Field, E$_{a}$, MV/m& 13.4& 29.1& 29.0 \\ \hline Ave. Dissipated
Power, W& 2& 11& 8 \\ \hline Ave. HOM-Power, W& 0.2& 0.5& 0.4 \\
\hline Ave. Cryogenic Power, W& 65& 42& 38 \\ \hline Total Ave. RF
Power, kW (*)& 17& 31& 30 \\




\hline Norm. rms Emittance, $\pi $ mm-mrad& 2.0& 2.0 & 2.0  \\
\hline Rms Bunch Area, $\pi  \quad ^{o}$MeV (805 MHz)& 0.5 & 0.5 &
0.5  \\ \hline \multicolumn{4}{p{366pt}}{\footnotesize (*)
Including 50{\%} RF power contingency.}
\end{tabular}
\end{center}
\end{table}
\clearpage
\newpage
\lhead{AGS Super Neutrino Beam Facility} \rhead{AGS Parameters}
\rfoot{April 15, 2003} \subsection{AGS Parameters}
\begin{table}[htbp]
\begin{center}
\caption{\label{tab:agsp}  AGS parameters. }
\begin{tabular}{|p{170pt}|p{170pt}|} \hline

~~~ Injection Energy & 1.2 GeV \\ \hline~~~ Extraction Energy & 28
GeV
\\ \hline~~~ Number of Injection Turns & 240 \\ \hline~~~ Stripping Efficiency & 98{\%}
\\ \hline~~~ Beam Bump  & 25 kW \\ \hline~~~ Electron Collection  & 25 Watts \\ \hline~~~ Number of
Protons & 8.9 $\times 10^{13}$ \\ \hline~~~ Harmonic Number &24 \\
\hline~~~ Filled Bucket & 23 \\ \hline~~~ Repetition Rate & 2.5 Hz
\\ \hline~~~ Strip Foil & carbon-carbon 300 $\mu$g /cm$^{2}$\\
\hline~~~ RF Frequency at Injection & 8.0 MHz
\\ \hline~~~  RF Frequency at Extraction & 12.0 MHz \\ \hline~~~  Peak RF
Voltage & 1.0 MV \\ \hline~~~ Transition Gamma  & 8.5 \\ \hline
~~~ Horizontal Tune & 8.7 \\ \hline~~~ Vertical Tune  & 8.9 \\
\hline~~~ Beta Maximum & 22.5 m
\\ \hline~~~  Beta Minimum & 11.5 m \\ \hline~~~  Horizontal Aperture & 7 cm\\ \hline~~~
Vertical Aperture & 7 cm
\\ \hline~~~  Power Supply & 4300 A, 25 kV\\
 \hline
\end{tabular}
\end{center}
\end{table}
\clearpage
\newpage
\lhead{AGS Super Neutrino Beam Facility} \rhead{Target/Horn
Parameters} \rfoot{April 15, 2003} \subsection{Target/Horn
Parameters}
\begin{table}[htbp]
\begin{center}
\caption{\label{tab:thp}  Target/Horn parameters. }
\begin{tabular}{|p{170pt}|p{170pt}|}
 \hline ~~~ Target Material & carbon-carbon \\ \hline~~~ Target Radius
& 6.0 mm \\ \hline~~~ Target Length & 80 cm \\ \hline~~~ Heat
Deposition  & 7.3 kJ/pulse\\ \hline~~~ Cooling Medium & He Gas \\
\hline~~~ Operating Peak Temperature & 800$^{\circ}$C \\ \hline~~~
Horn Small Radius & 70 mm
\\ \hline~~~ Horn Large Radius & 610 mm \\ \hline~~~ Horn Thickness
& 2.5 mm \\ \hline~~~ Horn Length & 217 cm
\\ \hline~~~ Horn Current & 250 kA \\ \hline~~~ Horn Peak Current Density  &690 kA/cm$^{2}$ \\
 \hline~~~
Repetition  Rate &  2.5 Hz \\ \hline~~~ Power Supply Wave Form &
Sinusoidal, base width 1.20 msec \\ \hline~~~ Power Supply Voltage
& 5 kV
\\ \hline~~~ Average Power &  700 kW \\
 \hline ~~~ Heat Deposition on Horn & 3 kJ/Pulse\\
\hline ~~~ Cooling Medium of Horn & Water \\ \hline
\end{tabular}
\end{center}
\end{table}
\clearpage
\newpage
\lhead{AGS Super Neutrino Beam Facility} \rhead{Decay Tunnel and
Shielding ...} \rfoot{April 15, 2003} \subsection{Decay Tunnel and
Shielding  Parameters}
\begin{table}[htbp]
\begin{center}
\caption{\label{tab:dtsp} The Decay tunnel~ and shielding
parameters.}
\begin{tabular}{|p{170pt}|p{140pt}|}
 \hline  ~~~ Decay Tunnel
Radius & 1 m \\ \hline ~~~ Decay Tunnel Length & 200 m
\\ \hline ~~~ Beam Dump Material & Steel \\ \hline ~~~ Beam Dump Length  &
9 m\\ \hline ~~~ Shielding of Target Bldg. & 0.6 m Steel and 4.3 m
of Heavy Concrete \\ \hline ~~~ Shielding of Decay Tunnel & 9 m
Soil
\\
 \hline
\end{tabular}
\end{center}
\end{table}
\clearpage
\newpage
\lhead{AGS Super Neutrino Beam Facility} \rhead{Conventional
Facilities ...} \rfoot{April 15, 2003} \subsection{Conventional
Facilities and Target Hill Parameters}
\begin{table}[htbp]
\begin{center}
\caption{\label{tab:cfthp}  Conventional facilities and target
hill parameters.}
\begin{tabular}{|p{230pt}|p{100pt}|}
 \hline ~~~ The Linac Tunnel Length & 120 m \\ \hline ~~~ Linac Tunnel
Diameter & 3 m  \\ \hline ~~~ Klystron Gallery Length & 120 m
\\ \hline ~~~ Klystron Gallery Dimension & 22 m x 12 m (W*H)\\ \hline ~~~ SRF
Testing and Assembly Building & 12 m x 36 m \\ \hline ~~~
Shielding of Linac Tunnel & 3 m Soil \\ \hline ~~~ Linac Service
Building & 12 m x 22 m
\\ \hline ~~~ Beam Transport Tunnel & 330 m \\ \hline ~~~ Beam
Transport Tunnel Diameter &3 m
\\ \hline ~~~ Shielding of Beam Transport Tunnel & 6 m Soil \\ \hline ~~~ Beam
Transport Service Buildings & 2x(250 m$^2$)\\ \hline ~~~ Beam
Elevation at Target & 43 m \\ \hline ~~~ Target Hill Height & 52
m\\ \hline ~~~ Target Hill Base Width & 154 m
\\ \hline ~~~   UP Hill Length to  Target & 200 m \\ \hline ~~~  Down Hill
Length from Target  & 220 m \\ \hline ~~~ Target Service Building
& 16.5 m x 15 m
\\ \hline ~~~ Electrical Substation & 6100 KVA \\ \hline ~~~ Water
Cooling Capacity  & 3.5 MW\\ \hline ~~~ Service Road & 250 m\\
\hline
\end{tabular}
\end{center}
\end{table}

\clearpage
\newpage

\lhead{AGS Super Neutrino Beam Facility} \rhead{Appendix B}
\rfoot{April 15, 2003}

\section{Appendix B: System Cost Estimate}
 \label{sec:appendixB}

\lhead{AGS Super Neutrino Beam Facility} \rhead{System Cost
Estimate} \rfoot{April 15, 2003}


\subsection{Linac Upgrade Cost }

\label{sec:linacupcost}
 \setcounter{table}{0}
 \setcounter{figure}{0}
 \setcounter{equation}{0}
\begin{table}[htbp]
\begin{center}
\caption{\label{tab:lupc}Linac upgrade cost. }
\begin{tabular}{|p{250pt}|p{100pt}|}

\hline

\textbf{Direct Costs} & \textbf{k{\$}}  \\

\hline \hline

Ionsource Power Supplies & 80 \\

\hline

Driver Stage RF Systems & 123 \\

\hline

RF Modulator PS's & 23 \\

\hline

High Power Cap Banks & 730 \\

\hline

Crowbar System & 76 \\

\hline

Low Level RF Amps  & 102 \\

\hline

Quadrupole Pulsers & 1016 \\

\hline

Misc. Cabling, etc. & 27 \\

\hline Labor & 325 \\ \hline
 \textbf{TOTAL DIRECT} & \textbf{2,542} \\
 \hline

\end{tabular}
\end{center}
\end{table}


\clearpage
\newpage
\lhead{AGS Super Neutrino Beam Facility} \rhead{SCL Cost}
\rfoot{April 15, 2003} \subsection{SCL Cost }

\label{sec:sclcost}
\begin{table}[htbp]
\begin{center}
\caption{\label{tab:cop}Cost ('02 {\$}) and other parameters.}
\begin{tabular}
{|p{123pt}|p{121pt}|p{121pt}|}
 \hline & LE-Section& ME- {\&} HE-Sections \\ \hline
 Frequency,MHz& 805 & 1,610 \\ \hline
 AC-to-RF Efficiency& \multicolumn{2}{|p{242pt}|} {0.45 for Pulsed Mode}  \\
  \hline Cryogenic Efficiency &\multicolumn{2}{|p{242pt}|}{0.0014 @ 2.1 $^{o}$K}  \\ \hline
Electricity Cost& \multicolumn{2}{|p{242pt}|}{0.05 {\$}/kWh}  \\
\hline Linac Availability& \multicolumn{2}{|p{242pt}|}{75 {\%} of
yearly time}
\\  \hline Normal
Conduct. Cost& \multicolumn{2}{|p{242pt}|}{20 k{\$}/m 50
k{\$}/quad}  \\ \hline Superconducting Cost& 200 k{\$}/m \par 360
k{\$}/module & 100 k{\$}/m \par 200 k{\$}/module \\
 \hline
 Cost of Klystron, in k{\$}:& & \\
  Power Source & 170 &  160 \\
  Trans. + PS   & 140 &  130\\
   Circul. +Loads & 95 & 80\\
   Controls & 70  & 60 \\
   Waveguides & 25 &  20\\
   \textbf{Total}&\textbf{500/Unit}& \textbf{450/Unit}\\

 \hline
 Cost of Refrigeration:&\multicolumn{2}{|p{242pt}|}{ }  \\
Refrigeration  &\multicolumn{2}{|p{242pt}|}{ 8.0 M{\$}} \\
 Control &\multicolumn{2}{|p{242pt}|} {1.0 M{\$}}\\
 Ancillary &\multicolumn{2}{|p{242pt}|}{1.0 M{\$}}\\
 Labor &\multicolumn{2}{|p{242pt}|}{6.0 k{\$}/m}\\
Transfer Lines&\multicolumn{2}{|p{242pt}|}{ 18.0 k{\$}/m}
\\ \hline
Electrical Distribution& \multicolumn{2}{|p{242pt}|}{0.14 {\$}/W
(AC) and 10 k{\$}/m}  \\ \hline Control&
\multicolumn{2}{|p{242pt}|}{10{\%} of the Total Cost above}  \\
\hline Tunnel Cost& \multicolumn{2}{|p{242pt}|}{15 k{\$}/m (Linac)
and 40 k{\$}/m (Klystr.)}  \\ \hline
\end{tabular}
\end{center}
\end{table}

\begin{table}[htbp]
\begin{center}
\caption{\label{tab:cop1}Cost estimate of the SCL project.}
\begin{tabular}
{|p{197pt}|p{54pt}|p{54pt}|p{54pt}|} \hline Linac Section& LE& ME&
HE \\ \hline Capital Cost '02 M{\$}: \par RF Klystron
\par Electric Distr. (**) \par Refrigeration Plant \par Warm
Structure (**) \par Cold Structure \par Control System \par Tunnel
(**)& ~~~~~\par 12.000 \par 0.435 \par (+)10.908 \par 0.733 \par
8.429
\par 3.250
\par 2.560&~~~~~~~ \par 16.200 \par 0.438 \par 3.017 \par 1.156 \par 4.796
\par 2.561 \par 2.564&~~~~~ \par 14.400 \par 0.413 \par 2.959 \par 1.040
\par 4.492 \par 2.330 \par 2.417 \\
\hline \textbf{Total Cost, '02 M{\$}}& 38.315& 30.731& 28.052 \\
\hline \textbf{Operation Cost, '02 k{\$}/y }(***) & 54.6& 64.9&
61.4 \\ \hline \multicolumn{4}{p{359pt}}{\footnotesize(**)
Includes 4.5m long matching insertion between LE and ME sections}
\\ \multicolumn{4}{p{359pt}}{\footnotesize(***) Includes only electric bill for
RF and Cryogenic Power converted to AC}  \\
\multicolumn{4}{p{359pt}}{\footnotesize(+) Includes 8.0M{\$} for
Main Cryogenic Station}
  \\
\end{tabular}
\end{center}
\end{table}

\begin{table}[htbp]
\caption{\label{tab:l3scl}Level 3 of WBS and cost (in 2002 k{\$})
of total SCL. }

\begin{tabular}
{|p{50pt}|p{176pt}|p{54pt}|p{54pt}|p{59pt}|} \hline
\multicolumn{2}{|p{226pt}|}{\textbf{The 1.2-GeV Linac}} &
\textbf{Material}& \textbf{Labor}& \textbf{Total} \\ \hline 2.1.3&
Low-Energy SCL& 29,116& 9,199& 38,316
\\ \hline 2.1.4& Medium-Energy SCL& 23,047& 7,685& 30,732
\\ \hline 2.1.5& High-Energy SCL& 20,991& 7,060& 28,051 \\
 \hline & \textbf{Total}& \textbf{73,154}& \textbf{23,944}&
\textbf{97,099}
\\ \hline
\end{tabular}
\end{table}


\begin{table}[htbp]
\caption{\label{tab:l5lescl}Levels 4 and 5 of WBS and cost (in
2002 k{\$}) of LE-section of SCL.}

\begin{tabular}
{|p{42pt}|p{108pt}|p{78pt}|p{54pt}|p{54pt}|p{59pt}|} \hline &
\textbf{Low-Energy SCL}& & \textbf{Material}& \textbf{Labor}&
\textbf{Total} \\  \hline 2.1.3.1& \textit{Cryo-Modules}& &
\textbf{6,743}& \textbf{1,686}& \textbf{8,429} \\ \hline &
2.1.3.1.1& Tanks& 2,248& 562& 2,810 \\ \hline & 2.1.3.1.2& RF
Cavities& 4,495& 1,124& 5,619 \\ \hline 2.1.3.2&
\textit{Refrigeration (*)}& & \textbf{8,726}& \textbf{2,182}&
\textbf{10,908} \\ \hline 2.1.3.3& \textit{Warm Insertions}& &
\textbf{586}& \textbf{147}& \textbf{733} \\ \hline & 2.1.3.2.1&
Quadrupoles& 352& 88& 440 \\ \hline & 2.1.3.2.2& Steering& 59& 15&
73 \\ \hline & 2.1.3.2.3& BPM& 59& 15& 73 \\ \hline & 2.1.3.2.4&
Vacuum Sys.& 117& 29& 147
\\ \hline 2.1.3.4& \textit{RF Power Supply}& & \textbf{9,600}&
\textbf{2,400}& \textbf{12,000} \\ \hline & 2.1.3.3.1& Klystrons&
2,880& 720& 3,600 \\ \hline & 2.1.3.3.2& Transmitters& 2,304& 576&
2,880 \\ \hline & 2.1.3.3.3& Circulators& 1,824& 456& 2,280 \\
\hline & 2.1.3.3.4& Controls& 1,344& 336& 1,680 \\ \hline &
2.1.3.3.5& Waveguides& 480& 120& 600 \\ \hline & 2.1.3.3.6&
Couplers& 768& 192& 960 \\ \hline 2.1.3.6& \textit{Electrical
Distribut.}& & \textbf{348}& \textbf{87}& \textbf{435} \\ \hline
2.1.3.7& \textit{Control}& & \textbf{2,600}& \textbf{650}&
\textbf{3,251} \\ \hline 2.1.3.8& \textit{Tunnel}& & \textbf{512}&
\textbf{2,048}& \textbf{2,560} \\ \hline 2.1.3& & \textbf{Total}&
\textbf{29,116}& \textbf{9,199}& \textbf{38,316}
\\ \hline
\end{tabular}
\end{table}

\begin{table}[htbp]
\caption{\label{tab:l5mescl}Levels 4 and 5 of WBS and cost (in
2002 k{\$}) of ME-section of SCL. }

\begin{tabular}
{|p{45pt}|p{109pt}|p{77pt}|p{54pt}|p{54pt}|p{59pt}|} \hline &
\textbf{Medium-Energy SCL}& & \textbf{Material}& \textbf{Labor}&
\textbf{Total} \\  \hline 2.1.4.1& \textit{Cryo-Modules}& &
\textbf{3,837}& \textbf{959}& \textbf{4,796} \\ \hline &
2.1.3.1.1& Tanks& 1,279& 320& 1,599 \\ \hline & 2.1.3.1.2& RF
Cavities& 2,558& 639& 3,197
\\ \hline 2.1.4.2& \textit{Refrigeration}& & \textbf{2,414}&
\textbf{603}& \textbf{3,017} \\ \hline 2.1.4.3& \textit{Warm
Insertions}& & \textbf{925}& \textbf{231}& \textbf{1,156} \\
\hline & 2.1.3.2.1& Quadrupoles& 555& 139& 694 \\ \hline &
2.1.3.2.2& Steering& 92& 23& 116 \\ \hline & 2.1.3.2.3& BPM& 92&
23& 116 \\ \hline & 2.1.3.2.4& Vacuum Sys.& 185& 46& 231 \\ \hline
2.1.4.4& \textit{RF Power Supply}& & \textbf{12,960}&
\textbf{3,240}& \textbf{16,200} \\ \hline & 2.1.3.3.1& Klystrons&
3,888& 972& 4,860 \\ \hline & 2.1.3.3.2& Transmitters& 3,110& 778&
3,888 \\ \hline & 2.1.3.3.3& Circulators& 2,462& 616& 3,078 \\
\hline & 2.1.3.3.4& Controls& 1,814& 454& 2,268 \\ \hline &
2.1.3.3.5& Waveguides& 648& 162& 810 \\ \hline & 2.1.3.3.6&
Couplers& 1,037& 259& 1,296 \\ \hline 2.1.4.5& \textit{Electrical
Distribut.}& & \textbf{350}& \textbf{88}& \textbf{438} \\ \hline
2.1.4.6& \textit{Control}& & \textbf{2,049}& \textbf{512}&
\textbf{2,561} \\ \hline 2.1.4.7& \textit{Tunnel}& & \textbf{513}&
\textbf{2,051}& \textbf{2,564} \\ \hline 2.1.4& & \textbf{Total}&
\textbf{23,047}& \textbf{7,685}& \textbf{30,732}
\\ \hline
\end{tabular}
\end{table}

\begin{table}[htbp]
\caption{\label{tab:l5hescl}Levels 4 and 5 of WBS and cost (in
2002 k{\$}) of HE-section of SCL.}

\begin{tabular}
{|p{48pt}|p{109pt}|p{76pt}|p{55pt}|p{54pt}|p{59pt}|} \hline &
\textbf{High-Energy SCL}& & \textbf{Material}& \textbf{Labor}&
\textbf{Total} \\  \hline 2.1.5.1& \textit{Cryo-Modules}& &
\textbf{3,594}& \textbf{898}& \textbf{4,492} \\ \hline &
2.1.3.1.1& Tanks& 1,198& 299& 1,497 \\ \hline & 2.1.3.1.2& RF
Cavities& 2,396& 599& 2,995
\\ \hline 2.1.5.2& \textit{Refrigeration}& & \textbf{2,367}&
\textbf{592}& \textbf{2,959} \\ \hline 2.1.5.3& \textit{Warm
Insertions}& & \textbf{832}& \textbf{208}& \textbf{1,040} \\
\hline & 2.1.3.2.1& Quadrupoles& 499& 125& 624 \\ \hline &
2.1.3.2.2& Steering& 83& 21& 104 \\ \hline & 2.1.3.2.3& BPM& 83&
21& 104 \\ \hline & 2.1.3.2.4& Vacuum Sys.& 166& 42& 208 \\ \hline
2.1.5.4& \textit{RF Power Supply}& & \textbf{11,520}&
\textbf{2,880}& \textbf{14,400} \\ \hline & 2.1.3.3.1& Klystrons&
3,456& 864& 4,320 \\ \hline & 2.1.3.3.2& Transmitters& 2,765& 691&
3,456 \\ \hline & 2.1.3.3.3& Circulators& 2,189& 547& 2,736 \\
\hline & 2.1.3.3.4& Controls& 1,613& 403& 2,016 \\ \hline &
2.1.3.3.5& Waveguides& 576& 144& 720 \\ \hline & 2.1.3.3.6&
Couplers& 922& 230& 1,152 \\ \hline 2.1.5.5& \textit{Electrical
Distribut.}& & \textbf{330}& \textbf{83}& \textbf{413} \\ \hline
2.1.5.6& \textit{Control}& & \textbf{1,864}& \textbf{466}&
\textbf{2,330} \\ \hline 2.1.5.7& \textit{Tunnel}& & \textbf{483}&
\textbf{1,934}& \textbf{2,417} \\ \hline 2.1.5& & \textbf{Total}&
\textbf{20,991}& \textbf{7,060}& \textbf{28,051}
\\ \hline
\end{tabular}
\end{table}

\clearpage
\newpage
\lhead{AGS Super Neutrino Beam Facility} \rhead{Conventional
Facilities Cost} \rfoot{April 15, 2003} \subsection{Conventional
Facilities and Target Hill Cost } \label{sec:cfcost}
\begin{table}[htbp]
\begin{center}
 \caption{ Conventional facilities and target hill
 cost.}
\begin{tabular}
 {|p{210pt}|p{90pt}|p{40pt}|p{55pt}|} \hline
\textbf{Conventional Construction}& Quantity& {\$}/unit& Cost
(k{\$}) \\ \hline Upgrade U-Line Shielding& 275'& 5500& 1512 \\
\hline Removals& LS& ~& 250 \\ \hline Fill& 726350 cu M& 18& 13075
\\ \hline Berm Stabilization& LS& ~& 800 \\ \hline Primary Beam
Tunnel {\&} Liner& 1100'& 1200& 1320 \\ \hline Decay Tunnel {\&}
Liner& 615'& 600& 369 \\ \hline Target Area Footings& 600 cu yds&
500& 300 \\ \hline Beam Dump Footings& 300 cu yds& 500& 150 \\
\hline Retaining Walls& 800 cu yds& 600& 480 \\ \hline Road
Improvements& LS& ~& 500 \\ \hline Target Area Shielding& 6200
tons& 350& 2170 \\ \hline Power/high {\&} low Voltage
Distribution& LS& ~& 4250 \\ \hline Water/Sanitary/Storm& LS& ~&
1350 \\ \hline Power Supply Houses& ~& ~& ~ \\ \hline Lower Beam
40 x 100& 4000 sq ft& 250& 1000 \\ \hline Upper Beam 20 x 80& 1600
sq ft& 250& 400 \\ \hline Target Building 50' x 55'& 2750 sq ft&
500& 1375 \\ \hline Near Detector Underground Facility& LS& ~&
6750
\\ \hline Near Detector Building 30' x40'& 1200& 300& 360 \\
\hline Cooling Tower 3.5 MW& LS& ~& 1850 \\ \hline Cooling Tower
.5 MW& LS& ~& 150 \\ \hline Monitoring Wells & 5& 25000& 125 \\
\hline \textbf{Conventional Total}& ~& ~& \textbf{38,536} \\
\hline
\end{tabular}
\label{tab:cfandthc1}
\end{center}
\end{table}

\begin{table}[htbp]
\begin{center}
 \caption{ Beam components costs.}
\begin{tabular}
 {|p{210pt}|p{90pt}|p{40pt}|p{55pt}|}
 \hline
\textbf{Beam Components}& Quantity& {\$}/unit& Cost (k{\$})
\\

\hline 39 Degrees Vertical Bend& 20 Dipoles & 100,000& 2000 \\
\hline 68.2 Degrees Horizontal Bend& 28 Dipoles& 100,000& 2800 \\
\hline Quads& 14& 90,000& 1260 \\ \hline H{\&}V Trim Magnets& 18&
45,000& 810 \\ \hline BPM's& 12& 25,000& 300 \\ \hline Current
Transformers& 2& 35,000& 70 \\ \hline Vacuum Components& 1200&
2000& 2400 \\ \hline Trim PS's& 18& 25,000& 450 \\ \hline Quad
PS's& 14& 75,000& 1050
\\ \hline Horiz. Bend PS& 1& 300,000& 300 \\ \hline Vert. Bend PS's&
3& 150,000& 450 \\ \hline PS {\&} Instr. Controls& 300 Devices&
8,000& 2400 \\ \hline Instrumentation -Flags/Loss Monitors/etc& 25
Devices& 30,000& 750 \\ \hline   Security Hardware& 3 Gates&
50,000& 150 \\ \hline Horn Power Supply& 1& ~& 3200 \\ \hline
Horns + spare horn1 {\&} Target& LS& ~& 538 \\ \hline
\textbf{Components Total}& ~& ~& \textbf{18,928} \\ \hline
\end{tabular}
\label{tab:cfandthc2}
\end{center}
\end{table}
\begin{table}[htbp]
\begin{center}
 \caption{ Installation cost.}
\begin{tabular}
 {|p{210pt}|p{90pt}|p{40pt}|p{55pt}|}
 \hline
\textbf{Installation}& Quantity& {\$}/unit& Cost (k{\$})
\\
  \hline Installation Materials&
100 Devices& 5000& 500 \\ \hline Remove/Install Beam Dump& 25,000
tons& 20& 500 \\ \hline Install Target Area Shielding& 6200 tons&
10& 62 \\ \hline Magnet Installation& 77& 10,000& 770 \\ \hline
Vacuum Installation& 1200 ft& 200& 240 \\ \hline Instr.
Installation& 40 Devices& 10,000& 400
\\ \hline Power {\&} Tray Installation& ~& 8 my& 800 \\ \hline
Security System Installation& ~& 6 my& 600 \\ \hline Horn
Installation& ~& 4 my& 400
\\ \hline \textbf{Installation Total}& ~& ~& \textbf{4,272} \\ \hline
\end{tabular}
\label{tab:cfandthc3}
\end{center}
\end{table}
\begin{table}[htbp]
\begin{center}
 \caption{Conventional facilities direct costs.}
\begin{tabular}
 {|p{210pt}|p{90pt}|p{40pt}|p{55pt}|}
 \hline
\textbf{Conventional Construction}& Cost (k{\$})
\\
 \hline
 Conventional & 38,536 \\ \hline
 Beam Components & 18,928 \\ \hline
 Installation & 4,272 \\
\hline \textbf{Total direct costs}& \textbf{61,736} \\ \hline
\end{tabular}
\label{tab:cfandthc4}
\end{center}
\end{table}



\clearpage
\newpage
\lhead{AGS Super Neutrino Beam Facility} \chead{}
\rhead{References} \rfoot{April 15, 2003}



\begin{thebibliography}{9}
\bibitem{agsbate} AGS{\_}BATE is a modification of the code BEAM. The code uses the
experimentally measured fields at the median plane of the AGS
magnets to ray trace the charged particles. The code provides also
transport matrices at any location along a trajectory and beam
parameters at any location along a closed orbit.
\end{thebibliography}
\end{document}